%% file: main.tex
\documentclass[a4paper,twocolumn,accepted=2026-03-25]{quantumarticle}
\pdfoutput=1

\usepackage{xcolor}
\usepackage{physics}
\usepackage{amssymb}
\usepackage{graphicx}
\usepackage{mathtools}
\usepackage{amsmath}
\usepackage{hyperref}
\usepackage{tikz}
\usetikzlibrary{calc}
\usetikzlibrary{arrows.meta}
\usepackage{xifthen}
\usepackage{xcolor} %
\usepackage{xparse}
\usepackage{cite}

\usepackage{amsfonts}
\usepackage{centernot}
\usepackage{dsfont}

\usepackage{amsthm}

\newtheorem{definition}{Definition}
\newtheorem{lemma}{Lemma}
\newtheorem{corollary}{Corollary}
\newtheorem{proposition}{Proposition}

\DeclarePairedDelimiter\floor{\lfloor}{\rfloor}
\DeclareMathOperator*{\argmin}{arg\,min}

\newcommand\equalhat{%
\let\savearraystretch\arraystretch
\renewcommand\arraystretch{0.3}
\begin{array}{c}
\stretchto{
    \scalerel*[\widthof{=}]{\wedge}
    {\rule{1ex}{3ex}}%
}{0.5ex}\\ 
=%
\end{array}
\let\arraystretch\savearraystretch
}

\hypersetup{
colorlinks = true,
urlcolor = blue,
linkcolor = blue,
citecolor = blue
}

\newcommand{\mpsmove}{left-right move}
\newcommand{\mpsmoves}{left-right moves}

\input{figures_in_paper_defs}

\colorlet{entcolfg}{blue!100}
\colorlet{entcolbg}{blue!20}

\colorlet{discolfg}{red!80!black}
\colorlet{discolbg}{red!10}

\colorlet{gencolfg}{cyan!75!black}
\colorlet{gencolbg}{cyan!10}

\def\flatgatesize{0.15}
\def\flatgateadvance{0.3}

\def\circuitRoundedCorners{1.5pt}

\newcommand{\TUM}{\affiliation{Technical University of Munich, TUM School of Natural Sciences, Physics Department, 85748 Garching, Germany}}
\newcommand{\MCQST}{\affiliation{Munich Center for Quantum Science and Technology (MCQST), Schellingstr. 4, 80799 M{\"u}nchen, Germany}}
\newcommand{\Nottingham}{\affiliation{School of Physics and Astronomy, University of Nottingham, Nottingham, NG7 2RD, UK}}
\newcommand{\CQNE}{\affiliation{Centre for the Mathematics and Theoretical Physics of Quantum Non-Equilibrium Systems, University of Nottingham, Nottingham, NG7 2RD, UK}}

\def\numdiag{{n_\mathrm{d}}}

\begin{document}

\author{Ra{\'u}l Morral-Yepes} \TUM \MCQST 
\author{Marc Langer} \TUM \MCQST
\author{Adam Gammon-Smith} \Nottingham \CQNE
\author{Barbara Kraus} \TUM \MCQST
\author{Frank Pollmann} \TUM \MCQST

\title{Disentangling strategies and entanglement transitions in unitary circuit games with matchgates}

\begin{abstract}
In unitary circuit games, two competing parties--an ``entangler" and a ``disentangler"--can induce an entanglement phase transition in a quantum many-body system. The transition occurs at a certain rate at which the disentangler acts. We analyze such games within the context of matchgate dynamics, which equivalently corresponds to evolutions of non-interacting fermions. We first investigate general entanglement properties of fermionic Gaussian states (FGS). We introduce a representation of FGS using a minimal matchgate circuit capable of preparing the state and derive an algorithm based on a generalized Yang-Baxter relation for updating this representation as unitary operations are applied. This representation enables us to define a natural disentangling procedure that reduces the number of gates in the circuit, thereby decreasing the entanglement contained in the system. 
We then explore different strategies to disentangle the systems and study the unitary circuit game in two different scenarios: with braiding gates, i.e., the intersection of Clifford gates and matchgates, and with generic matchgates.
For each model, we observe qualitatively different entanglement transitions, which we characterize both numerically and analytically.

\end{abstract}
\maketitle

\section{Introduction}

Phases of matter are a central concept in condensed matter physics, encompassing familiar examples like the phases of water as well as more exotic quantum phases such as spin liquids.
Understanding the stability of phases and the nature of phase transitions remains a fundamental goal in physics.  
In recent years, there has been growing interest in understanding which dynamical phases and phase transitions can arise in random quantum circuits~\cite{review_Fisher_2023}.
A widely studied framework in this context is the measurement-induced phase transition (MIPT), where measurements disrupt entanglement growth, driving a transition to an area-law phase at a critical measurement rate.
In generic systems, MIPTs describe the transition between volume-law and area-law entangled phases, typically characterized by a single critical point that separates the two regimes~\cite{MIPT1, MIPT2, MIPT3, MIPT4, MIPT5}. 
However, a qualitatively different behavior arises in monitored free fermion systems.
In such systems, even an infinitesimal measurement rate can destroy the volume-law phase~\cite{miptff1}.  
Nonetheless, certain conditions may stabilize phases with super-area-law scaling or give rise to a variety of distinct area-law entangled phases~\cite{miptff2, miptff12, miptff6, miptff25, miptff3, miptff5, miptff17, miptff7, miptff23, miptff4, miptff9, miptff24, miptff15, miptff26, miptff11, miptff13, miptff16, miptff8, miptff10, miptff18, miptff14, miptff19, miptff20, miptff21, miptff22, miptff27}.

An alternative to disentangling with measurements is the explicit construction of circuits consisting of two-qubit unitary gates that disentangle the state.  
For generic Haar-random states, however, this task remains exponentially hard in system size, even when the full state description is available. 
Identifying tractable instances of this problem connects naturally to the notion of entanglement complexity~\cite{entanglement_cooling, entanglement_cooling2, cooling3, cooling4, cooling5}.  
Notably, stabilizer states and fermionic Gaussian states can be efficiently disentangled using circuits of linear depth~\cite{Cliff4, PhysRevLett.120.110501, PhysRevApplied.9.044036, PRXQuantum.3.020328}.  
A key question in these settings is how to algorithmically determine optimal disentangling unitaries from a given state description.  
The disentangling problem has been studied in various settings, including matrix product states~\cite{Hauschild_2018, mansuroglu2025preparationcircuitsmatrixproduct}, Clifford-augmented matrix product states~\cite{camps1,camps2,camps3,camps4,camps5, liu2024classicalsimulabilitycliffordtcircuits}, quantum thermodynamics~\cite{PhysRevA.85.052121}, and using machine learning methods~\cite{tashev2024reinforcementlearningdisentanglemultiqubit, cemin2025learningstabilizenonequilibriumphases}.
Given the invertibility of unitaries, disentangling protocols are intrinsically connected to circuits which prepare the quantum state~\cite{state_prep1, state_prep2, state_prep3}.

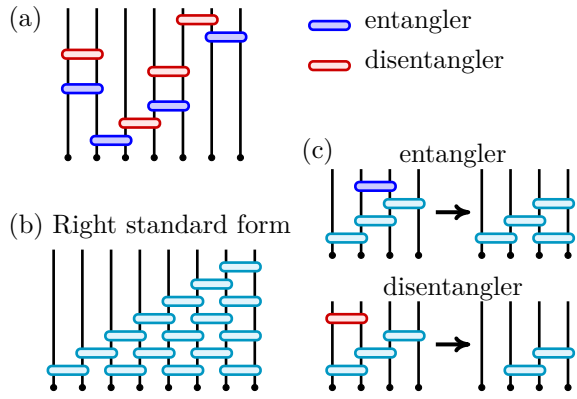
\begin{figure}[t!]
    \centering
    \begin{tikzpicture}
       \drawIntroductionFigure
    \end{tikzpicture}
    \caption{(a) Illustration of the unitary circuit game: Blue boxes represent random matchgates and red boxes are unitary gates chosen to disentangle the bond. (b) Right standard form (RSF) of fermionic Gaussian states: any pure FGS can be expressed as a matchgate circuit with this form. (c) Example of entangling and disentangling operations within the RSF formalism of FGS.}
    \label{fig: intro}
\end{figure}

The recently introduced framework of unitary circuit games~\cite{unitary_games} explores the competition between disentangling unitary dynamics and random unitary evolution.
In this setup, disentangling unitaries are applied with probability $p$, while random unitaries act with probability $1-p$, as illustrated in Fig.~\ref{fig: intro}a.
This framework can be interpreted as a model for the robustness of a disentangling protocol subject to gate imperfections.
Specifically, it captures the scenario where, at each step of the circuit, the applied gate may not correspond to the ideal disentangler.
Such imperfections may arise, for example, due to errors in selecting the correct disentangling gate.
The resulting competition yields qualitatively different behaviors depending on the underlying unitary ensemble.
In the Clifford unitary game, a phase transition separates an area-law phase, in which entanglement remains effectively suppressed, from a volume-law phase, where entanglement generation by random gates prevails.
In contrast, for Haar-random unitaries, no such transition is observed: the system invariably evolves toward a volume-law entangled state, irrespective of the disentangling rate~\cite{unitary_games}.

In this work, we investigate optimal disentangling strategies for fermionic Gaussian states (FGS) and study unitary circuit games within the framework of matchgate circuits.
Matchgates (MGs)~\cite{MG1} are a special class of two-qubit unitaries which, under the Jordan-Wigner transformation, correspond to fermionic evolutions generated by quadratic Hamiltonians~\cite{MG3,TeDi02}.
As a result, matchgate circuits, i.e., unitary evolutions composed exclusively of
MGs acting on nearest-neighbor qubits on a computational basis state and subsequent measurements in that basis, can be efficiently simulated on a classical computer~\cite{TeDi02,MG5}.
Such circuits generate pure FGS, a family of many-body states fully characterized by their two-point correlation functions.

We pursue two complementary approaches to disentangling FGS.
First, we consider a disentangling strategy that minimizes the von Neumann entanglement entropy at each step.  
Applying this method, we find that for braiding gates, the subset of matchgates corresponding to Clifford unitaries, the system enters an area-law phase at any finite disentangling rate.  
In contrast, for generic matchgates, the entropy minimization approach results in a volume-law phase, even at high disentangling rates.  
Second, we introduce a ``gate disentangler'' designed to minimize the number of matchgates required to generate a given FGS.  
To this end, we construct a representation of any pure FGS as a matchgate circuit, depicted in Fig.~\ref{fig: intro}b, which we denote the ``right standard form" (RSF).  
We develop efficient algorithms to manipulate states in RSF, enabling a systematic disentangling protocol.
Based on our results presented in Ref.~\cite{LaMo26}, such protocol is provably optimal.  
Using this framework, we study the unitary circuit game governed by the gate disentangler, uncovering an entanglement phase transition and characterizing its universal properties.

The remainder of this paper is organized as follows. 
In Sec.~\ref{sec:summary}, we present a summary of our findings. 
In Sec.~\ref{sec:fgs}, we provide an overview of FGS and MG circuits, introduce the ``right standard form" (RSF) as a circuit representation of FGS, and describe how to simulate matchgate circuits using this standard form.
In Sec.~\ref{sec:braiding}, we discuss the results of the unitary circuit game using braiding gates.
In Sec.~\ref{sec:unitary game generic}, we present the results of the unitary circuit game with generic random matchgates, comparing the two disentangling strategies.
Finally, we summarize our results and discuss open questions in Sec.~\ref{sec:discussion}. 
Technical details and additional proofs are provided in the appendices.

\section{Summary of main results}
\label{sec:summary}

\begin{table*}[t!]
    \centering
    \includegraphics{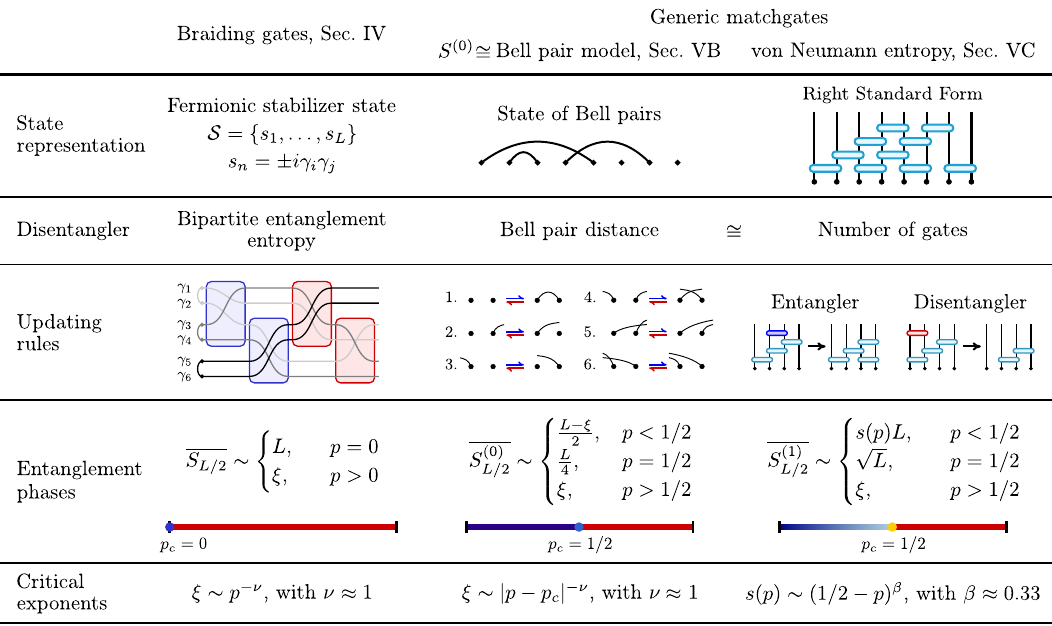}
    
    \caption{Summary of the models and results investigated in this work. We study three different variants of the unitary circuit game with matchgates. In the case of braiding gates, states are represented using the stabilizer formalism, where all stabilizers are quadratic in Majorana operators. The entangler acts by randomly permuting the Majorana operators within the stabilizers, and the disentangler minimizes the bipartite entanglement entropy across a given bond by choosing a suitable permutation of the Majorana operators. We find an area-law phase for any finite disentangling rate, with a diverging correlation length $\xi \sim p^{-\nu}$ and critical exponent $\nu \approx 1$ as $p \to 0$. Next, for generic matchgates sampled from the corresponding Haar measure, we analyze the Rényi-$0$ entropy. First we show that the dynamics can be mapped to a Bell pair model, where the system consists of either Bell pairs or disentangled qubits. The disentangler acts to reduce the physical distance between Bell pairs. This model exhibits a phase transition between a maximal volume-law and an area-law phase at $p_c = 1/2$, with the same critical exponent $\nu \approx 1$ as in the braiding gate model. Finally, for the von Neumann entropy, we introduce a circuit representation of FGSs, which we refer to as the ``right standard form'' (RSF). This representation plays a key role in our analysis and is provably optimal with respect to the number of matchgates required to generate a given state. Within this framework, the disentangler acts by reducing the total number of gates in the RSF. Using this approach, we observe a phase transition from a sub-maximal volume-law phase to an area-law phase at $p_c = 1/2$, with entanglement scaling as $\sqrt{L}$ at criticality, which differs from the result observed for the Rényi-0 entropy.}
    \label{fig: summary}
\end{table*}

In this section, we summarize the main findings of our work.
The central object of study is the unitary circuit game in which both the entangling and disentangling operations are restricted to matchgates (MGs).
We explore different subsets of MGs and various strategies for the disentangler, assuming full knowledge of the FGS, and focus on the entanglement properties of the dynamical steady state reached during the evolution.
A summary of our key results is presented in Table~\ref{fig: summary}, which we describe in more detail below.

\begin{itemize}
    \item \textit{Von Neumann disentangler for braiding states} (Sec.~\ref{sec:braiding}): We first consider braiding gates, a subset of unitaries generated by the intersection of Clifford and matchgates. Simulations of braiding gate circuits can be performed via a mapping to a Majorana loop model. In this setting, the disentangler is chosen to minimize the bipartite von Neumann entanglement entropy across a selected bond. We find that a volume-law phase occurs only for $p = 0$, while any finite disentangling rate leads to an area-law phase. This transition is characterized by a diverging correlation length $\xi \sim p^{-\nu}$ as $p \to 0$, with a critical exponent $\nu \approx 1$.
    \item \textit{Von Neumann disentangler for FGS} (Sec.~\ref{sec: von Neumann}): For general pure FGS, we implement a disentangler that numerically minimizes the bipartite von Neumann entropy. Simulations of this model indicate that the system remains in a volume-law phase for disentangling probabilities up to $p = 0.6$. However, this model exhibits significant finite-size effects, and the available numerics do not conclusively establish whether a phase transition to an area-law phase exist.    
    \item \textit{Gate disentangler for FGS}: To design a more effective disentangling strategy, we introduce a circuit representation of FGSs, which we denote the ``right standard form'' (RSF) (Sec.~\ref{subsec: circuit rep}). As we show in Ref.~\cite{LaMo26}, this representation is provably optimal in the sense that it requires the minimal number of matchgates to generate a given FGS. We present an algorithm to simulate matchgate circuits using the RSF and propose a disentangling algorithm based on reducing the number of gates in the RSF (Sec.~\ref{sec:algorithms}). We denote this kind of disentangler as ``gate disentangler''.
    \begin{enumerate}
        \item[a)] \textit{Bell pair model} (Sec.~\ref{sec: bell pair}): We start studying the behavior of the Rényi-0 entropy in the unitary circuit game with random MGs competing with the gate disentangler. We propose a Bell pair model that simplifies the simulations and that maps exactly to the original problem. In this model, Bell pairs are created and separated by the entangler, and brought together and destroyed by the disentangler. This system exhibits a sharp phase transition between maximal volume-law entanglement, $\overline{S_{L/2}^{(0)}} \sim L/2$, and an area-law phase at a critical probability $p_c = 1/2$. Near this point, the characteristic length scale diverges as $\xi \sim |p - p_c|^{-\nu}$, with critical exponent $\nu \approx 1$.
        \item[b)] \textit{Von Neumann entropy with gate disentangler} (Sec.~\ref{sec: von Neumann with gate disentangler}): We next examine the von Neumann entropy using the gate disentangler. Simulations are performed using the RSF representation, with the state updated via our entangling and disentangling algorithms. We observe a phase transition at $p_c = 1/2$, similar to the Bell pair model. However, the behavior differs significantly: in particular, in the volume-law phase the entanglement entropy scales as $\overline{S_{L/2}^{(1)}} \sim s(p) L$, with $s(p) \sim (1/2 - p)^\beta$ and a critical exponent $\beta \approx 0.33$.        
    \end{enumerate}
    \end{itemize}

These results demonstrate that different choices of gate sets and disentangling strategies lead to markedly different behaviors in the unitary circuit game.
In particular, minimizing entanglement entropies is sufficient to disentangle states, which are generated via braiding gates.
However, for general FGS, a more effective strategy is necessary.
In this case, the relevant figure of merit becomes the number of gates required to create the state.
Then, the disentangling procedure successfully yields an area-law phase provided the correct disentangling gate is chosen with probability greater than one half, i.e., $p > 1/2$.

\section{Fermionic Gaussian states and matchgate circuits}\label{sec:fgs}

In this section, we formally introduce matchgates and FGS. We begin by establishing our notation and outlining the general framework for matchgate circuits. Next, we review key algebraic properties of matchgates, including the generalized Yang-Baxter relation, which form the foundation for defining a standard form of matchgate circuits capable of generating any FGS. This standard form serves as an alternative and efficient representation of FGSs. We show how this formulation facilitates the simulation of matchgate circuits and provides a disentangling strategy that is provably optimal, as we prove in Ref.~\cite{LaMo26}.

\subsection{Notation and general definitions}
Throughout this paper, we consider a one-dimensional lattice of $L$ qubits with open boundary conditions. 
We will consider the evolution of this system under matchgate unitaries. 
Matchgates~\cite{MG1,MG2,MG5} are defined to be unitary gates acting on nearest-neighbor qubits of the form
\begin{equation}
    G(A,B) = \begin{pmatrix}
a_{11} & 0 & 0 & a_{12}\\
0 & b_{11} & b_{12} & 0\\
0 & b_{21} & b_{22} & 0\\
a_{21} & 0 & 0 & a_{22}
\end{pmatrix},
\end{equation}
where the matrices $A = (a_{ij})$ and $B = (b_{ij})$ satisfy $\det A = \det B$. 
A general matchgate can be expressed as the unitary generated by a Hamiltonian of the form
\begin{align}
\begin{split}
       H =& \alpha_1 Z_i + \beta_1 Z_{i+1}+ \alpha_2 X_i X_{i+1}+ \\ &+ \beta_2 Y_i Y_{i+1}+\alpha_3 X_i Y_{i+1} + \beta_3 Y_i X_{i+1}, 
\end{split}
\end{align}
where the subscripts indicate the qubits on which the Pauli operators act.
Matchgates are equivalent to free fermionic nearest-neighbor gates via a Jordan-Wigner transformation~\cite{MG3,TeDi02}.
Equivalently, any unitary generated by a free fermion Hamiltonian $H$, $U = e^{iH}$, corresponds to a circuit of matchgates acting on nearest-neighbors. 
The Jordan-Wigner transformation is a unitary, one-to-one transformation between the space of $L$ fermionic modes, with Majorana operators $\{\gamma_i\}_{i=1}^{2L}$, and the space of $L$ qubits.
Such transformation is given by
\begin{align}
    \begin{split}
        \gamma_{2j-1} &= Z_1 Z_2\dots  Z_{j-1} X_j,\\
        \gamma_{2j} &= Z_1 Z_2 \dots  Z_{j-1} Y_j .
    \end{split}
\end{align}
Gaussian unitary operations $U$ are in one-to-one correspondence with special orthogonal transformations, where $R\in\text{SO}(2L)$ is given by~\cite{TeDi02}
\begin{equation}
    U^\dagger \gamma_i U = \sum_{j=1}^{2L} R_{ij} \gamma_j.
\end{equation}
In the following, we will interchangeably refer to one or the other representation.

A pure state $\ket{\psi}$ is said to be a fermionic Gaussian state (FGS) if it can be generated by a circuit of matchgates acting on a computational basis state. Mixed FGS arise as reduced states of pure FGS. Given an FGS on $L$ qubits, one can define its antisymmetric covariance matrix $\Gamma$ via
\[
\Gamma_{kl} = \frac{i}{2} \langle [\gamma_k, \gamma_l]\rangle_\rho.
\]
The FGS corresponding to $\Gamma$ is a pure state iff $\Gamma \Gamma^T = \mathds{1}$. The covariance matrix corresponding to the reduced state on consecutive qubits $\{k_1, \ldots, k_s\}$ is given by
$\Gamma_{\{2k_1-1, 2k_1, \ldots, 2k_s-1,2k_s\}}$~\footnote{To obtain the reduced state of non-consecutive qubits, these qubits first need to be swapped into a consecutive order to the first position of the chain~\cite{FrLeBr12}.}, where $\Gamma_{\{a_1,\ldots,a_l\}}$ is a matrix composed of the $a_1,\ldots,a_l$-th rows and columns of $\Gamma$.
At the heart of many efficient simulation algorithms for FGS is Wick's theorem~\cite{WicksThm,Bach_1994}, which states that the expectation value of a Majorana string operator $\gamma_{k_1} \ldots \gamma_{k_s}$ can be efficiently calculated via the relation
\[
i^s \langle \gamma_{k_1} \ldots \gamma_{k_s}\rangle_\rho = \operatorname{Pf}(\Gamma_{\{k_1,\ldots,k_s\}}),
\]
where $\operatorname{Pf}$ denotes the \emph{Pfaffian} of a matrix, which in turn can be efficiently evaluated. The covariance matrix of an FGS can be brought to the so-called fermionic Williamson normal form~\cite{BOTERO200439},
\[
R \;\! \Gamma\;\! R^{\mathrm{T}}  =\bigoplus_{k=1}^L \lambda_k J_2 \; 
\]
where $R\in \text{SO}(2L)$, and
\[J_2 = \begin{pmatrix}0&-1\\1&0\end{pmatrix}.\]
The numbers $-1 \leq \lambda_k \leq 1$ are called the Williamson eigenvalues. An FGS is pure iff all Williamson eigenvalues satisfy $\vert\lambda_k\vert = 1$. 

For the Schmidt decomposition of bipartite pure FGS, the following result is known. Given two regions of consecutive qubits $A = \{A_1, \ldots, A_k\}$ and $B = \{B_1, \ldots, B_l\}$, any pure bipartite FGS $\ket\psi_{\mathrm{AB}}$ can be written as $\ket\psi_{\mathrm{AB}} =U_\mathrm{A} \otimes U_\mathrm{B} \ket{\phi}_{\mathrm{AB}}$, where $U_\mathrm{A}$ and $U_\mathrm{B}$ are matchgate circuits, and $\ket{\phi}_\mathrm{AB}$ is a tensor product of computational basis states and states of the form $\cos{\alpha_i}\ket{00}_{\mathrm{A}_i \mathrm{B}_i}+\sin{\alpha_i}\ket{11}_{\mathrm{A}_i \mathrm{B}_i}$~\cite{BOTERO200439}. This can be seen by conjugating $\Gamma$ with the rotation $R_A \oplus R_B$, where $R_A$ ($R_B$) is the rotation that takes the reduced covariance matrix $\Gamma_A$ ($\Gamma_B$) to its Williamson normal form. The numbers 
$\{\alpha_i\}$ are related to the Williamson eigenvalues $\{\lambda_i\}$ of the reduced state on any party via $\cos(2\alpha_i) = \lambda_i$. Consequently, the $n$-th Rényi entanglement entropy of a bipartite FGS~\cite{PeCh99,Pe03} can be calculated knowing the Williamson eigenvalues of a reduced state using the formula
\begin{equation}
    \label{eq:renyi_entropy}
    S^{(n)} = \frac{1}{1-n}\sum_i \log_2\left[\left(\frac{1+\lambda_i}{2}\right)^n+\left(\frac{1-\lambda_i}{2}\right)^n\right].
\end{equation}
Moreover, the zeroth Rényi entropy, i.e., the logarithm of the Schmidt rank, is given by the number of Williamson eigenvalues different from $1$, and is therefore always a positive integer.

\subsection{Algebraic identities for matchgate circuits and fermionic Gaussian states}

Here, we introduce two properties fulfilled by matchgates. These will be used in the following sections in order to reduce the number of gates required to create a given FGS.

The first one, the Yang-Baxter equation, was first introduced in the context of statistical mechanics~\cite{Yang, BAXTER1972193}. Matchgates have been proven to satisfy a generalized Yang-Baxter relation~\cite{Yang_Baxter_circuit_compression}. Specifically, given any three matchgates $ U_{i-1,i} $, $ U'_{i, i+1} $, and $ U''_{i-1, i} $, there exist three other matchgates $ V_{i,i+1} $, $ V'_{i-1, i} $, and $ V''_{i, i+1} $ such that
\begin{equation}
    U_{i-1,i}U'_{i, i+1}U''_{i-1, i} = V_{i,i+1}V'_{i-1, i}V''_{i, i+1}.
\end{equation}
Here and in the following, the subscripts denote on which qubit lines the gate acts on. When dropping gate labels, we can denote this relation graphically as
\[\vcenter{\hbox{\drawYBMoveOnly}} \; .\]
As shown in Refs.~\cite{Yang_Baxter_circuit_compression,Yang_Baxter_circuit_compression_algorithm}, this generalized Yang-Baxter relation enables the compression of any MG circuit acting on $L$ qubits to a maximum of $ L(L-1)/2 $ MGs.

The second property can be observed when applying matchgates to a computational basis state.
Specifically, given two matchgates $U_{i-1, i}$ and $U'_{i, i+1}$, we can always find two other matchgates $V_{i-1, i}$ and $V'_{i, i+1}$ such that
\begin{equation}
    U_{i-1, i}U'_{i, i+1}\ket{0
    00}=V_{i, i+1}V'_{i-1, i}\ket{000}.
\end{equation}
Graphically, we represent this relation as
\[\vcenter{\hbox{\drawMPSMoveOnly}} \; ,\]
where the three dots on the bottom represent the state $\ket{000}$~\footnote{Note that similar relations hold when replacing the state $\ket{000}$ with an arbitrary computational basis state.}. In the following, we will refer to this relation as ``\mpsmove'', as we move the gate from left to right and vice versa. Note that the Yang-Baxter relation is an equality of unitaries, while the \mpsmove~can only be applied when acting on a product state. In appendix~\ref{appendix_Yang_Baxter}, we provide a proof of both these properties and present a method to compute the $ V $ matrices based on the given $ U $ matrices.

\subsection{Representations of Fermionic Gaussian States} \label{subsec: circuit rep}

On the one hand, fermionic Gaussian states can be entirely represented by their covariance matrix. This representation serves as a basis for many efficient simulation algorithms~\cite{FGSSurace}. On the other hand, states can be represented as a circuit acting on a product state, i.e., as a possible way of generating them. Since MG circuits can always be decomposed into at most $L(L-1)/2$ individual MGs~\cite{PhysRevLett.120.110501, PhysRevApplied.9.044036, PRXQuantum.3.020328,Yang_Baxter_circuit_compression,Yang_Baxter_circuit_compression_algorithm, braccia2025optimalhaarrandomfermionic}, this representation is always efficient for FGSs. Building on this observation, we introduce a slightly more efficient circuit representation via circuits in \emph{right standard form} (RSF, see below), which requires at most $\floor{L^2/4}$ MGs. This representation is optimal in terms of the number of gates required to generate the state~\cite{LaMo26}, and thus leads to a natural choice for a disentangling algorithm, as we will explain below. In the following, we introduce this form and explain why it can be used to represent any pure FGS.

We define an RSF to be a list of $0 \leq \numdiag \leq \floor{L/2}$ pairs of integers $((k_i,l_i))_{i=1}^{\numdiag}$, satisfying $1\leq k_i \leq k_{i+1} -2 \leq L-1$, and $1\leq l_i\leq L-k_i$. A circuit $U$ is in such an RSF layout, if it is given as a product
\[ U = D^{(1)} \ldots D^{(k_\numdiag)}\]
of \emph{diagonals}, i.e., sequences of the form
\[ D^{(i)} = U^{(i,l_i)}_{k_i+l_i-1, k_i+l_i} \ldots U^{(i,1)}_{k_i, k_i+1}\] of gates $U^{(ij)}$. The superscripts here serve as an index to the given diagonals (gates). To give an example, the circuit
\[\drawIncompleteRSFExampleAMainText\]
is in the RSF $((1,5), (4,4), (6,1))$, meaning that there are diagonals of gates starting at bonds 1, 4, and 6, with each diagonal consisting of 5, 4, and 1 MGs respectively. This RSF defines the FGS $\ket{\psi}=U\ket{0\ldots0}$. Notice that this definition implies that, for instance, one cannot have a gate between qubits 1 and 2 in the third layer of the circuit. We may also define a \textit{left standard form}, where diagonals go from right to left, yielding a mirror image of the RSF.

\begin{figure*}[t!]
    \centering
    \begin{tikzpicture}
       \drawExampleActionFigure
    \end{tikzpicture}
    \caption{Example of the application of the (a) absorption algorithm, where the dark blue gate is absorbed into the RSF, and (b) disentangling algorithm, where the light gray gate is removed from the RSF, by applying a gate (depicted in red) in the second bond. The numbers above the arrows indicate which step of the algorithm is applied (see main text).}
    \label{fig: absorption algorithm}
\end{figure*}
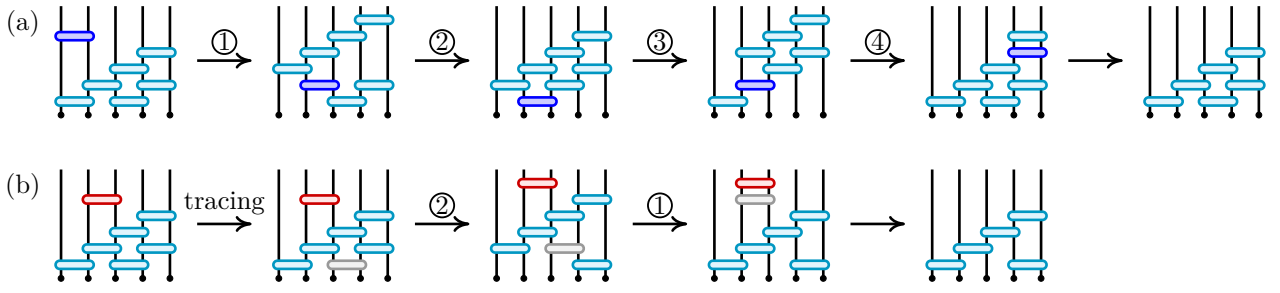

That RSF matchgate circuits can represent an arbitrary pure FGS can be seen as follows. Starting from the fact that any FGS $\ket \psi$ can be written as a sequence of MGs $U^{(i)}$ acting on a computational basis state $\ket{x}$, we show that each intermediate state $\ket{\psi^i} = \prod_{j\leq i} U^{(j)}\ket{x}$ is represented by a circuit in RSF, using repeated application of the Yang-Baxter and \mpsmoves~(an example is given in Fig.~\ref{fig: absorption algorithm}a). In particular, the absorption algorithm presented in the following section allows for an efficient update of the RSF circuit describing the state $\ket{\psi^{i+1}}$, given the RSF circuit describing $\ket{\psi^i}$ and the gate $U^{(i+1)}$. The details of how the RSF changes when adding gates are given in Appendix~\ref{appendix_FGS_circuit}. In Ref.~\cite{LaMo26} we provide a method to extract an RSF circuit directly from the covariance matrix, and we prove that it yields circuits that are indeed optimal in terms of the number of necessary gates.

As we show in the next subsection, circuits in RSF admit a natural choice for a disentangling algorithm based on reducing the number of MGs required to generate the state.
This differs fundamentally from strategies based on minimizing entropic quantities such as Rényi entropies.
We will focus on this disentangling strategy in the following.

The representation of a state as an RSF provides simple and stable access to certain quantities, while some others get hidden. For example, when the gates composing the RSF are generic MGs, the Rényi entropy for $n=0$ is directly accessible, since each gate maximally increases the bond dimension. In contrast, computing the Rényi entropy for $n > 0$ requires first evaluating the covariance matrix and subsequently applying Eq.~\eqref{eq:renyi_entropy} in the standard manner.

\subsection{Absorption and disentangling algorithms}
\label{sec:algorithms}
In the following, we describe two algorithms for updating RSF circuits.
First, we present the so-called ``absorption algorithm,'' which is used to update RSF circuits when a matchgate is applied (i.e., added) on a given bond. 
Then, we introduce a disentangling algorithm that, given a bond, identifies a gate that reduces the number of gates in the RSF, if such a gate exists (i.e., it removes the gate).

To see that applying a MG to a circuit in RSF can again be transformed into a circuit in RSF requires also to determine the gates that can be absorbed into existing gates and the gates that can be removed. Analyzing all possible cases is tedious, but straightforward. For the sake of completeness, but also because we believe some cases are very illuminating, we present the details here and in App.~\ref{appendix_FGS_circuit}.

\subsubsection{Absorption algorithm}

\label{abs_algorithm}

Here, we describe a sequence of steps illustrating how a gate $A$ acting on qubits $i-1,~i$ can be absorbed into a state represented by a RSF circuit. After each step, one checks whether the gate $A$ can be absorbed by another gate acting on the same pair of qubits. For brevity, we omit these additional checks. In the circuit diagrams below, the gate $A$ is shown in dark blue, while all other gates are depicted in light cyan.%

\begin{enumerate}
    \item If the gate $A_{i-1,i}$ is locally part of a sequence of gates $A_{i-1,i} U_{i,i+1} U'_{i-1,i},$ perform the Yang-Baxter move and replace the sequence with
    $V_{i,i+1} V'_{i-1,i} A'_{i,i+1}.$ 
    \begin{center}
        \begin{tikzpicture}
    
       \drawAbsorbingone
    \end{tikzpicture}
    \end{center}
    Then, relabel $A \leftarrow A'$ and repeat step 1. 

    Otherwise, go to step 2.
    \item If the gate $A_{i-1,i}$ is locally part of a sequence $A_{i-1,i}U_{i,i+1}\ket{000},$ use the \mpsmove~and replace this sequence with $V_{i,i+1} A'_{i-1,i}\ket{00}.$ 
    \begin{center}
    \begin{tikzpicture}
       \drawAbsorbingtwo
    \end{tikzpicture} 
    \end{center}
    Then, relabel $A\leftarrow A'$. Go to step 3.
    
    \item  If the gate $A_{i,i+1}$ is locally part of a sequence $U_{i-1,i}A_{i,i+1}\ket{000},$ use the \mpsmove~and replace this sequence with $A'_{i,i+1}V_{i-1,i}\ket{000}.$ 
    \begin{center}
        \begin{tikzpicture}
       \drawAbsorbingthree
    \end{tikzpicture} 
    \end{center}
    Then, relabel $A\leftarrow A'$. Go to step 4.

     \item If the gate $A_{i-1,i}$ is locally part of a sequence $U_{i-1,i} U'_{i,i+1} A_{i-1,i},$ perform the Yang-Baxter move and replace the sequence with
    $A'_{i,i+1}V_{i-1,i} V'_{i,i+1}.$ 
    \begin{center}
        \begin{tikzpicture}
            \drawAbsorbingfour
        \end{tikzpicture}
    \end{center}
    Then, relabel $A \leftarrow A'$ and repeat step 4. 
    
    Otherwise, terminate the algorithm.
     
\end{enumerate}

This algorithm updates RSF circuits into different RSF circuits under the application of a matchgate. An example of the application of this algorithm is shown in Fig.~\ref{fig: absorption algorithm}a.

\subsubsection{Disentangling algorithm}
\label{disent_alg}

In this section, we introduce a disentangling algorithm based on RSF circuits.
The core idea is to take as input an FGS in RSF and a chosen bond of the chain, and to return a matchgate.
This matchgate, produced by the disentangling algorithm, is required to satisfy two properties:
First, when applied to the selected bond, it should reduce the entanglement in the state in a well-defined and quantifiable way.
Second, repeated application of such disentangling gates across various bonds should eventually drive the system towards a product state.
We note that many reasonable disentangling strategies satisfy the first property but fail to guarantee the second.
For example, as discussed in more detail in appendix~\ref{appendix_disentanglers}, numerical evidence shows that selecting the matchgate that minimizes the bipartite von Neumann entropy across a fixed partition does not lead to a product state after repeated application of the algorithm for a finite number of iterations.
Therefore, in this section we present a disentangling algorithm explicitly designed to reduce the number of gates required to represent the state as an RSF.
Since this representation is known to be optimal in terms of gate count~\cite{LaMo26}, such a disentangler is provably optimal for fully disentangling the state and transforming it into a product state.

To find a disentangling gate acting on bond $i$ given a circuit in RSF, the idea is to rewrite, if possible, an RSF circuit $U$ using Yang-Baxter and \mpsmoves~into another RSF circuit $V$ and a gate $A$ such that $U\ket{0\ldots0} = A_{i,i+1} V\ket{0\ldots0}$.
If this rewriting is possible, it reduces the number of gates in $V$ by one compared to $U$ (since the YB and left-right moves cannot reduce the total number of gates).
The disentangling algorithm then returns the matchgate $A^\dagger$.
If such a rewriting is not possible, the algorithm terminates without returning a disentangling gate, since the number of gates of the RSF cannot be further reduced acting on this bond.

On an algorithmic level, the rewriting can be performed by applying the steps of the absorption algorithm in reverse.
However, before doing so, it is necessary to determine whether the circuit can be rewritten in the desired form $U\ket{0\ldots0} = A_{i,i+1} V\ket{0\ldots0}$.
This is addressed by first identifying the ``target gate", i.e., the gate in the RSF representation of $U$ that must be removed to obtain the RSF representation of $V$.
To this end, an auxiliary identity gate is inserted on bond $i$ and subsequently absorbed using the standard absorption algorithm.
If the algorithm terminates by absorbing the auxiliary gate into another gate, that gate is identified as the target gate, and it can then be extracted from $U$ by applying the absorption algorithm in reverse, yielding the gate $A$ and the circuit $V$.
An illustration of this procedure is provided in Fig.~\ref{fig: absorption algorithm}b.

Let us note here that there are situations in which the absorption algorithm can be applied in reverse order in two distinct, nonequivalent ways. In such cases, we must choose the option that retraces the path followed by the auxiliary identity gate. The alternative reversal would move the gate to a different bond than the one where we apply the disentangling gate. To give an example, in the RSF circuit 
\begin{center}
    \drawFourQubitRSF
\end{center}
the top right gate can be reached by placing a gate on qubits one and two or in qubits three and four.

Finally, there is a particular case in which the absorption algorithm with the auxiliary gate does not identify the correct target gate. This happens when the following situation is encountered:
\begin{center}
    \begin{tikzpicture}
       \node at (2.5,2.7) {$B_{i-1,i} C_{i,i+1}U_{i-1,i}\ket{000} = $};
       \drawSpecialCaseOne; \node at (5.7, 2.6) {,};
    \end{tikzpicture} 
\end{center}
where the red gate $B$ is the auxiliary identity gate and $C$ and $U$ are the gates composing the RSF. In this case, the absorption algorithm applied to the red $B$ gate would lead to choose the gate $C$ as the target gate. However, this gate cannot be removed by applying the absorption algorithm in reverse. Instead, by applying the left-right move we observed that the gate $U$ is the one that can be removed. Therefore, $U$ has to be marked as the target gate by the disentangling algorithm.

\section{Unitary circuit game with braiding gates}\label{sec:braiding}

We start with a discrete subset of matchgates.
This subset, known as braiding gates~\cite{bravyi}, consists of unitaries that are both matchgates and Clifford gates~\footnote{Clifford gates are unitary operations that map Pauli strings, i.e., tensor products of Pauli operators, onto other Pauli strings.}. 
Notably, this set forms a matchgate 3-design~\cite{Wan_2023}, meaning that the first three moments of the uniform distributions over braiding gates and general matchgates coincide.
In the free-fermion picture, braiding gates correspond to unitaries that permute Majorana operators, i.e., unitaries satisfying $U^\dagger \gamma_i U = \pm\gamma_{\tau(i)}$ for all $i$, where $\tau$ is a permutation.
In terms of the covariance matrix $\Gamma$, braiding gates permute its rows and columns up to a sign.
The evolution under braiding gates can be simulated efficiently using the standard stabilizer formalism~\cite{Cliff1, Cliff2, Cliff3, Cliff4}.
However, by employing the Majorana representation, the model can also be mapped analytically onto a two-dimensional loop model, similar to approaches used in systems with measurements~\cite{PRXQuantum.2.030313, miptff2, miptff6, miptff11, Klocke_2023}.
This mapping reduces computational cost while preserving the exact results.
In Appendix~\ref{appendix_braiding} we provide details of the loop model used to perform the simulations.

\begin{figure}[t!]
    \centering
    \includegraphics[width=1\columnwidth]{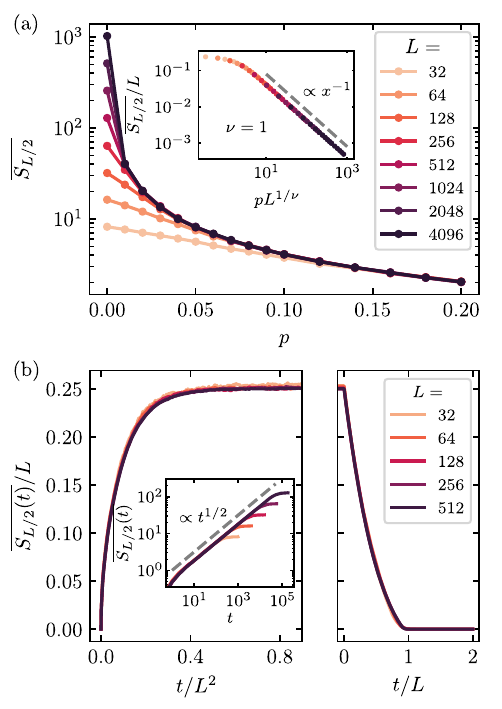}
    \caption{Numerical results of the braiding gate model. (a) Averaged half-chain entanglement entropy normalized by the system size in the steady state of the unitary game as a function of the disentangling probability $p$. The results indicate an area law phase for every nonzero probability. The inset shows the data collapse for critical exponent $\nu=1$. (b) Evolution of the averaged half-chain entanglement entropy $\overline{S_{L/2}}$ as a function of time $t$ with a random braiding evolution (left panel) followed by a disentangling evolution (right panel), for different system sizes $L$. The time axis is normalized by $L^2$ and by $L$ in the left and right panels respectively to indicate the convergence time. The inset of the left panel shows the diffusive spread of the entanglement entropy in the $p=0$ case, with $\overline{S_{L/2}}\propto \sqrt{t}$.}
    \label{fig: braiding results}
\end{figure}

Let us now turn to the unitary circuit game. 
The entangling operation is a unitary randomly sampled from the set of braid gates.
The disentangling operation consists in selecting the braid gate that implements the permutation of Majorana operators which maximally reduces the bipartite entanglement entropy across a given bond.
Note that, unlike in generic FGS, all Rényi entropies are equal in braiding states, as all nonzero Schmidt values are identical.
Regarding our numerical simulations, for each disentangling probability $p$, we perform 200 circuit realizations of the game. For each realization, we average the steady-state value of the entanglement entropy over $10^3$ time steps, where each time step comprises $L$ entangling or disentangling moves on randomly chosen bonds.

Figure~\ref{fig: braiding results}a shows the numerical results for the averaged half-chain entanglement entropy $\overline{S_{L/2}}$ as a function of the disentangling probability $p$.
These results clearly indicate that the system is in an area-law phase for any $p>0$, in contrast to the volume-law phase observed when no disentangling is applied,
\begin{equation}
    \overline{S_{L/2}}\sim\begin{cases}
      L, \qquad p=0\\
      \xi, \qquad p>0
    \end{cases}\,,
\end{equation}
where $\xi$ is a characteristic length scale that diverges as a power law when approaching $p = 0$.
This behavior differs from that observed in unitary circuit games with Clifford gates or  random gates sampled from the Haar measure of $U(4)$, where a volume-law phase persists over a finite range of $p$~\cite{unitary_games}.
In the inset of Fig.~\ref{fig: braiding results}a, we show a data collapse of $\overline{S_{L/2}}/L$ versus $pL^{1/\nu}$.
We numerically find a good fit for $\nu = 1$, indicating that the entanglement entropy diverges as $p^{-\nu}$ as $p$ approaches zero.
A brief review of finite-size scaling and critical exponents is provided in Appendix~\ref{appendix_finite_size}.

The absence of an extended volume-law phase can be attributed to the different rates at which entanglement is generated by random evolution and reduced by disentangling operations.
The numerical results in Fig.~\ref{fig: braiding results}b show an entangling evolution followed by a disentangling evolution, averaged over $10^3$ trajectories.
In the left panel, we show the entanglement growth for a system with $p = 0$ (no disentangling gate applied) as a function of time for various system sizes.
This results in diffusive entanglement growth, as illustrated in the inset.
This is the expected behavior for random matchgate circuits and arises from the diffusive spreading of Majorana operators during random evolution~\cite{Nahum_2017, dias2021diffusiveoperatorspreadingrandom}.
Moreover, as shown in those works, the thermalization time scales quadratically with system size, $T_{\text{eq}} \propto L^2$, and the entropy saturates to a value that follows a volume law, i.e., proportional to the size of the subsystem.
In contrast, disentangling occurs on a timescale linear in the system size, $T_{\text{disent}} \propto L$, as demonstrated by the data collapse in the right panel of Figure~\ref{fig: braiding results}b.
This behavior can be understood in the Majorana picture: in the worst case, a stabilizer can be composed of two Majorana operators corresponding to qubits separated by a distance proportional to $L$, requiring $L$ disentangling operations to move the Majorana operators together. Since there are $L$ Majorana pairs, the entire state can be disentangled with $L^2$ operations, i.e., in linear time with respect to system size.

\section{Unitary circuit games with generic matchgates}\label{sec:unitary game generic}

In this section, we present the results for the general unitary game, in which the entangling evolution is given by random Haar matchgates, i.e., matchgates sampled from the Haar measure of $\text{SO}(4)$.
The selection of the disentangling gate, however, is nontrivial and the results depends sensitively on the specific disentangling procedure used.
Here, we explore two different choices for the disentangler.

First, we consider a brute-force numerical minimization of the bipartite von Neumann entanglement entropy.
This procedure is analogous to that employed for the braiding gate model in Sec.~\ref{sec:braiding}.
Unlike in that case, however, both the gate set and the entropy values are continuous here, requiring a fully numerical minimization.
Second, we examine the \textit{gate disentangler}, which employs the algorithm defined in Section~\ref{disent_alg} to identify a matchgate that reduces the number of gates in the RSF.
In Appendix~\ref{appendix_disentanglers}, we numerically compare the performance of various disentanglers and find that the gate disentangler provides the most effective strategy for fully disentangling the state.
A formal proof of the optimality of the gate disentangler is presented in Ref.~\cite{LaMo26}.
Note that, while the von Neumann disentangler focuses exclusively on bipartite properties of the system, the gate disentangler uses information about the optimal matchgate circuit that prepares the state.

The unitary circuit game with the gate disentangler yields qualitatively different behavior for the Rényi entropy at $n=0$ and for $n>0$.
To study the case $n=0$, we introduce a Bell pair model that maps exactly onto the behavior of the Rényi-0 entropy in its discrete formulation, and we analyze the associated phase transition.
Subsequently, we investigate the phase transition in the von Neumann entanglement entropy, using the simulation methods developed in Section~\ref{subsec: circuit rep}.

\subsection{Von Neumann disentangler}
\label{sec: von Neumann}

We begin by considering a disentangling strategy that minimizes the bipartite von Neumann entanglement entropy at a given bond.
To achieve this, we parametrize a generic matchgate as~\cite{Kraus_2001,matchgates_disentangling,murao}
\begin{equation}
    U_{\text{MG}}=\left(e^{i\phi_3 Z}\otimes e^{i\phi_4 Z}\right)e^{i(\alpha X\otimes X + \beta Y\otimes Y)}\left(e^{i\phi_2 Z}\otimes e^{i\phi_1 Z}\right).
\end{equation}
This decomposition highlights that the parameters $\phi_3$ and $\phi_4$ correspond to local rotations and therefore do not alter entanglement.
As a result, determining the disentangling gate requires optimizing only the parameters $\alpha$, $\beta$, $\phi_1$, and $\phi_2$.
The Rényi-$n$ disentangler gate is defined by the minimization
\begin{equation}
\label{eq: minimization}
    \argmin_{\alpha,\beta, \phi_1,\phi_2}S^{(n)}_{x}(U_{\text{MG}}\ket{\psi}),
\end{equation}
where $S^{(n)}_{x}$ denotes the Rényi entropy for the bipartition at bond $x$.
In the numerical results presented below, we focus on the von Neumann disentangler ($n=1$), although similar outcomes are observed for other values with $n > 0$.
We note that determining the disentangling gate involves a numerical minimization procedure that depends on the chosen optimization method and numerical precision.
However, we find that this dependence has negligible effect on the steady-state properties of quantities such as the von Neumann entropy.

\begin{figure}[t!]
    \centering
    \includegraphics[width=1\columnwidth]{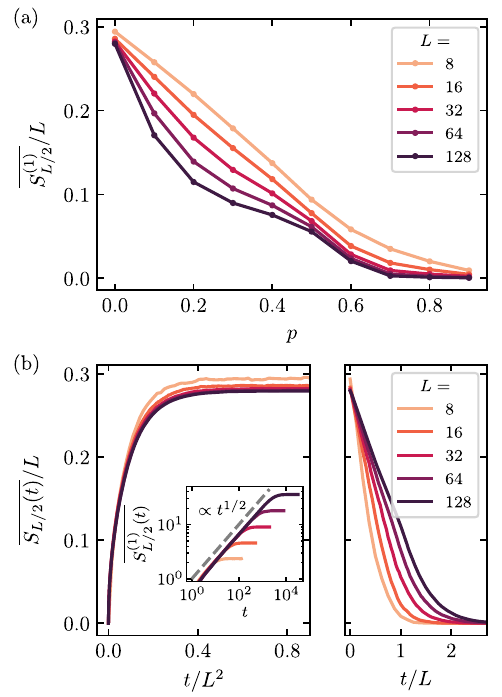}
    \caption{Numerical results for the unitary game with von Neumann disentangler. (a) Averaged half-chain von Neumann entanglement entropy normalized by the system size in the steady state of the unitary game as a function of the disentangling probability $p$. We find a decay with system size for any nonzero disentangling probability, but with the system sizes achieved we do not observe convergence. (b) Evolution of the averaged half-chain von Neumann entropy $\overline{S_{L/2}^{(1)}}$ as a function of time $t$ with a random matchgate evolution (left panel) followed by a disentangling evolution with von Neumann disentangler (right panel), for different system sizes $L$. The time axis is normalized by $L^2$ and by $L$ in the left and right panels respectively. The inset of the left panel shows the diffusive spread of the entanglement entropy in the $p=0$ case, with $\overline{S_{L/2}^{(1)}}\propto \sqrt{t}$.}
    \label{fig: renyi disentangler}
\end{figure}

Figure~\ref{fig: renyi disentangler}a shows the behavior of the averaged half-chain von Neumann entropy $\overline{S_{L/2}^{(1)}}/L$ as a function of the disentangling probability $p$ for various system sizes $L$.
Each data point is obtained by averaging over 100 trajectories, with each trajectory further averaged over 100 time steps within the steady state.
At $p=0$, the results show volume-law scaling, consistent with a random FGS.
For $p>0$, we observe that $\overline{S_{L/2}^{(1)}}/L$ decreases with increasing system size.
However, due to finite-size effects, it remains unclear whether this quantity saturates to a finite value or vanishes in the thermodynamic limit.
We also note that for all disentangling probabilities $p<1$ considered, the Rényi-0 entropy in the steady state has maximum rank, as discussed in Appendix~\ref{appendix_disentanglers}.

The numerical results in Fig.~\ref{fig: renyi disentangler}b show an entangling evolution followed by a disentangling evolution, averaged over 100 trajectories. 
In the left panel, we show that for an evolution without disentangling gates, $p=0$, the von Neumann entropy grows diffusively as $t^{1/2}$ and saturates to a volume law phase.
These results are in direct analogy with those obtained for braiding gates in Sec.~\ref{sec:braiding}, and reproducing the behavior observed in Refs.~\cite{Nahum_2017, dias2021diffusiveoperatorspreadingrandom}.
However, the disentangling evolution shown in the right panel is different than that observed for braiding gates (cf. Fig.~\ref{fig: braiding results}b).
In this case, the disentangling time increases with system size and does not collapse to a single curve.
This explains the different behavior of the unitary games with respect to the braiding gate model.
We leave an analysis of the scaling of the disentangling time with system size as an open question for further research.

Additional numerical results for this model are presented in Appendix~\ref{appendix_vonNeumann_more}, where we provide evidence that the system remains in a volume-law phase for disentangling probabilities up to at least $p=0.6$.
For larger values of $p$, given the system sizes and numerical precision accessible in our simulations, we cannot determine whether the system continues to exhibit volume-law behavior or instead transitions to an area-law phase.
Nonetheless, our numerical analysis does not reveal any signatures of critical behavior.
In particular, we do not observe any physical quantity exhibiting a divergence or scaling behavior typically associated with a phase transition.

\subsection{Bell pair model for the Rényi-0 entropy}
\label{sec: bell pair}
In the previous models, we considered the unitary game with a disentangler that minimizes the bipartite entanglement entropy.
However, when the entangling gates are sampled from the Haar measure on $\text{SO}(4)$, this strategy does not yield clear phase distinctions with respect to entanglement.
In the following, we consider the gate disentangler, that minimizes the number of gates required to create the state of the system in the RSF, therefore fulfilling the property (property 2 from above) that repeated applications of this disentangler drives the system to a product state. 

In this section, we focus on the behavior of the Rényi\mbox{-}0 entropy in the unitary circuit game with a gate disentangler.
The simulation is performed by representing the state in an RSF and evolving it using the absorption and disentangling algorithms outlined in Sec.~\ref{subsec: circuit rep}.
Since we are solely interested in $S^{(0)}$, we do not focus on the specifics of the gates themselves, but rather on the presence or absence of a gate at a given position. For generic gates, this information is sufficient to determine the Rényi-0 entanglement entropy~\cite{Nahum_2017}.
This problem is discrete, where each entangling and disentangling step involves moving a gate using Yang-Baxter to preserve the RSF. Each of these steps requires $\mathcal{O}(L)$ basic operations.
To simplify the simulation, we introduce a Bell pair model that reproduces the exact behavior of the Rényi-0 entanglement entropy, but with each step requiring only $\mathcal{O}(1)$ operations.
In the following, we briefly describe this Bell pair model. 
An analytical proof that this model maps exactly to the dynamics of the generic model for the Rényi-0 entropy is provided in Appendix~\ref{appendix:telephone}.

We consider a qubit system with $L$ sites arranged in a 1D configuration. Each qubit $s_1$ can either be in the state $\ket{0}$ or entangled with another qubit $s_2$ in a Bell state $\ket{\Phi^+}_{s_1, s_2} = (\ket{00} + \ket{11})/\sqrt{2}$. The entanglement for a bipartition $A / B$ of the system corresponds to the number of Bell pairs for which one qubit resides in $A$ and the other in $B$. A possible configuration for a system with 8 sites is
\begin{equation*}
    \drawExampleStringState\,,
\end{equation*}
where each line corresponds to a Bell pair. Note that this state can be created by the circuit
\begin{equation*}
    \Bellpairmodelexample\,,
\end{equation*}
where the gates in the bottom layer create the state $\ket{\Phi^+}$, while the other gates are SWAP gates. This circuit construction establishes a natural correspondence between RSF circuits and Bell pair states.

\begin{figure}[t!]
    \begin{tikzpicture}
    \figureExplanationOfStrings
    \end{tikzpicture}
    \caption{Entangling (left to right) and disentangling (right to left) rules for the Bell pair model.
    In rules 5 and 6, the entangler increases the distance of the longest Bell pair at the expense of shortening the other pair, while the disentangler does the opposite.
    If none of the rules apply, the state is not modified.}
    \label{fig: rules classical}
\end{figure}
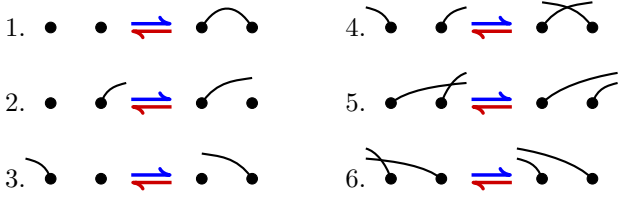

Now, let us consider the unitary game. In each step, either the entangler or the disentangler is randomly assigned a bond to act on. The allowed operations include creating or removing a Bell pair or applying a SWAP gate. The updating rules for both the entangler and the disentangler are summarized in Fig.~\ref{fig: rules classical}.
If possible, the entangler increases the entanglement across the bond by either creating a new Bell pair (case 1) or applying a SWAP (cases 2, 3, and 4).
Additionally, the qubits are swapped if the length of the longest (or tied-for-longest) pair attached to the two nodes can be increased (cases 5 and 6).
The disentangler performs the exact opposite of the entangler.

\begin{figure}[t!]
    \centering
    \includegraphics[width=0.99\columnwidth]{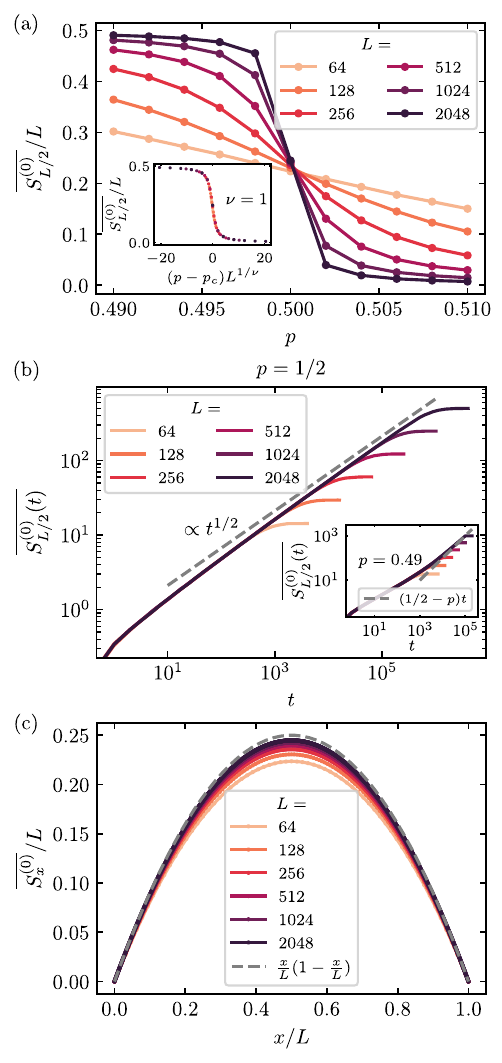}
    \caption{Numerical results for the unitary circuit game with gate disentangler for the Rényi-0 entanglement entropy, simulated through the equivalent Bell pair model. (a) Averaged half-chain Rényi-0 entropy in the steady state of the unitary game as a function of the disentangling probability $p$. The inset shows the data collapse for critical exponent $\nu=1$. (b) Evolution of the half-chain entanglement entropy $\overline{S^{(0)}_{L/2}}(t)$ as a function of time $t$ at the critical point $p=1/2$ and inside the volume law phase at $p=0.49$ (inset). (c) Rényi-0 entanglement profile at the critical point. The dashed line corresponds to the thermodynamic limit result.}
    \label{fig: gate disentangler classical}
\end{figure}

We now turn to the numerical results. Each data point consists of $10^3$ realizations of the circuit, with averages in the steady state performed over $2L$ time steps for each realization. Figure~\ref{fig: gate disentangler classical}a shows the averaged half-chain Rényi-0 entropy, normalized by the system size, $\overline{S_{L/2}^{(0)}}/L$, in the steady state as a function of the disentangling probability $p$.
We observe a phase transition between a volume law phase, where the circuit is maximally entangled ($\overline{S_{L/2}^{(0)}} \rightarrow L/2$), and an area law phase, where the entropy becomes constant and independent of the system size.
The critical point of this transition is located at $p_c = 1/2$. At this point, a sub-maximal volume law phase emerges, where the entropy converges to $\overline{S_{L/2}^{(0)}} \rightarrow L/4$.
In summary, we find the following scaling behavior for the entropy:
\begin{equation}
\overline{S_{L/2}^{(0)}} \sim
    \begin{cases}
       \frac{L-\xi}{2}, \qquad p < 1/2 \\
      \frac{L}{4}, \qquad p = 1/2 \\
      \xi, \qquad p > 1/2
    \end{cases}\,,
\end{equation}
where $\xi$ is a characteristic length scale.
We perform a data collapse around this point, assuming $\xi \sim |p - p_c|^{-\nu}$, and obtain an excellent collapse of the data for a critical exponent $\nu \approx 1$ (see Appendix~\ref{appendix_finite_size} for further details). 
This critical exponent was also observed in the context of unitary circuit games with Clifford unitaries~\cite{unitary_games}.

In Fig.~\ref{fig: gate disentangler classical}b, we show the time evolution of the entropy as a function of time, both at the critical point and in the volume law phase (inset), averaged over $10^3$ trajectories.
At the critical point ($p_c = 1/2$), we observe that the growth of entanglement follows a diffusive behavior. In contrast, in the volume law phase, the entropy grows ballistically at long times, with a velocity given by $v_E = 1/2 - p$.

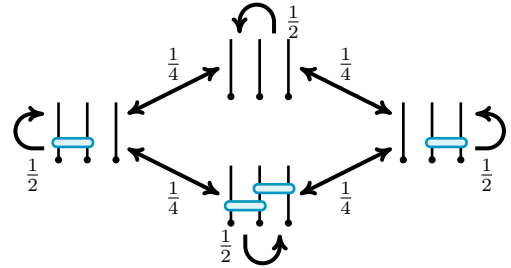
\begin{figure}[ht!]
    \centering
    \begin{tikzpicture}
       \drawMarkovChainFigure
    \end{tikzpicture}
    \caption{Markov chain representing the unitary game with gate disentangler at the critical probability $p=1/2$.}
    \label{fig: Markov chain}
\end{figure}

Finally, in Fig.~\ref{fig: gate disentangler classical}c we show the average entanglement profile $\overline{S_x^{(0)}(L)}$ for the Rényi-0 entropy for different system sizes. Here and in the following, we call the entanglement profile $S_x^{(0)}(L)$ the bipartite entanglement entropy evaluated at each bond~$x$ of the system~\footnote{Note that since several distinct RSF circuits can lead to the same Rényi-$0$ entanglement profile, a model based in that, such as the surface growth model considered in Ref.~\cite{unitary_games}, cannot be mapped exactly to this unitary circuit game.}.
The dashed line corresponds to the asymptotic behavior of the entanglement profile for increasing system size, which is given by
\begin{equation}
\label{eq:profile_critical}
    \frac{\overline{S_x^{(0)}(L)}}{L} = \frac{x}{L}\left(1-\frac{x}{L}\right).
\end{equation}
In Appendix~\ref{appendix_asymptotics}, we present a rigorous derivation showing that Eq.~(\ref{eq:profile_critical}) gives indeed the entanglement profile in the thermodynamic limit. Here, we provide a brief outline of the argument. First, note that the unitary circuit game model can be understood as a Markov chain~\cite{levin2017markov} over all possible RSF circuits (or Bell pair configurations). For instance, consider the case of $L=3$ with four possible RSF circuits. By following the rules of the game, the transition probabilities can be determined, as illustrated in Fig.~\ref{fig: Markov chain}. At the critical point ($p_c = 1/2$), the chain is both irreducible (i.e., every state can be reached from any other state) and symmetric (i.e., the transition probability from state A to state B is the same as from state B to state A). Consequently, the stationary state of the system is a uniform distribution over all RSF circuits. To derive Eq.~(\ref{eq:profile_critical}), one can solve certain counting problems to obtain a closed-form expression for both the number of configurations at each system size $L$, and the expected Rényi-0 entropy at each bond. In the thermodynamic limit, as defined in App.~\ref{appendix_asymptotics}, Eq.~(\ref{eq:profile_critical}) can then be deduced from this.

In summary, the Bell pair model exhibits a phase transition between volume law and area law entanglement, with a critical point at $p_c = 1/2$. At this critical point, the entanglement scales according to a volume law. 
This transition is reminiscent of the classical and Clifford unitary circuit games discussed in Ref.~\cite{unitary_games}, where the volume law also reaches a maximum value of $L/2$, and the characteristic length scale diverges with a critical exponent $\nu = 1$. 
However, the behavior at the critical point in the Bell pair model differs, as it exhibits a volume law scaling, in contrast to the square-root scaling observed in those models.

\subsection{Von Neumann entropy with gate disentangler}
\label{sec: von Neumann with gate disentangler}

\begin{figure}[t!]
    \centering
    \includegraphics[width=1\columnwidth]{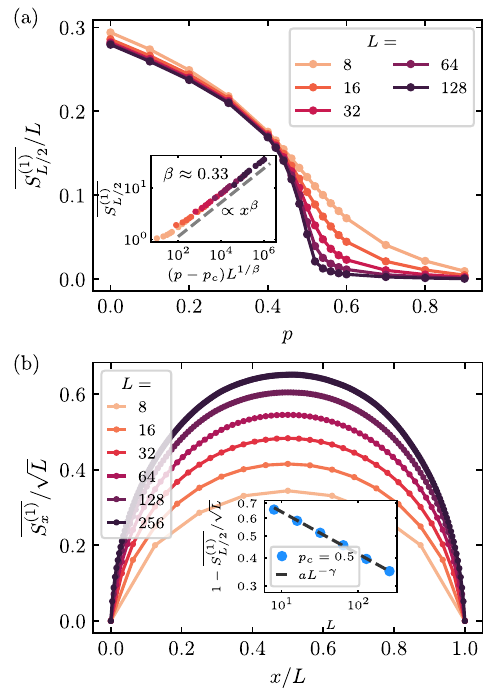}
    \caption{Numerical results of the unitary circuit game with gate disentangler for the von Neumann entanglement entropy. (a) Averaged half-chain von Neumann entropy in the steady state $S_{L/2}^{(1)}$ of the unitary game normalized by the system size as a function of the disentangling probability $p$. The inset shows the data collapse for critical exponent $\beta\approx0.33$. (b) Averaged von Neumann entanglement profile normalized by $\sqrt{L}$. The inset shows the fitting for the sub-leading behavior at the half-chain for $S^{(1)}_{L/2}=\sqrt{L}-aL^{1/2-\gamma}$, with $a \approx 0.99$ and $\gamma \approx0.19$}
    \label{fig: gate disentangler von Neumann}
\end{figure}

To analyze the results of the unitary game in terms of other Rényi entropies, it is necessary to track the individual matchgates in the RSF throughout the simulation. 
This approach goes beyond the Bell pair model previously discussed. 
Therefore, to simulate the unitary circuit game we utilize the RSF, and use the absorption to apply the entangling random matchgates and the disentangling algorithm to remove gates from the RSF. To extract information about the von Neumann entropy, we run the entire circuit that generated the state to obtain the covariance matrix. 
From this, we can calculate the Williamson eigenvalues and thus any Rényi entropy using Eq.~\eqref{eq:renyi_entropy}. 
All operations described here are at most polynomial in the system size, making it feasible to simulate large systems.
For each system size and probability, we perform $10^3$ realizations of the circuit and average for $10^3$ time steps in the steady state.

We observe that the results of the Rényi-0 entanglement entropy, shown in the previous section, already ensure an area law phase for $p>1/2$ (since $S^{(0)}$ fulfills an area law, and $S^{(0)}\geq S^{(n)}$ for all $n$). Now we will be interested in the behavior within the volume law phase. 
The numerical results depicted in Fig.~\ref{fig: gate disentangler von Neumann}a show the existence of a sub-maximal volume law phase for $p<1/2$, and an area law phase for $p>1/2$,
\begin{equation}
\overline{S_{L/2}^{(1)}}\sim
    \begin{cases}
       s(p)L, \qquad p<1/2\\
      \sqrt{L}(1-aL^ {-\gamma}), \qquad p=1/2 \\
      \xi, \qquad p>1/2
    \end{cases}\,.
\end{equation}
The inset of Fig.~\ref{fig: gate disentangler von Neumann}a shows a data collapse assuming the power law behavior $s(p)\sim (1/2-p)^\beta$, indicating a critical exponent $\beta\approx0.33$ in the volume law phase.
Similar results are found for any Rényi entropy $S^{(n)}$ with $n>0$, with similar values for the critical exponent $\beta$.
Note that, in comparison to the model with a von Neumann disentangler from Fig.~\ref{fig: renyi disentangler}a, the prefactor of the volume law $s(p)$ here is larger than the one obtained in that model for small values of $p$. However, as argued in App.~\ref{appendix_vonNeumann_more}, for larger values of $p$ the gate disentangler is able to stabilize an area law, in contrast with the von Neumann disentangler.

The critical point exhibits a different behavior compared to $S^{(0)}$. In this case, $\overline{S^{(1)}_{L/2}}/L \rightarrow 0$ as $L \rightarrow \infty$, indicating a sub-volume scaling.
In Fig.~\ref{fig: gate disentangler von Neumann}b we show the von Neumann entanglement entropy profile for increasing system size, normalized by $\sqrt{L}$. 
From direct observation one cannot determine whether this is the correct scaling, due to the large finite size effects.
However, we propose the qualitative function $S^{(1)}_{L/2}=\sqrt{L}-aL^{1/2-\gamma}$, including some power-law second-to-leading order correction to the entanglement entropy.
A numerical fitting to the available data yields an excellent result (inset of Fig.~\ref{fig: gate disentangler von Neumann}b), with parameters $\gamma \approx 0.19$ and $a\approx 0.99$.
Further research is needed to fully characterize this critical point.

\section{Discussion}\label{sec:discussion}

Motivated by the investigation of entanglement phase transitions within the framework of unitary circuit games~\cite{unitary_games}, we considered the problem of implementing such games with matchgates.
To this end, we addressed two main challenges: First, the problem of finding optimal disentangling unitaries given full access to the fermionic Gaussian state; second, the investigation of unitary circuit games in two settings, one involving braiding gates (the intersection of matchgates and Clifford gates) and the other involving generic matchgates.

For braiding gates, we showed that a disentangler minimizing the von Neumann entanglement entropy is sufficient to fully disentangle the system.
Moreover, in the corresponding unitary circuit game, we found that any finite disentangling rate is capable of keeping the entanglement growth induced by the random evolution controlled, and therefore in an area law phase.
This demonstrates the robustness of the disentangling process against errors during the evolution.

For general fermionic Gaussian states, we studied the construction of optimal disentangling matchgates.
We introduced a novel representation of FGS in terms of structured matchgate circuits, which we named ``right standard form" (RSF).
These RSF circuits, which will be further studied in the complementary paper~\cite{LaMo26}, provide a practical tool for studying unitary circuit games with FGS: They enable efficient updates under gate application by absorbing gates into RSF circuits, and offer a systematic way to identify disentangling operations by essentially inverting the absorption procedure.
Such disentangling gates are optimal in the sense that they reduce maximally the minimal number of matchgates required to create the state.

Then, we analyzed different versions of the unitary circuit game with general matchgates.
When employing a strategy based on numerical minimization of the von Neumann entropy, we observed that the system remains in a volume law entangled phase even for large disentangling probabilities, and we found no evidence of a phase transition to an area law phase.
In contrast, when using the gate disentangler derived from the RSF, we uncovered a richer phase diagram.
By studying a simple model based on Bell pairs, we exactly reproduced the behavior of the Rényi-0 entropy, allowing simulations at large system sizes.
In this setting, we identified a sharp phase transition between volume law and area law entanglement at a critical probability $p_c = 1/2$, and we analytically characterized the critical point, where the entanglement entropy still scales as a volume law.
For the von Neumann entropy under the gate disentangler dynamics, we provided numerical evidence for a critical point also at $p_c = 1/2$, with consistent scaling $S_{L/2}^{(1)} \propto \sqrt{L}$ at leading order.
Overall, the existence of an extended area-law phase in the unitary circuit game can be understood as the region where an imperfect disentangler still succeeds: even if disentangling gates are applied correctly only a fraction $p$ of the time, as long as $p$ is large enough, the system eventually reaches an area-law phase.

The introduction of the RSF framework for FGS opens many new avenues.
From an application perspective, it would be natural to study measurement-induced phase transitions using similar methods.
Furthermore, investigating the stationary distribution of the circuit parameters under various types of dynamics could provide deeper analytical insights into entanglement properties and other quantities.
On the technical side, the development of more numerically stable update algorithms for RSF circuits would be highly valuable.
Several open questions remain regarding unitary circuit games with matchgates.
In the case of the gate disentangler, a more complete characterization of the critical state and the universality of the transition is an important direction for future work.
It also remains an open question whether an area-law phase exists for the von Neumann disentangler strategy, and how this disentangler behaves in the thermodynamic limit when acting on random FGS.
Finally, a promising direction for future research is the extension of unitary circuit games to settings where players aim to optimize physical quantities beyond entanglement entropy, such as energy, following strategies similar to those proposed in~\cite{Erbanni_2024}. Exploring such strategies could uncover novel dynamical phases and broaden the scope of the unitary circuit game framework.
Furthermore, one could consider an extension of the unitary circuit games to a multiplayer setting (with more than two players), extending strategies from the theory of quantum games~\cite{Benjamin_2001} to our random circuit scenario. 

\begin{acknowledgments}
We thank S. L. Sondhi for his contributions to the early development of the project and for insightful discussions. We are grateful to Beatriz Dias and Sheng-Hsuan Lin for insightful discussions. We thank Poetri Sonya Tarabunga for suggesting a simpler proof for the asymptotic form of the thermodynamic limit profile with the gate disentangler. R.M-Y. thanks Marco Lastres, Lukas Haller, Fabian Pichler, and Bernhard Jobst for helpful comments. A.G-S.~was supported by the UK Research and Innovation (UKRI) under the UK government’s Horizon Europe funding guarantee [grant number EP/Y036069/1]. This work was supported by the Deutsche Forschungsgemeinschaft (DFG, German Research Foundation) under Germany’s Excellence Strategy EXC-2111-390814868, TRR 360 (project-id 492547816), FOR 5522 (project-id 499180199), and the Munich Quantum Valley, which is supported by the Bavarian state government with funds from the Hightech Agenda Bayern Plus. M.L. and B.K. acknowledge funding from the BMW endowment fund and the Horizon Europe programmes HORIZON-CL4-2022-QUANTUM-02-SGA via the project 101113690 (PASQuanS2.1) and HORIZON-CL4-2021-DIGITAL-EMERGING-02-10 under grant agreement No. 101080085 (QCFD).

\textbf{Data and materials availability:} Raw data and simulation codes are available on Zenodo~\cite{zenodo}.
\end{acknowledgments}

\begin{appendix}

\section{Relations between matchgates}
\label{appendix_Yang_Baxter}

\newcommand{\ID}{\mathds{1}}

Matchgates are known to satisfy a generalized Yang-Baxter relation \cite{Yang_Baxter_circuit_compression}. Using this relation alone, it has been shown that any MG circuit can be compressed into a circuit of at most $L(L-1)/2$ gates \cite{Yang_Baxter_circuit_compression,Yang_Baxter_circuit_compression_algorithm}. As shown in this work, when also taking into account that MG satisfy the \mpsmove, more efficient circuit layouts for representing states can be determined. In this appendix, we give a simple proof of both the Yang-Baxter relation for matchgates, as well as the \mpsmove. The former is based on an Euler decomposition of the orthogonal matrix corresponding to a general MG circuit on three qubits, whereas the latter follows from a decomposition of arbitrary FGS on three qubits. Specifically, for the latter we show that any FGS on three qubits can be written as $U_{1,2}V_{2,3}\ket{000}$ (or as $U'_{2,3}V'_{1,2}\ket{000}$) with $U$, $U'$ and $V$, $V'$ MGs. We remark that with generic quantum gates, it is known that any state can be represented with either layout. The novelty of our result is that the same statement holds when restricting to FGS states and MGs. One of the reasons why we present very detailed proofs is that as such they can be directly translated into numerical algorithms to perform both the Yang-Baxter and the \mpsmove.\par\medskip

To prove the Yang-Baxter property, we will show the following more general statement: For an arbitrary MG circuit $V$ acting on three qubits, there exist three matchgates $U, U'$ and $U''$ s.t. 
\begin{equation}
V_{1,2,3} = U_{2,3} U'_{1,2} U''_{2,3}. \label{eq:three_qubit_mg_decomp}
\end{equation}
Similarly, there exists a decomposition of $V$ into three MG acting on qubits $(1,2), (2,3)$ and $(1,2)$. In the following, we only show Eq.~(\ref{eq:three_qubit_mg_decomp}), since the other decomposition can be constructed by just relabeling $1\leftrightarrow3$. Denote by $R$ the $6\times 6$ orthogonal matrix corresponding to $V$, i.e., the matrix satisfying 
\begin{equation*}
V^\dagger \gamma_i V = \sum_{j=1}^{6} R_{ij} \gamma_j. %
\end{equation*}
Note that when given $V$ as a matrix, one can compute $R$ via $R_{ij} = \Tr (V^\dagger \gamma_i V \gamma_j)$. On the other hand, if the MGs are given as $V = \exp(\mathrm{i} H)$, with the quadratic Hamiltonian
\[
    H = \mathrm{i} \sum_{kl} h_{kl}\gamma_k\gamma_l,
\]
where $h$ is an antisymmetric matrix, then the corresponding orthogonal matrix is given by $R = \exp(4h)$~\cite{MG5}.

We will now construct a particular Euler decomposition of $R$, from which one can read off three orthogonal $4\times 4$ matrices $R^{(1)}, R^{(2)}, R^{(3)}$ s.t.
\begin{equation}R = (\ID_2 \oplus R^{(1)}) \, (R^{(2)} \oplus \ID_2 ) \, (\ID_2 \oplus R^{(3)}),\label{eq:yb_from_euler_decomp_orth}\end{equation}
where $\ID_k$ denotes the $k$-dimensional identity matrix.

Let $c_1 = (c_{1,1},c_{1,2},c_{1,3},c_{1,4}, c_{1,5},c_{1,6})^\mathrm{T}$ denote the first column of $R$. When applying a rotation $E^{(5)}_{5,6}$ of the form
\[E^{(i)}_{j,j+1} = \ID_{j-1} \oplus \begin{pmatrix}\cos\alpha_i & -\sin\alpha_i \\ \sin\alpha_i&\cos\alpha_i \end{pmatrix} \oplus \ID_{6 - j -1},\]
with $\cos{\alpha_5} = c_{1,5} \; (c_{1,5}^2+c_{1,6}^2)^{-\frac12}$ and $\sin{\alpha_5} = -c_{1,6} \; (c_{1,5}^2+c_{1,6}^2)^{-\frac12}$, we get 
\[
E^{(5)}_{5,6} c_1 = (c_{1,1},c_{1,2},c_{1,3},c_{1,4}, \sqrt{c_{1,5}^2+c_{1,6}^2},0)^{\mathrm{T}}.
\]
Repeating this argument we can determine four additional rotations such that 
\[ E^{(1)}_{1,2} \, E^{(2)}_{2,3} \, E^{(3)}_{3,4} \, E^{(4)}_{4,5} \, E^{(5)}_{5,6} \, c_1 = \begin{pmatrix}1&0&0&0&0&0 \end{pmatrix}^\mathrm{T}.\]
When applying this chain of rotations to $R$, the first column of the resulting matrix is orthogonal to all the other columns, and therefore
\[ E^{(1)}_{1,2} \, E^{(2)}_{2,3} \, E^{(3)}_{3,4} \, E^{(4)}_{4,5} \, E^{(5)}_{5,6} \, R = \ID_1 \oplus S,\]
where $S$ is an orthogonal $5\times 5$ matrix. Using a similar argument, we get another four elementary rotations, and an orthogonal $4\times4$ matrix $R^{(3)}$ s.t.
\[E^{(6)}_{2,3} \, E^{(7)}_{3,4} \, E^{(8)}_{4,5} \, E^{(9)}_{5,6} \, (\ID_1 \oplus S) = \ID_2 \oplus R^{(3)}. \]
Since $E^{(8)}_{4,5} \, E^{(9)}_{5,6}$ and $E^{(1)}_{1,2} \, E^{(2)}_{2,3}$ commute, we can reorder the complete sequence of rotations to read
\begin{align*}
&E^{(6)}_{2,3} \, E^{(7)}_{3,4} \, E^{(8)}_{4,5} \, E^{(9)}_{5,6} \, E^{(1)}_{1,2} \, E^{(2)}_{2,3} \, E^{(3)}_{3,4} \, E^{(4)}_{4,5} \, E^{(5)}_{5,6} = \\
&E^{(6)}_{2,3} \, E^{(7)}_{3,4} \, E^{(1)}_{1,2} \, E^{(2)}_{2,3}  \, E^{(8)}_{4,5}  \, E^{(9)}_{5,6} \, E^{(3)}_{3,4} \, E^{(4)}_{4,5} \, E^{(5)}_{5,6}.
\end{align*}
In the last line, notice that the first four factors act only on the first four basis vectors, while the remaining five factors act only on the last four basis vectors. We can thus collect their respective actions into two rotations $R^{(2)\,\mathrm{T}} \oplus \ID_2 $ and $\ID_2 \oplus R^{(1)\,\mathrm{T}}$. In total, this gives
\[
(\ID_2 \oplus R^{(2)\,\mathrm{T}}) \, (R^{(1)\,\mathrm{T}} \oplus \ID_2) \, R = \ID_2 \oplus R^{(3)},
\]
i.e., the decomposition of $R$ leading to the decomposition in Eq.~(\ref{eq:yb_from_euler_decomp_orth}). Up to an irrelevant~\footnote{The global phase is irrelevant since we deal only with entanglement generated by MG circuits. One could however easily determine the global phase for an exact decomposition, since the involved unitaries are of size $8\times 8$.} global phase, each $R^{(i)}$ uniquely identifies a MG. For completeness, we outline how one could determine these MGs. The first step is to find again an Euler decomposition of the arbitrary $4\times 4$ rotation matrix, $R$ into elementary rotations. Each elementary rotation $E_{j,j+1}(2 \alpha)$ then corresponds to a single-qubit or nearest neighbor MG $\exp(-\alpha \gamma_j \gamma_{j+1})$. Finally, these MGs can be multiplied to give the MG $U$ corresponding to $R$.
\par \medskip

Let us now provide a proof of the {\mpsmove} property. We show an equivalent statement, namely that any pure FGS on three qubits can be generated by two MGs acting a computational basis state. In  Ref.~\cite{Br05_fgs_on_three_qubits}, it has been shown that the set of pure three-qubit fermionic states~\footnote{Fermionic states are eigenstates of the parity operator $Z^{\otimes L}$, where $Z$ is the Pauli-$Z$ matrix. Even and odd parity refers to an eigenvalue of $+1$ or $-1$.} and FGS coincide. Therefore, we show that any even parity fermionic state $\ket\psi$ on three qubits can be generated by two MGs, $U$ and $V$, s.t. 
\begin{equation}\ket{\psi} = U_{2,3} V_{1,2}\ket{000}.\label{eq:ex_u_v_for_lr_move}\end{equation} 
A proof for the odd parity states, as well as the reverse order of the MGs, is analogous. An arbitrary even parity fermionic state can be written as
\[\ket{\psi} = \mu(\alpha\ket{000} + \beta\ket{011}) + \nu(\gamma\ket{110} + \delta \ket{101}),\]
with $\vert\alpha\vert^2 + \vert\beta\vert^2 = \vert\gamma\vert^2 + \vert\delta\vert^2 =\vert \mu\vert^2 + \vert \nu\vert^2 = 1$. Introduce the matrices
\[ A = \begin{pmatrix}\alpha^*  &  \beta^* \\ -\beta & \alpha \end{pmatrix}, \quad\mathrm{and}\quad B = \begin{pmatrix} \gamma&-\delta\\\delta^*&\gamma^* \end{pmatrix}. \]
Clearly, $A$ and $B$ are unitary and have the same determinant, therefore $G(A,B)$ is a MG, and
\[ G(A,B)_{2,3} \ket{\psi} = \mu\ket{000} + \nu\ket{110}. \]
Finally, with the unitary
\[ C = \begin{pmatrix} \mu^*&\nu^*\\-\nu&\mu \end{pmatrix}, \]
we get
\[\ket{000} = G(C,C)_{1,2} \, G(A,B)_{2,3} \ket{\psi}.\]
This demonstrates the existence of MGs $U$ and $V$ in Eq.~\eqref{eq:ex_u_v_for_lr_move}. By relabeling qubits $1\leftrightarrow3$ we get the alternative decomposition.

\section{General circuit for pure FGS}
\label{appendix_FGS_circuit}

In this appendix, we state again the formal definition of right standard form (RSF) circuits. Furthermore, we prove that the application of the absorption algorithm as presented in the main text always yields a RSF circuit. Specifically, in Lemma~\ref{lemma: absorbing} we show that when representing a state $\ket\psi$ with an RSF circuit, for any $i$ and any matchgate $U$, the state $U_{i,i+1}\ket\psi$ can also be represented as an RSF circuit. The RSF of the latter can be computed efficiently, given the RSF of $\ket\psi$. As a corollary, we get that all fermionic Gaussian states on $L$ qubits can be represented as a circuit of at most $\floor{L^2/4}$ gates acting on a computational basis state (see Corollary~\ref{corollary:maximalRSF}). \par\medskip

The RSF is a special layout a circuit can have. When referring to the layout of a circuit, we mean in which order gates are applied on which qubits. For instance, consider a circuit given by a sequence of $K$ gates $(U^{(i)})_{i=1}^K$ together with the instructions that the $i$-th gate acts on qubits $m_i,m_i+1$ for each $i$~\footnote{To be precise, we consider circuit layouts to be equivalent if the gates act on distinct qubits and therefore can be exchanged. For example, the layouts $U_{1,2}V_{3,4}$ and $V_{3,4}U_{1,2}$ are equivalent.}. The action of the whole circuit is then given by
\[
\prod_{i=1}^K U^{(i)}_{m_i, m_i+1} = U^{(K)}_{m_K,m_K+1} \ldots U^{(1)}_{m_1,m_1+1}.
\]

\begin{figure}[t!]
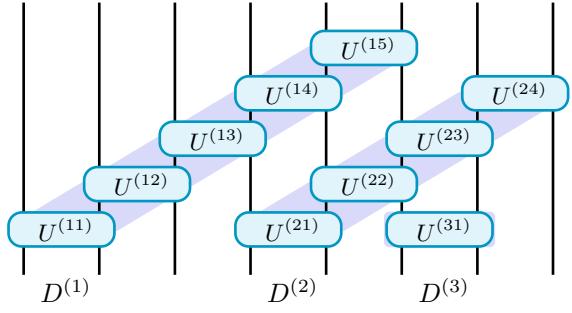

    \centering
    \drawIncompleteRSFExampleALabeled
    \caption{An example circuit in right standard form labeled by $((1,5),(4,4),(6,1))$ consisting of the three diagonals $D^{(1)}$, $D^{(2)}$, $D^{(3)}$.}
    \label{fig:labeled RSF circuit example}
\end{figure}

\begin{definition}[Right standard form (RSF)]
Let $\numdiag$ be an integer, with $0\leq \numdiag \leq L/2$, and $((k_i,l_i))_{i=1}^{\numdiag}$ be a list of $\numdiag$ pairs of integers such that $1\leq k_i \leq k_{i+1} -2 \leq L-1$, and $1\leq l_i\leq L-k_i$. A circuit is in the right standard form labeled by the sequence $((k_i,l_i))_{i=1}^{\numdiag}$, if it is a concatenation 
\[ D^{(1)} D^{(2)} \ldots D^{(n_d)} \]
of $\numdiag$ \emph{diagonals}, given by
\[ D^{(i)} = U^{(i,l_i)}_{k_i+l_i - 1, k_i+l_i} \ldots U^{(i,1)}_{k_i, k_i+1},
\]
where $U^{(i,j)}_{a,a+1}$ are nearest neighbor two qubit gates labeled by $i$ and $j$, and acting on qubits $a$ and $a+1$.
\end{definition}
An example of a RSF circuit with labeled gates and diagonals is presented in Fig.~\ref{fig:labeled RSF circuit example}. We remark that the condition $k_i \leq k_{i+1} - 2$ ensures that the first gate in each of the diagonals act on distinct qubits (for instance, in Fig.~\ref{fig:labeled RSF circuit example}, the gates $U^{(11)},U^{(21)}$ and $U^{(31)}$ act on qubits $(1,2)$, $(4,5)$ and $(6,7)$ respectively). Considering how the diagonals are combined into the circuit, one can view the first gate of each diagonal to act in parallel. The same is true for the second, third, and subsequent gates of each diagonal. That is, a circuit in RSF can also be composed of $n_\mathrm{l} = \max_{i} l_i$ layers of the form
\[
\prod_{i=1}^{\numdiag} U^{(i,j)}_{k_i +j - 1, k_i+ j},
\]
$j=1,\ldots, n_\mathrm{l}$, where we set $U^{(ij)} = \ID$ if $j > l_i$ to conform with the fact that any given layer is not necessarily densely populated with gates (as is e.g. the case with any layer in Fig.~\ref{fig:labeled RSF circuit example}). \par\medskip

Let us now prove that any FGS $\ket\psi = U \ket{0\ldots0}$ can be represented with an RSF circuit. The idea is to decompose $U$ into individual MGs $U^{(i)}$, and define intermediate states via $\ket{\psi^{i+1}} = U^{(i)}\ket{\psi^i}$. Clearly, the initial state $\ket{\psi^0} = \ket{0\ldots0}$ is represented by the empty RSF. The following lemma ensures that any intermediate state, including the final one, can be represented again by an RSF circuit.

\begin{lemma}
\label{lemma: absorbing}
    Consider a state defined with an RSF matchgate circuit acting on $\ket{0}^{\otimes L}$. When applying another matchgate to any pair of consecutive qubits, the resulting state can again be described by a — potentially different — RSF circuit. The new circuit may be obtained using the absorption algorithm described in Sec.~\ref{abs_algorithm}.
\end{lemma}
\begin{proof} We start by remarking that the case of an empty RSF is trivial. Suppose the circuit is given in non-empty RSF $((k_i,l_i))_{i=1}^{\numdiag}$, and the additional gate is acting on qubits $q,q+1$. We will apply the absorption algorithm from Sec.~\ref{abs_algorithm} to the additional gate. To this end, we consider every possible position of the applied gate $q$ relative to the parameters $(k_1,l_1)$ of the first diagonal. Depending on their relation, we either explain directly how the new RSF is obtained, or we reduce the problem to absorbing a gate into an RSF with $\numdiag -1$ diagonals. Thus, repeating the following argument at most $\numdiag$ times shows how the new RSF can be obtained. Let us assume $l_1 \geq 2$, that is, the first diagonal contains at least two gates and acts at least on qubits $k_1, k_1+1$, and $k_1+2$. The case $l_1=1$ is a trivial modification of the following.
\begin{enumerate}
    \item \label{item:proof absorbing commute} $q \geq k_1 +l_1 + 1$. In this case, the additional gate commutes with the first diagonal since they act on different qubits: 
    \begin{center}$\quad\quad$\drawAbsorbProofCaseA\end{center}
    The problem then reduces to absorbing an additional gate on qubits $q,q+1$ into a circuit in RSF  $((k_i,l_i))_{i=2}^{\numdiag}$.
    
    \item $q = k_1 +l_1$. None of the steps of the absorption algorithm can be applied. That is, it terminates immediately. The gate will be attached to the first diagonal, giving the circuit 
    \begin{center}$\quad\quad$\drawAbsorbProofCaseB\end{center}
    in RSF $((k_1, l_1 + 1), \ldots, (k_\numdiag, l_\numdiag))$.

    \item $q = k_1 + l_1 - 1$. The gate can be combined with the last gate of the diagonal:
    \begin{center}$\quad\quad$\drawAbsorbProofCaseC\end{center}
    The form of the circuit is not modified.

    \item \label{item:proof absorbing yb reduction} $k_1 < q < k_1 + l_1 - 1$. Step $1$ of the absorption algorithm can be applied:
    \begin{center}$\quad\quad$\drawAbsorbProofCaseD\end{center}
    After doing so, the problem reduces to absorbing a gate on qubits $q+1,q+2$ into a smaller circuit in RSF  $((k_i,l_i))_{i=2}^{\numdiag}$.

    \item \label{item:proof absorbing many substeps} $q = k_1$. Here, one needs to take into account the parameters $(k_2,l_2)$ of the second diagonal and consider two sub-cases. In the first one, $k_2 > k_1 + 2$. Here, the algorithm performs step $1$ once, then step $3$, then combines two gates:
    \begin{center}$\quad\quad$\drawAbsorbProofCaseE\end{center}
    The resulting circuit remains in the initial RSF. In the second case, $k_2 = k_1 + 2$, the algorithm applies in sequence steps $1$, $2$ and $3$:
    \begin{center}$\quad\quad$\drawAbsorbProofCaseF\end{center}
    From here on, again several cases need to be considered: If $l_1 - 1 > l_2$, step 4 is applied $l_2$ times, then the marked gate can be absorbed into a another gate: 
    \begin{center}$\quad\quad$\drawAbsorbProofCaseG\end{center}
    The circuit remains in the initial RSF. If conversely, $l_1 - 1 \leq l_2$, after $l_1 - 1$ applications of step 4 no absorption is possible, and the algorithm terminates:
    \begin{center}$\quad\quad$\drawAbsorbProofCaseH\end{center}
    The circuit is then in RSF $((k_1, l_2 +2 ), (k_2, l_1 - 1), (k_3, l_3),\ldots, (k_\numdiag, l_\numdiag))$.

    \item \label{item:proof absorbing add left} $q = k_1 -1$. Step $2$ of the algorithm will be applied, since the check that the requirements for step $1$ are not given. After step $2$, no further steps are possible:
    \begin{center}$\quad\quad$\drawAbsorbProofCaseI\end{center}
    The resulting circuit is in RSF $((k_1 - 1, l_1 + 1), \ldots, (k_\numdiag, l_\numdiag))$.

    \item $q \leq k_1 - 2 $. None of the steps in the algorithm can be applied, it therefore terminates immediately. The newly obtained circuit 
    \begin{center}$\quad\quad$\drawAbsorbProofCaseJ\end{center}
    is already in RSF $((q,1), (k_1, l_1), \ldots (k_\numdiag, l_\numdiag))$.

\end{enumerate}
In two of the cases, a recursion argument is used, and the gates needs to be absorbed into an RSF circuit with $\numdiag -1$ diagonals. Repeated recursion will therefore end after at most $\numdiag$ steps.
\end{proof}

This lemma also ensures that all states occurring are always represented by an RSF circuit. Note that there exists an RSF $((2i - 1, L - 2i + 1))_{i=1}^\numdiag$ with $\numdiag = \floor{L/2}$ that has the maximal number of gates any RSF on $L$ qubits can have. Absorbing any additional gate into this circuit gives the same RSF. For the representation of generic states, we therefore get the following: 
\begin{corollary}
\label{corollary:maximalRSF}
    Any pure fermionic Gaussian state can be generated by at most $\floor{L^2 /4}$ matchgates acting on $\ket{0}^{\otimes L}$.
\end{corollary}

We conclude this section with a few remarks. Firstly, our result gives a slight improvement over the previously known $L(L-1)/2$ necessary gates~\cite{Yang_Baxter_circuit_compression,Yang_Baxter_circuit_compression_algorithm} to represent an arbitrary FGS. Secondly, it is important to note that the proofs in this section only rely on the usage of Yang-Baxter and \mpsmoves. The results obtained therefore hold also for any other gate set satisfying these properties. Finally, when a circuit is given in RSF, one cannot find another circuit with fewer gates by only using Yang-Baxter and \mpsmoves: This follows from a result in Ref.~\cite{LaMo26}, where we show that in the generic case, when $\ket\psi=U\ket{0}^{\otimes L}$ with an RSF matchgate circuit $U$, there cannot exist another MG circuit $V$ with fewer gates than $U$ such that $\ket\psi = V \ket{0}^{\otimes L}$ \footnote{In non-generic cases, for instance if all gates commute, it is possible to find circuits with fewer gates. This is, however, due to additional properties of the gates, and cannot be achieved using the Yang-Baxter and \mpsmoves{}, and consecutive gate combination alone.}.

\section{Majorana braiding model}
\label{appendix_braiding}

In this appendix, we summarize the mapping of the braiding model onto a Majorana loop model, which serves as the foundation for our numerical simulations presented in Sec.~\ref{sec:braiding}. 
Further details about this map can be found in Refs.~\cite{miptff11, Klocke_2023}.

We employ the formalism of fermionic stabilizer states, introduced in~\cite{bravyi}.
Consider a system with $L$ qubits. An operator $s$ is said to be a stabilizer of a state $\ket{\psi}$ if its action on the state is trivial, $s\ket{\psi}=\ket{\psi}$.
The state $\ket{\psi}$ is said to be a fermionic stabilizer state if it has $L$ stabilizers of the form $\pm i\gamma_i\gamma_j$, where $1\leq i,j \leq 2L$, with the constraint that each Majorana operator is used exactly once in all $L$ stabilizers.
In this framework, the entanglement entropy is easily accessible~\cite{fattal2004entanglementstabilizerformalism}: For a bipartition of the lattice $A=\{1,\dots,n\}$ and $B=\{n+1,\dots,L\}$, we count how many stabilizer pairs of Majoranas cross the boundary, considering that Majoranas $\gamma_{2i-1}$ and $\gamma_{2i}$ belong to site $i$. The entanglement entropy is then given by half the number of such boundary-crossing pairs.
Notice that the sign of the stabilizers does not affect the value of the entanglement entropy. Therefore, we will not include it in the following discussion.
 
\begin{figure}[ht]
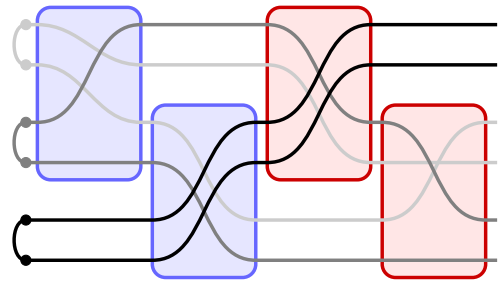

    \centering
    \drawLoopFigure
    \caption{Illustrative evolution of Majorana pairs corresponding to each fermionic stabilizer. The blue boxes represent random braiding gates, while red boxes denote disentangling gates. Time evolves from left to right. Each Majorana pair is depicted as a line with a distinct shade of gray.}
    \label{fig:with_loops}
\end{figure}

Now, we describe the updating rule of the stabilizers when applying random braiding and disentangling gates.
A random braiding gate $U_{i,i+1}$ acts on the four Majoranas contained in bond $i$, namely $\gamma_{2i-1}$, $\gamma_{2i}$, $\gamma_{2i+1}$, and $\gamma_{2i+2}$, by randomly permuting their positions.  
Formally, this corresponds to applying a permutation $\tau$ that rearranges the four affected Majoranas while leaving all others unchanged, substituting the indices in the stabilizers with $\tau(k)$.  
For example, the first two unitaries in Fig.~\ref{fig:with_loops} perform the updates in the stabilizers
\begin{equation*}
    \begin{matrix} i\gamma_1\gamma_2 \\ i\gamma_3\gamma_4 \\ i\gamma_5\gamma_6 \end{matrix}
\quad\xrightarrow{U_{1,2}}\quad
\begin{matrix} i\gamma_2\gamma_3 \\ i\gamma_1\gamma_4 \\ i\gamma_5\gamma_6 \end{matrix}
\quad\xrightarrow{U_{3,4}}\quad
\begin{matrix} i\gamma_2\gamma_5 \\ i\gamma_1\gamma_6\\ i\gamma_3\gamma_4 \end{matrix}\,,
\end{equation*}
where the first permutation is $\tau_1 = (321)$ and the second permutation is $\tau_2=(35)(46)$.

A disentangling operation in bond $i$ is implemented by choosing a permutation of the Majoranas $\gamma_{2i-1}$, $\gamma_{2i}$, $\gamma_{2i+1}$, and $\gamma_{2i+2}$ that reduces the entanglement entropy. 
This is accomplished by arranging the Majoranas such that the distance between the paired Majoranas in each stabilizer is minimized.
As an example, the last two unitaries in Fig.~\ref{fig:with_loops} perform the disentangling updates in the stabilizers
\begin{equation*}
    \begin{matrix} i\gamma_2\gamma_5 \\ i\gamma_1\gamma_6\\ i\gamma_3\gamma_4 \end{matrix}
\quad\xrightarrow{U'_{1,2}}\quad
\begin{matrix} i\gamma_4\gamma_5 \\ i\gamma_3\gamma_6 \\ i\gamma_1\gamma_2 \end{matrix}
\quad\xrightarrow{U'_{3,4}}\quad
\begin{matrix} i\gamma_4\gamma_5 \\ i\gamma_5\gamma_6\\ i\gamma_1\gamma_2 \end{matrix}
\end{equation*}
After the final unitary, each stabilizer consists of the two Majorana operators corresponding to one site, and therefore the system is in a product state.

\section{Review of finite-size scaling and critical exponents}
\label{appendix_finite_size}

In this appendix, we provide a brief review about the finite-size scaling hypothesis and critical exponents.
A detailed discussion about this topic can be found in~\cite{Sandvik_2010}.

The idea behind finite-size scaling is to define a characteristic length scale $\xi$, usually denoted as correlation length, that diverges in the critical point.
When the system size is much larger than this length, $L\gg\xi$, then the system effectively acts as it would in the thermodynamic limit, with $\xi$ being the natural length scale. Instead, when $L\ll\xi$, the finite size of the system becomes relevant and thus $L$ becomes the natural length scale of the system.
The scaling hypothesis tells that any quantity that is singular in the critical point has the form
\begin{equation}
    Q(p, L) = L^\sigma \tilde{F}(\xi / L),
\end{equation}
where $\sigma$ is the scaling of the quantity $Q$ at the critical point. Assuming that the characteristic length diverges as a power law in the critical point, $\xi\sim|p-p_c|^{-\nu}$, we can write
\begin{equation}
    Q(p, L) = L^\sigma F(|p-p_c|L^{1/\nu}).
\end{equation}
Therefore, plotting $Q(p,L)/L^\sigma$ as a function of $|p-p_c|L^{1/\nu}$ should yield a single curve for the data with different system sizes. 

As an example, consider the phase transition in the Bell pair model discussed in Sec.~\ref{sec: bell pair} of the main text, where we numerically find the scaling
\begin{equation}
\overline{S_{L/2}^{(0)}}\sim
    \begin{cases}
       \frac{L-\xi}{2}, \qquad p<p_c\\
      \frac{L}{4}, \qquad p=p_c \\
      \xi, \qquad p>p_c
    \end{cases}\,.
\end{equation}
Here, we observe that at the critical point we have $\sigma = 1$. Therefore, we expect the function $F$ to have the asymptotic behavior
\begin{equation}
F(x)\propto
    \begin{cases}
       \frac{1-|x|^{-\nu}}{2}, \qquad x\rightarrow -\infty\\
      \text{const}, \qquad x=0 \\
      x^{-\nu}, \qquad x\rightarrow\infty
    \end{cases}\,.
\end{equation}
This is the form observed in the inset of Fig.~\ref{fig: gate disentangler classical}a of the main text for $\nu\approx1$.

\section{Free fermion disentanglers}
\label{appendix_disentanglers}

\begin{figure*}[ht]
    \centering
    \includegraphics{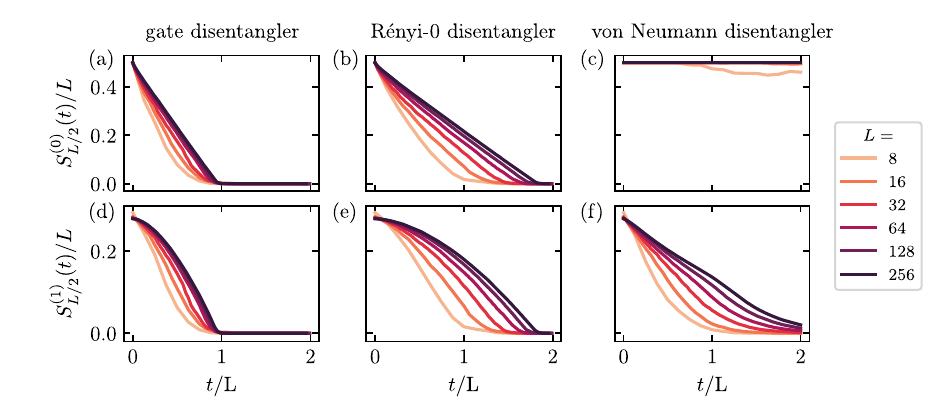}
    \caption{Evolution of Rényi-0 entropy $S^{(0)}$ (first row) and von Neumann entropy $S^{(1)}$ (second row) for a disentangling evolution starting from a random FGS with gate disentangler, Rényi-0 disentangler, and von Neumann disentangler. All quantities are normalized by the system size $L$. Each line is the average of 100 runs, and each run is evolved for $2L$ times steps, where each time step corresponds to applying $L$ disentangling gates in random bonds.}
    \label{fig:disentanglers}
\end{figure*}

In Section~\ref{sec:unitary game generic} we study the results of the unitary circuit game with general matchgates for different choice of disentanglers. In this appendix, we assess the performance of three different disentanglers when acting on a random FGS: the gate disentangler, the Rényi-0 disentangler, and the von Neumann disentangler. In particular, we would like to find disentangling strategies that not only reduce the entanglement entropy in some quantifiable way, but also that repeated applications of such algorithm in different bonds bring the state into a product state. 

The gate disentangler reduces the number of matchgates required to prepare the state in RSF, using the disentangling algorithm described in Sec.~\ref{disent_alg}. In contrast, the Rényi-0 and von Neumann disentanglers minimize the corresponding entropy at the given bipartition, as described by Eq.~\eqref{eq: minimization}. However, it is important to note that the Rényi-0 entropy is integer-valued for FGS. As a result, the minimization process involves identifying a fine-tuned minimum for specific parameters of the matchgate. Consequently, numerical minimization methods are often insufficient for finding the optimal gate. Instead, we employ the disentangling algorithm used by the gate disentangler to determine the optimal gate, and then we apply it only if it successfully reduces the Rényi-0 entropy at the bipartition (and otherwise we just apply the identity).

The numerics in this appendix are performed in the following way: we start with a product state with $L$ qubits and apply $L^3$ random matchgates in random bonds using the absorption algorithm. Then, we use the resulting state in RSF to disentangle the state with the three different disentanglers. The von Neumann disentangler is performed based only on information of the covariance matrix, while the others use the RSF to find the disentangling gate using the disentangling algorithm. We run such simulation for system sizes up to $L=256$ and for time $2L$, with each time step consisting of $L$ disentangling gates acting on random bonds. We perform 100 runs of the simulation.

The results of such simulations are shown in Fig.~\ref{fig:disentanglers}. The first row shows how the Rényi-0 entropy is reduced for different disentanglers and system sizes. Among the disentanglers, the gate disentangler achieves the fastest disentangling, with the disentangling time converging to $t_{\text{disent}}\approx L$. The Rényi-0 disentangler is slower but still fully disentangles the state within a time $t_{\text{disent}}\approx 2L$. In both cases, the Rényi-0 entropy decreases linearly over time. In contrast, the von Neumann disentangler fails to reduce the Rényi-0 entropy during the disentangling process. This highlights that the von Neumann disentangler focuses on minimizing the magnitude of the singular values rather than driving them to exactly zero, as achieved by the other two disentanglers.

The shorter disentangling time for the gate disentangler indicates that in some cases the depth of the circuit can be reduced without changing the bipartite Rényi-0 entropy. For example, consider an RSF with generic gates for $L=4$, and suppose that we act with a disentangling gate in the first bond:
\begin{center}$\quad\quad$\drawDisentanglingExampleI\end{center}
In this case, the Rényi-0 disentangler would not apply any gate, since its profile is maximal and therefore it can be reduced only in the central bond. In contrast, the gate disentangler can remove one gate, leading to the RSF
\begin{center}$\quad\quad$\drawDisentanglingExampleII \end{center}
which still has a maximal profile, but is generated by one gate fewer.

The second row of Fig.~\ref{fig:disentanglers} shows the evolution of the von Neumann entanglement entropy as a function of time for the different disentanglers. While the disentangling times for the gate disentangler and the Rényi-0 disentangler remain identical to those observed for the Rényi-0 entropy, the decay of entanglement is not linear over time. In the initial stages, the von Neumann disentangler achieves a more rapid reduction in entropy. However, as time progresses, both the gate disentangler and the Rényi-0 disentangler outperform the von Neumann disentangler. This indicates that strategies with better short-term results may not necessarily be optimal in the long term.

We checked that the results presented in this appendix corresponding to the von Neumann disentangler are qualitatively equivalent for other choices of Rényi-$n$ disentanglers, with $n>0$.

\section{Volume-law phase for the unitary game with von Neumann disentangler}
\label{appendix_vonNeumann_more}

\begin{figure}[t!]
    \centering
    \includegraphics[width=1\columnwidth]{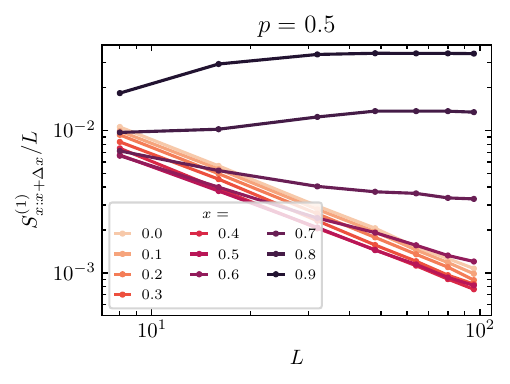}
    \caption{Averaged contribution to the half-chain von Neumann entanglement entropy given by the Williamson eigenvalues in the interval $[x, x+dx)$ with $dx=0.1$ at disentangling probability $p=0.5$. Each point is obtained by averaging 100 trajectories and 100 time steps into the steady state of each.}
    \label{fig: renyi disentangler partition}
\end{figure}

To determine the existence of a volume-law phase in the unitary game with von Neumann disentangler, we examine the contribution of the Williamson eigenvalues to the entanglement entropy.
In particular, we consider the quantity
\begin{equation}
    S^{(n)}_{x:y} = \frac{1}{1-n}\sum_{x\leq\lambda<y} \log_2\left[\left(\frac{1+\lambda}{2}\right)^n+\left(\frac{1-\lambda}{2}\right)^n\right],
\end{equation}
where we only sum the contribution to the Rényi entropy of the Williamson eigenvalues in the interval $[x,y)$. We observe that by definition $ S^{(n)}_{0:1} =  S^{(n)}$. In Fig.~\ref{fig: renyi disentangler partition} we show the behavior of $S^{(1)}_{x:x+\Delta x}/L$ for $\Delta x=0.1$ and different values of $x$ at $p=0.5$. These numerical results indicate that the contribution of small Williamson eigenvalues decays with system size. However, the contribution of eigenvalues close to 1 remains constant with increasing system size. This indicates that the steady state is still in a volume law phase, while explaining the behavior observed in Fig.~\ref{fig: renyi disentangler}, where the half-chain entanglement entropy normalized by the system size decays with increasing system size. A similar behavior is observed for other disentangling probabilities up to $p=0.6$. 

Now, we can compare the results of the von Neumann disentangler (Fig.~\ref{fig: renyi disentangler}) with the results for the gate disentangler (Fig.~\ref{fig: gate disentangler von Neumann}a) of the main text. 
We observe that the von Neumann disentangler is more effectively reducing the entanglement entropy for low values of $p$, where we find a volume law phase in both cases. 
Instead, beyond $p=0.5$ the von Neumann disentangler still has a volume law phase, while the gate disentangler is able to induce an area law phase.

Determining the existence of an extended area law phase for the von Neumann disentangler remains inconclusive based on our numerical results.
Specifically, the results for disentangling probabilities close to $1$ could be consistent with either a volume law characterized by a very small prefactor or an area law.

\section{Equivalence of RSF circuits and Bell pair configurations}\label{appendix:telephone}

When simulating the circuit game with the gate disentangler in the volume-law phase, each entangling or disentangling gate requires performing a sequence of $\mathcal{O}(L)$ Yang-Baxter or \mpsmoves{}. To enable the simulation of larger systems, we introduce an equivalent model that provably yields the same evolution of the $S^{(0)}$ entropy. Crucially, each update in this model can be executed with only $\mathcal{O}(1)$ operations. Additionally, this model simplifies several counting problems that will arise in the next section, where we analytically investigate properties of the circuit game's critical point. 

In the following, we first introduce this model and then prove its exact equivalence to the unitary circuit game with the gate disentangler, when restricted to the dynamics of the $S^{(0)}$ entropy. Note that one cannot find here an equivalent model based solely on the entanglement profile, such as the surface growth model considered in Ref.~\cite{unitary_games}. This is because distinct RSF circuits can give rise to the same entanglement profile. For instance, the RSF circuits labeled by $((1,2),(3,1))$ and $((1,3),(3,1))$ generically give rise to the same Rényi-0 entanglement profile~\cite{Nahum_2017}. By applying a suitable gate on qubits 2 and 3, in the former RSF, qubits 1 and 2 can be completely disentangled from qubits 3 and 4, whereas the best one can do in the other case is to reduce the Rényi-0 entropy by 1 (see also Appendix~\ref{appendix_disentanglers}). \par \medskip

The simplified model can be described as follows: Consider $L$ qubits arranged on a line. In a valid configuration, each qubit is either in the state $\ket{0}$ or forms a Bell pair with exactly one other qubit. All such configurations can be uniquely labeled by a partition of the set $\{1,\ldots,L\}$ into subsets of size $1$ and $2$. The former correspond to qubits in the state~$\ket{0}$, while the latter represent qubits in the Bell state $\ket{\Phi^+}$. For example, the state $\ket{\Phi^+}_{1,5}\ket{\Phi^+}_{2,3}\ket{\Phi^+}_{4,7}\ket{0}_{6}\ket{0}_8$ corresponds to the partition $\{\{1,5\}, \{2,3\}, \{4,7\}, \{6\}, \{8\}\}$, and can be represented graphically as 
\[\drawExampleStringState \; .\]
The entanglement of a state $\ket{D}$, labeled by the partition $D$, with respect to the bipartition $1\ldots m \,\vert\, m+1\ldots L$, is given by the number of Bell pairs that cross the boundary. That is, 
\[S^{(0)}_m (\ket{D})  = \# \{\{s_1,s_2\} \in D \;\vert\; s_1\leq m < s_2\}.\]
To obtain the dynamics of this Bell pair model, we impose some update rules as explained in the main text. These include the creation of a Bell pair on adjacent qubits initially in the $\ket{00}$ state, its inverse operation, and swapping of adjacent qubits. \par\medskip

We will now construct a mapping, $\pi$, between RSF circuit layouts and the Bell pair configurations. A straightforward way to do so is as follows: Given an RSF circuit, replace each gate in the first layer with a gate $G$ satisfying $G\ket{00} = \ket{\Phi^+}$, and replace all subsequent gates with SWAP gates. Acting with this modified circuit on the initial state $\ket{0}^{\otimes L}$ defines the state $\ket{\pi(C)}$. For example, the circuit 
\[\Bellpairmodelexample\]
produces the Bell pair configuration illustrated earlier. We claim that the mapping $\pi$ introduced here is a bijection, and provide justification for that below. We also need to show that the entanglement entropies $S_m^{(0)}(C\ket{0}^{\otimes L})$ and $S_m^{(0)}(\ket{\pi(C)})$ are equal for any RSF circuit $C$ and for all bonds $m$. Finally, we need to demonstrate that the update rules governing each model induce consistent dynamics: Specifically, if applying the absorbing algorithm at bond $b$ transforms $C$ into $\tilde{C}$, then the Bell pair update rules at bond $b$ must transform the state $\ket{\pi(C)}$ into $\ket{\pi(\tilde{C})}$.

To prove that $\pi$ is bijective, we first show that the cardinalities of the sets of RSF circuit layouts and Bell pair configurations on $L$ qubits coincide. This can done by finding appropriate recursive definitions of the respective sets. We then leverage the fact that these definitions are very similar in their structure to construct $\pi$ in a more comprehensible way, from which its bijective nature follows immediately. We then present Propositions~\ref{prop:bijection} and~\ref{prop:dynamics}, which guarantee that $\pi$ preserves both the $S^{(0)}$ entanglement structure and the corresponding dynamics of the two models.

We begin by motivating how a recursive definition of RSF circuits (and Bell pair configurations) can be obtained. Consider an arbitrary RSF circuit labeled by $((k_i,l_i))_{i=1}^\numdiag$ acting on $L+1$ qubits. If the first diagonal does not begin on the first qubit, i.e., $k_1 \geq 2$, then by omitting the first qubit, one obtains an RSF circuit on $L$ qubits. In the alternative case, where $k_1 = 1$, a diagonal starts at the first qubit, and removing this initial diagonal yields an RSF circuit on $L-1$ qubits. Importantly, these two cases are mutually exclusive: an RSF circuit that satisfies the first condition cannot satisfy the second, and vice versa.

A recursive definition of RSF circuits can be formulated as follows: Any RSF circuit on $L+1$ qubits is either (i) an RSF circuit on $L$ qubits extended by adding one qubit to the left, or (ii) an RSF circuit on $L-1$ qubits extended by adding two qubits to the left and introducing a non-empty diagonal starting on those two qubits. In the second case, the length of this additional diagonal can be any integer between $1$ and $L$. The total number $T(L)$ of RSF circuits on $L$ qubits thus satisfies the recurrence relation
\begin{equation}
    T(L+1) = L \; T(L-1) + T(L), \label{eq:telephone_recursion}
\end{equation}
with initial values
\[T(0) = 1, \; T(1) = 1,\; T(2) = 2,\; T(3) = 4, \ldots\]
The sequence $T(L)$ is known as the sequence of \emph{telephone numbers} or \emph{involution numbers}~\cite{origin_of_telephone}. These numbers will play a central role in Appendix~\ref{appendix_asymptotics}, where we analyze the entanglement profile at the critical point of the unitary circuit game.

We now turn to Bell pair configurations and show that they can be generated recursively in a manner analogous to RSF circuits. Consider a Bell pair configuration state $\ket{D}$ on $L+1$ qubits. There are two distinct possibilities: either (i) the first qubit is in the state $\ket{0}$, and the remaining $L$ qubits form an arbitrary Bell pair configuration; or (ii) the first qubit is entangled with one of the remaining $L$ qubits, in which case the other $L-1$ qubits form an arbitrary Bell pair configuration. The total number of Bell pair configurations on $L$ qubits is thus also given by $T(L)$.

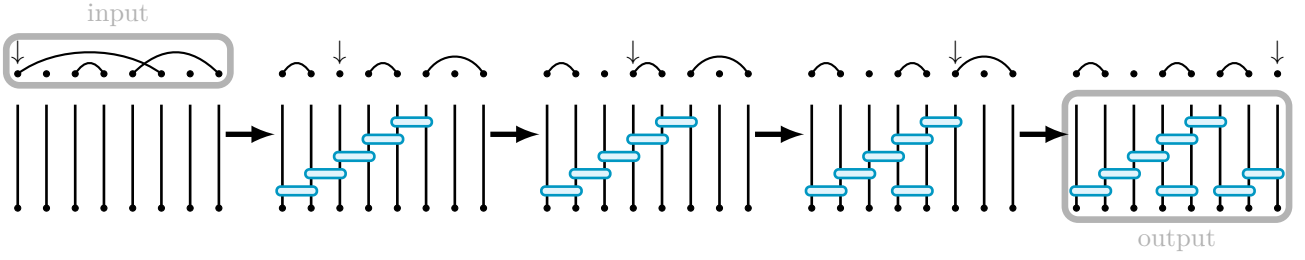
\begin{figure*}[ht]
    \begin{tikzpicture}
    \begin{scope}[xshift = 0cm]
    	\begin{scope}[scale=0.38]
            \draw[black!30, line width=2.5pt, rounded corners = 5pt] (-0.4,-0.4) rectangle node[above,yshift=0.3cm]{input} (7.4,1.3);
			\draw[black, thick] (0,0) .. controls (1.333,1) and (3.666,1) .. (5,0);
			\draw[black, thick] (2,0) .. controls (2.333,0.5) and (2.666,0.5) .. (3,0);
			\draw[black, thick] (4,0) .. controls (5,1) and (6,1) .. (7,0);
			\foreach \k in {0,1,2,3,4,5,6,7} {\filldraw[black] (\k,0) circle (3pt); }
			\node[anchor=north] at (0,1.5) {$\downarrow$};
		\end{scope}
		\begin{scope}[yshift=0.2cm, rotate=90, scale=0.76]
			\circuitInitXAt{-2.3}
			\foreach\p in {0,1,2,3,4,5,6,7}{\circuitQubitLine{-2.6}{-0.8}{-\p}
						\draw[fill] ($0.5*(-5.2, -\p)$) circle[radius=1.5pt];}	
		\end{scope}
	\end{scope}
	\begin{scope}[xshift = 3.5cm]
    	\begin{scope}[scale=0.38]
			\draw[black, thick] (0,0) .. controls (0.333,0.5) and (0.666,0.5) .. (1,0);
			\draw[black, thick] (3,0) .. controls (3.333,0.5) and (3.666,0.5) .. (4,0);
			\draw[black, thick] (5,0) .. controls (5.5,0.8) and (6.5,0.8) .. (7,0);
			\foreach \k in {0,1,2,3,4,5,6,7} {\filldraw[black] (\k,0) circle (3pt); }
			\node[anchor=north] at (2,1.5) {$\downarrow$};
		\end{scope}
		\begin{scope}[yshift=0.2cm, rotate=90, scale=0.76]
			\circuitInitXAt{-2.3}
			\foreach\p in {0,1,2,3,4,5,6,7}{\circuitQubitLine{-2.6}{-0.8}{-\p}
						\draw[fill] ($0.5*(-5.2, -\p)$) circle[radius=1.5pt];}
			\circuitMultiGate[\flatgatesize][gencolbg][gencolfg]{}{-0}{-1}{}\circuitAdvanceXBy{\flatgateadvance}
			\circuitMultiGate[\flatgatesize][gencolbg][gencolfg]{}{-1}{-2}{}\circuitAdvanceXBy{\flatgateadvance}
			\circuitMultiGate[\flatgatesize][gencolbg][gencolfg]{}{-2}{-3}{}\circuitAdvanceXBy{\flatgateadvance}
			\circuitMultiGate[\flatgatesize][gencolbg][gencolfg]{}{-3}{-4}{}\circuitAdvanceXBy{\flatgateadvance}
			\circuitMultiGate[\flatgatesize][gencolbg][gencolfg]{}{-4}{-5}{}		
		\end{scope}
	\end{scope}
	\begin{scope}[xshift = 7cm]
    	\begin{scope}[scale=0.38]
			\draw[black, thick] (0,0) .. controls (0.333,0.5) and (0.666,0.5) .. (1,0);
			\draw[black, thick] (3,0) .. controls (3.333,0.5) and (3.666,0.5) .. (4,0);
			\draw[black, thick] (5,0) .. controls (5.5,0.8) and (6.5,0.8) .. (7,0);
			\foreach \k in {0,1,2,3,4,5,6,7} {\filldraw[black] (\k,0) circle (3pt); }
			\node[anchor=north] at (3,1.5) {$\downarrow$};
		\end{scope}
		\begin{scope}[yshift=0.2cm, rotate=90, scale=0.76]
			\circuitInitXAt{-2.3}
			\foreach\p in {0,1,2,3,4,5,6,7}{\circuitQubitLine{-2.6}{-0.8}{-\p}
						\draw[fill] ($0.5*(-5.2, -\p)$) circle[radius=1.5pt];}
			\circuitMultiGate[\flatgatesize][gencolbg][gencolfg]{}{-0}{-1}{}\circuitAdvanceXBy{\flatgateadvance}
			\circuitMultiGate[\flatgatesize][gencolbg][gencolfg]{}{-1}{-2}{}\circuitAdvanceXBy{\flatgateadvance}
			\circuitMultiGate[\flatgatesize][gencolbg][gencolfg]{}{-2}{-3}{}\circuitAdvanceXBy{\flatgateadvance}
			\circuitMultiGate[\flatgatesize][gencolbg][gencolfg]{}{-3}{-4}{}\circuitAdvanceXBy{\flatgateadvance}
			\circuitMultiGate[\flatgatesize][gencolbg][gencolfg]{}{-4}{-5}{}		
		\end{scope}
	\end{scope}
	\begin{scope}[xshift = 10.5cm]
    	\begin{scope}[scale=0.38]
			\draw[black, thick] (0,0) .. controls (0.333,0.5) and (0.666,0.5) .. (1,0);
			\draw[black, thick] (3,0) .. controls (3.333,0.5) and (3.666,0.5) .. (4,0);
			\draw[black, thick] (5,0) .. controls (5.5,0.8) and (6.5,0.8) .. (7,0);
			\foreach \k in {0,1,2,3,4,5,6,7} {\filldraw[black] (\k,0) circle (3pt); }
			\node[anchor=north] at (5,1.5) {$\downarrow$};
		\end{scope}
		\begin{scope}[yshift=0.2cm, rotate=90, scale=0.76]
			\circuitInitXAt{-2.3}
			\foreach\p in {0,1,2,3,4,5,6,7}{\circuitQubitLine{-2.6}{-0.8}{-\p}
						\draw[fill] ($0.5*(-5.2, -\p)$) circle[radius=1.5pt];}
			\circuitMultiGate[\flatgatesize][gencolbg][gencolfg]{}{-0}{-1}{}
			\circuitMultiGate[\flatgatesize][gencolbg][gencolfg]{}{-3}{-4}{}\circuitAdvanceXBy{\flatgateadvance}
			\circuitMultiGate[\flatgatesize][gencolbg][gencolfg]{}{-1}{-2}{}\circuitAdvanceXBy{\flatgateadvance}
			\circuitMultiGate[\flatgatesize][gencolbg][gencolfg]{}{-2}{-3}{}\circuitAdvanceXBy{\flatgateadvance}
			\circuitMultiGate[\flatgatesize][gencolbg][gencolfg]{}{-3}{-4}{}\circuitAdvanceXBy{\flatgateadvance}
			\circuitMultiGate[\flatgatesize][gencolbg][gencolfg]{}{-4}{-5}{}		
		\end{scope}
	\end{scope}
	\begin{scope}[xshift = 14cm]
    	\begin{scope}[scale=0.38]
			\draw[black, thick] (0,0) .. controls (0.333,0.5) and (0.666,0.5) .. (1,0);
			\draw[black, thick] (3,0) .. controls (3.333,0.5) and (3.666,0.5) .. (4,0);
			\draw[black, thick] (5,0) .. controls (5.333,0.5) and (5.666,0.5) .. (6,0);
			\foreach \k in {0,1,2,3,4,5,6,7} {\filldraw[black] (\k,0) circle (3pt); }
			\node[anchor=north] at (7,1.5) {$\downarrow$};
		\end{scope}
		\begin{scope}[yshift=0.2cm, rotate=90, scale=0.76]
            \draw[black!30, line width=2.5pt, rounded corners = 5pt] (-0.6,0.2) rectangle node[below,yshift=-0.8cm]{output} (-2.8,-3.7);
			\circuitInitXAt{-2.3}
			\foreach\p in {0,1,2,3,4,5,6,7}{\circuitQubitLine{-2.6}{-0.8}{-\p}
						\draw[fill] ($0.5*(-5.2, -\p)$) circle[radius=1.5pt];}
			\circuitMultiGate[\flatgatesize][gencolbg][gencolfg]{}{-0}{-1}{}
			\circuitMultiGate[\flatgatesize][gencolbg][gencolfg]{}{-3}{-4}{}
			\circuitMultiGate[\flatgatesize][gencolbg][gencolfg]{}{-5}{-6}{}\circuitAdvanceXBy{\flatgateadvance}
			\circuitMultiGate[\flatgatesize][gencolbg][gencolfg]{}{-1}{-2}{}
			\circuitMultiGate[\flatgatesize][gencolbg][gencolfg]{}{-6}{-7}{}\circuitAdvanceXBy{\flatgateadvance}
			\circuitMultiGate[\flatgatesize][gencolbg][gencolfg]{}{-2}{-3}{}\circuitAdvanceXBy{\flatgateadvance}
			\circuitMultiGate[\flatgatesize][gencolbg][gencolfg]{}{-3}{-4}{}\circuitAdvanceXBy{\flatgateadvance}
			\circuitMultiGate[\flatgatesize][gencolbg][gencolfg]{}{-4}{-5}{}		
		\end{scope}
	\end{scope}
    \draw[-{latex}, line width=2pt] (2.75, -0.8) -- (3.4, -0.8);
    \draw[-{latex}, line width=2pt] (6.25, -0.8) -- (6.9, -0.8);
    \draw[-{latex}, line width=2pt] (9.75, -0.8) -- (10.4,-0.8);
    \draw[-{latex}, line width=2pt] (13.25,-0.8) -- (13.9,-0.8);
	\end{tikzpicture}
	\caption{An example application of the algorithm given to find the RSF circuit layout corresponding to a given Bell pair configuration state. The arrow above the Bell pair configuration state indicates the value of the variable $k$ as defined in the text, which is increased in subsequent steps of the algorithm: If the qubit indexed by $k$ is in a Bell pair with another qubit, this pair is swapped to consecutive qubits, a corresponding diagonal is added to the RSF circuits and $k$ is increased by $2$. Otherwise, no gates are added, and the variable is increased by $1$.}
	\label{fig:example_alg_bell_to_rsf}
\end{figure*}

We now construct a mapping $\tilde \pi$ from Bell pair configurations to RSF circuit layouts, and later demonstrate that $\tilde \pi = \pi^{-1}$. Given a Bell pair configuration state $\ket{D}$, we define the corresponding RSF circuit layout $C = \tilde \pi(D)$ algorithmically. Introduce a variable $k$, initialized to $k=1$, start with an empty RSF circuit, and iterate the following steps until $k = L$:
\begin{enumerate}
    \item If qubit $k$ is in the state $\ket{0}$, increment $k$ by one.
    \item If qubit $k$ is entangled with another qubit $k+l$, with $l \geq 1$, swap that qubit to position $k+1$ by applying $l-1$ SWAP operations. Append a diagonal with parameters $(k, l)$ to the RSF circuit; that is, extend the current RSF circuit labeling to $((k_1, l_1), \ldots, (k, l))$. Then, increment $k$ by two.
\end{enumerate}
Figure~\ref{fig:example_alg_bell_to_rsf} illustrates the application of this algorithm to a simple example.

Let us now argue that $\tilde \pi$ is indeed the inverse of $\pi$. Consider for this the RSF circuit layout $((k_1,l_1),\ldots,(k_\numdiag, l_\numdiag)) = \tilde \pi(D)$. We show that applying $\pi$, as given above, yields the Bell pair configuration $\ket{D}$. The application of $\pi$ amount to creating Bell pairs in positions $(k_i, k_i+1)$, and then applying the remaining SWAP gates corresponding to the remaining gates in all the diagonals. Although the order in which the gates are applied is given by the circuit, one still can choose to apply gates for instance layer by layer, or diagonal by diagonal. In the latter case, on has to start with the gates in the last diagonal $(k_\numdiag, l_\numdiag)$, followed by the gates in the second-to-last diagonal, and so on. This procedure effectively corresponds running the above algorithm in reverse, hence showing that $\ket{\pi(\tilde\pi(D))} = \ket{D}$.

The recursive identification of RSF circuits and Bell pair configuration states will be crucial for proving the following two propositions. The first one, Proposition~\ref{prop:bijection}, shows that the Rényi-$0$ entanglement profile of a typical RSF circuit $C$ and the one of the corresponding state $\ket{\psi(C)}$ are equal: The mapping introduced here is, strictly speaking, a correspondence between Bell pair configurations and RSF circuit layouts, rather than specific RSF circuit instances. The equivalence in $S^{(0)}$ entropy holds in the generic case where each gate in the RSF circuit increases $S^{(0)}$ maximally. To elaborate, let $(U_t)_{t=1}^{n_\text{g}}$ denote any ordering of the circuit's gates, and define the intermediate states as $\ket{\psi^t} = \prod_{t'\leq t} U_{t'}\ket{0\ldots0}$. The maximal increase condition then requires that, for each $t$,
\begin{equation}\label{eq:app_entanglement_increases_maximally}
S^{(0)}_m \big(\!\ket{\psi^{t+1}}\!\big) = 1 + \min
    \big{\{} S^{(0)}_{m-1} \big(\!\ket{\psi^t} \!\big), S^{(0)}_{m+1} \big(\!\ket{\psi^t} \!\big) \big{\}},
\end{equation}
where $m$ denotes the bond on which $U_t$ acts (i.e., the bond between qubits $m$ and $m+1$)~\cite{Nahum_2017}.

\begin{proposition}
\label{prop:bijection}
The mapping $\pi$ between RSF circuits $C$ and Bell pair states $\ket{\pi(C)}$, labeled by the corresponding partitions $\pi(C)$, satisfies the following: For any RSF circuit $C$, 
\begin{equation}\label{eq:single_string_entanglement}
S^{(0)}_m \big(\ket{\pi(C)}\big) = S^{(0)}_m(C \ket{0}^{\otimes L})
\end{equation}
for $0 \leq m \leq L$, provided that each gate in $C$ maximally increases the Rényi-0 entropy (the case considered in this paper).
\end{proposition}

\begin{proof}
We use the recursive definitions of RSF circuits and Bell pair configurations to prove Proposition~\ref{prop:bijection} by induction. First, the cases of an empty RSF circuit (i.e., a completely unentangled Bell pair configuration) and a single diagonal labeled $(k,l)$ (representing a Bell pair configuration with a single entangled pair between qubits $k$ and $k+l$) are trivial. Now, suppose the correspondence for the $S^{(0)}$ entropy holds for all circuits and Bell pair configurations on $L$ qubits (and thus also on $L-1$ qubits). There are two possible ways to construct an RSF circuit (or Bell pair configuration) on $L+1$ qubits. In the first case, by adding a single qubit to the left, the correspondence of $S^{(0)}$ follows immediately. We thus focus on the second case, where an additional non-empty diagonal is added to an RSF circuit on $L-1$ qubits.

Suppose $C$ is an arbitrary RSF circuit on $L-1$ qubits, labeled $3, \ldots, L+1$. To the left of this circuit, we add two qubits, labeled $1$ and $2$. Now, for each $l = 1, \ldots, L$, define an RSF circuit $C_l$ by adding an additional diagonal of length $l$, starting at the first bond between qubits $1$ and $2$. For $l = 1$, we have
\begin{align*}S^{(0)}_m(C_1\ket{0}^{\otimes L+1}) = & \\ 
S^{(0)}_m(\ket{\pi(C_1)}) =& 
\begin{cases} 
	1, & m=1,\\ 
	0, & m=2, \\ 
	S^{(0)}_m(\ket{\pi(C)}), & m\geq 3
\end{cases}
\end{align*}
for all bonds $1 \leq m \leq L$: For $m \geq 3$, this follows from the induction hypothesis, while the values at $m = 1$ and $m = 2$ can be easily determined.

We now employ an additional inductive argument over $l$ to show that $S^{(0)}_m(C_l \ket{0}^{\otimes L+1}) = S^{(0)}_m(\ket{\pi(C_l)})$ for all $l > 1$ and $m$. Note that $C_{l-1}$ and $C_l$ differ by only a single gate applied on bond $l$. Therefore, their entanglement entropies match at every bond $m \neq l$. At bond $l$, the additional gate causes the entanglement entropy to increase maximally, as specified by the condition in~\eqref{eq:app_entanglement_increases_maximally}. On the other hand, the states $\ket{\pi(C_{l-1})}$ and $\ket{\pi(C_l)}$ differ only by a SWAP gate applied on bond $l$, so the entanglement entropies can differ at most at bond $l$.

The $S^{(0)}$ entanglement entropy of $\ket{\pi(C_{l-1})}$ at bond $l$ can differ from the entropy at bond $l+1$ by either $0$, or $\pm 1$. To complete the proof, we now identify which Bell pair configurations are compatible with each of these three cases and show that the entanglement entropy increases maximally when applying the SWAP gate to obtain $\ket{\pi(C_l)}$.  In $\ket{\pi(C_{l-1})}$, by construction, there is a Bell pair shared between qubits $1$ and $l$, so we have $S^{(0)}_{l-1}(\ket{C_{l-1}}) = S^{(0)}_{l}(\ket{C_{l-1}}) + 1$. The three possible cases for bond $l+1$ are:
\begin{itemize}
    \item $S_{l+1}(\ket{\pi(C_{l-1})}) = S_{l}(\ket{\pi(C_{l-1}}) + 1$. In this case, there must be a Bell state shared between qubits $l+1$ and $b$, with $b > l+1$. Applying then the SWAP on qubits $l,l+1$, there are Bell states shared between qubit pairs $1, l+1$ and $l, b$, thus the entanglement at bond $l$ increases by~$2$. When depicting qubits $l$ and $l+1$, we can visualize the change at bond $l$ as
    \begin{center}\begin{tikzpicture}[scale=0.67]
    \foreach \k in {0,1,3,4} {\filldraw[black] (\k,0) circle (3pt);}
    \draw[-{latex}, black, line width=1.5pt] (1.5,0) -- (2.5,0);
	\draw[black, thick] (-0.5,0.5) .. controls (-0.2,0.5) and (0,0.3) .. (0,0);
	\draw[black, thick] (1.5,0.5) .. controls (1.2,0.5) and (1,0.3) .. (1,0);
	\draw[black, thick] (2.5,0.5) .. controls (3,0.5) and (4,0.3) .. (4,0);
	\draw[black, thick] (4.5,0.5) .. controls (4,0.5) and (3,0.3) .. (3,0);
    \end{tikzpicture}.\end{center}
    \item $S_{l+1}(\ket{\pi(C_{l-1})}) = S_{l}(\ket{\pi(C_{l-1})})$. In this case, the state of qubit $l+1$ must be $\ket{0}$. After applying the SWAP gate, 
    \begin{center}\begin{tikzpicture}[scale=0.67]
    \foreach \k in {0,1,3,4} {\filldraw[black] (\k,0) circle (3pt);}
    \draw[-{latex}, black, line width=1.5pt] (1.5,0) -- (2.5,0);
	\draw[black, thick] (-0.5,0.5) .. controls (-0.2,0.5) and (0,0.3) .. (0,0);
	\draw[black, thick] (2.5,0.5) .. controls (3,0.5) and (4,0.3) .. (4,0);
    \end{tikzpicture},\end{center}
    there is exactly one additional Bell pair shared across the bond $l$, hence the entanglement entropy increases by $1$.
    \item $S_{l+1}(\ket{\pi(C_{l-1}})) = S_{l}(\ket{\pi(C_{l-1})}) - 1$. Here, there exists a qubit $a$, $1<a<l$, that shares a Bell pair with qubit qubit $l+1$. When performing the SWAP,
    \begin{center}\begin{tikzpicture}[scale=0.67]
    \foreach \k in {0,1,3,4} {\filldraw[black] (\k,0) circle (3pt);}
    \draw[-{latex}, black, line width=1.5pt] (1.5,0) -- (2.5,0);
	\draw[black, thick] (-0.5,0.6) .. controls (-0.2,0.4) and (0,0.3) .. (0,0);
	\draw[black, thick] (-0.5,0.4) .. controls (0,0.35) and (1,0.3) .. (1,0);
	\draw[black, thick] (2.5,0.4) .. controls (2.8,0.35) and (3,0.3) .. (3,0);
	\draw[black, thick] (2.5,0.6) .. controls (3,0.55) and (4,0.3) .. (4,0);
    \end{tikzpicture},\end{center}
    the entanglement entropy does not change.
\end{itemize}
In each case, the entanglement increases corresponding to Eq.~(\ref{eq:app_entanglement_increases_maximally}). The claim then follows from this correspondence.
\end{proof}

Having established the equivalence of RSF circuits and Bell pair configurations, and the corresponding entanglement profiles, it is clear that one can induce update rules on the Bell pair configurations. Given a state $\ket{D}$, this can essentially be done by updating the corresponding RSF circuit $\pi^{-1}(D)$, and then computing again the new state $\ket{D'}$. Here, we prove that the same dynamics can be obtained in a simpler way, namely by applying the update rules defined in Fig.~\ref{fig: rules classical}. Specifically, we show that when applying a gate to an RSF circuit $C$ and obtaining a new RSF circuit $\tilde C$ via the absorption algorithm, the Bell pair update rules applied to $\ket{\pi(C)}$ yield the state $\ket{\pi(\tilde C)}$, and that the same correspondence holds for disentangling operations. For completeness, we briefly comment here on how the Bell pair update rules shown in Fig.~\ref{fig: rules classical} are to be interpreted, focusing only on the entangling (left-to-right) direction. Rule 1 states that when the state $\ket{00}$ is encountered across bond $b$, it should be replaced with a Bell pair. In terms of partitions, this updates $\{b\}, \{b+1\}$ to $\{b, b+1\}$. Rule 2 replaces the configuration $\{b\}, \{b+1, c\}$, where $c > b+1$, with $\{b, c\}, \{b\}$. Rules 3 and 4 are analogous to this case. Rule 5 updates a configuration $\{b, c_1\}, \{b+1, c_2\}$, where the crossing lines imply $c_1 < c_2$, to $\{b, c_2\}, \{b+1, c_1\}$, and rule 6 is interpreted analogously. The correspondence between the dynamics of both models is summarized in the following proposition.

\begin{proposition}
\label{prop:dynamics}
Consider an RSF circuit $C$ and a generic matchgate $U$ acting on bond $b$. Let $\tilde{C}$ denote the RSF circuit obtained by applying the absorption algorithm for $U$ to $C$. Furthermore, let $V$ denote the transformation on the Bell pair partitions induced by applying the entangling rules shown in Fig.~\ref{fig: rules classical} on bond $b$. Then,
\begin{equation}
    V(\pi(C)) = \pi(\tilde{C}),
\end{equation}
i.e., the Bell pair states $\ket{V(\pi(C))}$ and $\ket{\pi(\tilde{C})}$ are identical.
\end{proposition}

Proving Proposition~\ref{prop:dynamics} is a straight-forward, but tedious task, due to the number of different configuration that one has to consider. A simplification in our proof relies again on the recursive identification of RSF circuits and Bell pair configurations.

\begin{proof}
    Our strategy for the proof is as follows: For each update rule, we identify all Bell pair configurations compatible with the initial state of the rule. Using the mapping $\pi^{-1}$, we determine the corresponding RSF circuits. We then show that applying the absorption algorithm yields RSF circuits that, when mapped through $\pi$, reproduce the Bell pair configurations obtained via the update rule. We finally argue that if none of the update rules apply, the RSF circuit remains unchanged.

    We begin with a simplification step that enables us to ignore irrelevant diagonals and qubits. From the construction of $\pi$, the first diagonal with parameters $(k_1, l_1)$ always corresponds to a Bell pair between qubits $k_1$ and $k_1 + l_1$ in the associated Bell pair state. This identification generally does not hold for the second diagonal $(k_2, l_2)$ due to potential qubit swaps, which complicate tracking Bell pairs. However, when a gate is applied on bond $b$, only two consecutive diagonals may differ between the initial and final RSF circuits (see the proof of Lemma~\ref{lemma: absorbing}). In particular, all preceding diagonals either commute with the gate or can be pushed past it via Yang-Baxter moves, which modify only the gates, but not the diagonal structure of the RSF. Therefore, we may ignore those earlier diagonals and qubits via suitable relabeling, as detailed below.

    To that end, we perform a sequence of reductions that effectively relabel and remove irrelevant qubits and diagonals, preserving the property that the relevant qubits (initially labeled $b$ and $b+1$) remain adjacent. Update rules can be applied after these reductions, and the original labeling can be restored afterward. Suppose the RSF circuit $C$ is labeled by $((k_1, l_1), \ldots, (k_\numdiag, l_\numdiag))$, with $(k_1, l_1)$ the first diagonal. By construction, we have $\{k_1, k_1 + l_1\} \in \pi(C)$. If $k_1 \geq b$, or if the circuit is empty, no reduction is necessary. Otherwise, we consider two cases:
    \begin{itemize}
        \item $k_1 + l_1 < b$: In the RSF picture, the gates in the diagonal $(k_1, l_1)$ trivially commute with a gate applied on bond $b$, $b+1$ (cf. item~\ref{item:proof absorbing commute} in Lemma~\ref{lemma: absorbing}). Hence, we may disregard this diagonal, along with qubits $k_1$, $k_1+1$. In the Bell pair picture, the corresponding Bell pair $\{k_1, k_1 + l_1\}$ can be ignored. We relabel the remaining qubits as $1, \ldots, L-2$, and $(k_1, l_1)$ now labels the former second diagonal.

        \item $k_1 + l_1 > b + 1$: In this case, the first diagonal contains at least one gate acting on qubits $b, b+1$ and one on $b+1, b+2$. Applying the absorption algorithm on bond $b$, $b+1$ implies doing a Yang-Baxter move (cf. item~\ref{item:proof absorbing yb reduction} in Lemma~\ref{lemma: absorbing}). This reduces to absorbing a gate on $b+1$, $b+2$ into an RSF subcircuit, allowing us to disregard the original first diagonal. The corresponding Bell pair $\{k_1, k_1 + l_1\}$ can likewise be ignored, and qubits are relabeled as before.
    \end{itemize}
    We can repeat this reduction until neither $k_1 + l_1 < b$ nor $k_1 + l_1 > b + 1$, so we can assume one of the following holds:
    \begin{itemize}
        \item $k_1 < b$ and $k_1 + l_1 = b$,
        \item $k_1 < b$ and $k_1 + l_1 = b + 1$,
        \item $k_1 = b$,
        \item $k_1 = b + 1$, or
        \item $k_1 \geq b + 2$.
    \end{itemize}
    Additionally, all qubits $q < k_1$ are in the state $\ket{0}$.

    We now verify that the Bell pair update rules and RSF dynamics correspond under $\pi$. First, note that the configurations depicted in Fig.~\ref{fig: rules classical} (both left and right of the arrows) are exhaustive: any pair of neighboring qubits in a Bell pair configuration matches one of these. We examine each case in the left column (i.e., entangling operations):
    \begin{itemize}
        \item \textbf{Case 1}: The initial state on qubits $b, b+1$ is $\ket{00}$, and is updated to $\ket{\Phi^+}$. Since $k_1 \geq b + 2$, no gate acts on $b, b+1$ in the RSF circuit $C$, so the absorption algorithm simply adds the gate. Thus, $\pi(\tilde{C}) = V(\pi(C))$.

        \item \textbf{Case 2}: We have $\{b+1, c\} \in \pi(C)$ for $c > b+1$. From the above alternatives, we infer that the first diagonal starts at qubit $k_1 = b+1$, and by item~\ref{item:proof absorbing add left} in Lemma~\ref{lemma: absorbing}, it follows that $\pi(\tilde{C}) = V(\pi(C))$.

        \item \textbf{Cases 3, 4 and 6}: The only compatible alternative is $k_1 < b$ and $k_1 + l_1 = b$. We therefore have $\{k_1, b\} \in \pi(C)$, i.e., $(k_1, b)$ is the first diagonal on the RSF circuit, which gets updated to $(k_1, b+1)$ after applying the gate. In each case, the updated RSF and Bell pair configurations coincide.

        \item \textbf{Case 5}: We have $k_1 = b$ by the above alternatives. There exist $c_1, c_2 > b+1$ s.t. $\{b,c_1\},\{b+1, c_2\}\in\pi(C)$. Moreover, $c_1 < c_2$, since otherwise this would correspond to the right figure in case 5. For the first diagonal, $(b, l_1)$, we have $l_1 = c_1 - b$. Following the construction of $\pi$, one can see that there must be a second diagonal $(k_2, l_2)$ in the RSF circuit with $k_2 = b+2$ and $l_2 = c_2 - b - 2$. From $c_1 \leq c_2 - 1$, we get $l_1 - 1 \leq l_2$. Following the relevant substeps in item~\ref{item:proof absorbing many substeps} in the proof of Lemma~\ref{lemma: absorbing}, we see that these two diagonals in the updated RSF circuit $\tilde C$ read $(b, l_2+2)$ and $(b+2, l_1 -1)$.
        Let us now investigate $\pi(\tilde C)$. Since $l_2 + 2 = c_2 - b$, we have $\{b, c_2\}\in\pi(\tilde C)$. Furthermore, the second diagonal, $(b+2, l_1 -1)$, corresponds to a Bell pair labeled by $\{b+1, b+1+l_1-1\} = \{b+1, c_1\}$ (this can again be seen by ``undoing the swaps'' from the first diagonal $(b, l_2+2)$). All the other diagonals and Bell pairs respectively will not be modified, and hence $V(\pi(C)) = \pi(\tilde C)$.
    \end{itemize}
    It remains to verify the configurations on the right side of the arrows in Fig.~\ref{fig: rules classical}, corresponding to disentangling operations. These are handled analogously: each can be viewed as resulting from applying a gate on bond $b$ to a circuit $C'$ to obtain $C$. Hence, applying another gate on $b$ leaves the RSF unchanged. This confirms that the Bell pair and RSF dynamics agree in all cases.
\end{proof}

\onecolumngrid

\section{Thermodynamic limit of Rényi-0 entropy at criticality}
\label{appendix_asymptotics}

\newcommand{\entzero}{S^{(0)}}
\newcommand{\aventzero}{\bar S^{(0)}}
\newcommand{\xvar}{\alpha}
\newcommand{\entlimit}{s}
\newcommand{\enterror}{\varepsilon}
\newcommand{\errbound}{b}

The unitary circuit games with matchgates and a gate disentangler yield a phase transition between stable volume- and area-law entanglement phases. 
This phase transition can be detected through the behavior of the Rényi-0 entanglement entropy, which can be directly obtained from the RSF representation.
In this appendix, we investigate the behavior of the Rényi-0 entanglement entropy at the critical point, $p_c=1/2$. We derive analytically that the average Rényi-0 entanglement profile, when rescaled and normalized w.r.t. the system size $L$ (see Corollary~\ref{thm:asymtotic_phone_theorem}), takes the form \[\frac1L \aventzero_{\alpha L}(L) = \alpha(1-\alpha), \] for $0\leq\alpha\leq 1$ in the thermodynamic limit $L\to\infty$. Therefore, the average entanglement entropy still obeys a volume law, i.e., $\bar S^{(0)} \sim L$. To arrive at this result, we also provide exact expressions for $\aventzero_m(L)$ for any bond $0\leq m\leq L$ at a finite $L$ (see Proposition~\ref{thm:finite_phone_theorem}). \par\medskip

Before we state our result formally, let us recall some definitions. After evolving FGSs for a sufficiently long time with the unitary circuit game, these states will be distributed according to the \emph{stationary distribution}, denoted by $\mathcal{D}_L$. The average Rényi-$0$ entropy for system size $L$ at bond $m$ over this distribution is given by 
\[\aventzero_{m} (L) =  \int\limits_{\psi \sim \mathcal{D}_L} \!\!\mathrm{d}\psi \, \entzero_{m}(\ket\psi). \]
Recall furthermore that the number of distinct RSF circuit layouts (and Bell pair configurations, respectively) on $L$ qubits is given by the telephone number $T(L)$ (see Eq.~\ref{eq:telephone_recursion} in Appendix~\ref{appendix:telephone}). We will show the following proposition:

\begin{proposition}\label{thm:finite_phone_theorem} For any finite system size $L$, and any bond $m$ with $0<m<L$, the average Rényi-$0$ entropy of states distributed according to the stationary distribution of the unitary circuit game at the critical point $p=\frac12$ is given by
\[
\aventzero_m(L) = m(L-m) \frac{T(L-2)}{T(L)}.
\]
\end{proposition}

To do the thermodynamic limit, we evaluate the average entanglement entropy at the bond $\xvar L$ for $0<\xvar<1$, and then perform the limit $L\to \infty$. Theorem~3 in Ref.~\cite{origin_of_telephone_lemma} states that $T(L)/T(L-1)\sim \sqrt{L}$. Therefore, it holds that $T(L-2)/T(L) \sim 1/L$, and we immediately get the following corollary. 

\begin{corollary}\label{thm:asymtotic_phone_theorem}
For $\xvar\in[0,1]$, for states distributed according to the stationary distribution of the unitary circuit game at the critical point $p=\frac12$, it holds that
\[ \frac{1}{L}\aventzero_{\xvar L}(L) \xrightarrow{L\to\infty} \xvar(1-\xvar). \]
\end{corollary}

We will now present a proof of Proposition~\ref{thm:finite_phone_theorem}. As we have explained in the previous appendix, the value of $S^{(0)}(L)$ generically depends only on the RSF circuit layout and not on the parameters of the gates (see Eq.~\eqref{eq:app_entanglement_increases_maximally}). We therefore only need to determine the distribution of RSF circuit layouts at the critical point, to then derive the expression for $\aventzero_m$ from this distribution. Note furthermore that from the arguments in Appendix~\ref{appendix:telephone}, it follows that the stationary distribution of the RSF circuit layout and the corresponding Bell pair states is equal for both models. As a result, both models have the same Rényi-0 entanglement profile and we can use them interchangeably. In the following, we omit writing the $L$ dependence of $\aventzero$.

\begin{proof}[Proof of Proposition~\ref{thm:finite_phone_theorem}]

Firstly, we argue that the stationary distribution of RSF circuits in the unitary circuit game with the gate disentangler at $p=1/2$ is uniform over all possible RSF circuit layouts. Equivalently, this corresponds simply to a uniform distribution of all possible Bell pair diagrams. As explained in the main text, the game on circuit configurations can be viewed as a Markov chain. Corresponding to this Markov chain is a square transition matrix with a number of rows equal to the number of possible RSF circuit layouts, which contains the probabilities of moving from one RSF circuit to another one in a single time step. At the critical point $p=\frac12$, this transition matrix is symmetric. Indeed, if for instance an entangling move on bond $b$ leads from a configuration $A$ to a configuration $B$, then a disentangling move on bond $b$ leads from configuration $B$ to $A$. Since the bonds are selected uniformly random, and since at the critical point, the entangler and disentangler act equally likely, the transition probabilities from $A$ to $B$ and from $B$ to $A$ are equally large \footnote{Note that sometimes obtaining an RSF circuit $B$ from another RSF circuit $A$ can be done by applying a single gate at one of several bonds. For instance, the RSF labeled $((1,3),(3,1))$ on four qubits can be obtained from the RSF labeled $((1,2),(3,1))$ by either placing a gate on the first or on the third bond. It still holds that when placing a gate on any bond $b$ changes $A$ to $B$, then a disentangling move applied on $b$ changed $B$ to $A$. Therefore the transition probabilities are still symmetric.}.
Furthermore, the Markov chain is irreducible, i.e., it is possible to reach any configuration from any other configuration with a non-zero probability in a finite number of steps. It is a known fact that such chains have a unique stationary distribution \cite{levin2017markov}. The symmetry of the transition matrix implies that it is doubly stochastic, and hence the uniform distribution over RSF circuits is the stationary state.\par\medskip

We now proceed to derive an exact expression for $\aventzero_m$ for an arbitrary bond $0<m<L$ for a finite $L$. The corresponding counting problem can be simplified when using Bell pair configurations rather than RSF circuits. For each $0\leq k \leq m$, we count the number $N_m(k)$ of Bell pair configurations such that there are exactly $k$ Bell pairs with one qubit in $\{1,\ldots,m\}$ and the other qubit in $\{m+1, \ldots L\}$, so that the entanglement entropy for this partition is exactly~$k$. Note that $0 \leq k \leq \min(m, L-m)$ due to the sizes of the subsystems. Since each of the $T(L)$ configuration appears with equal probability (see above), we then have
\[ \aventzero_m = \frac{1}{T(L)} \sum_{k=0}^{\min(m,L-m)} k N_m(k). \]

Obtaining $N_m(k)$ can be done in the following way: Choose $k$ qubits within the first subsystem of size $m$. There are $\binom{m}{k}$ distinct possibilities to do so. Next, choose $k$ qubits within the other subsystem of size $L-m$, giving another $\binom{L-m}{k}$ possibilities. 
For each choice of those qubits, exactly $k$ Bell pairs need to be shared across the bipartition. For this, there are $k!$ possible assignments to do so (e.g. qubit $1$ can be entangled with any of $k$ qubits in the other subsystem, qubit $2$ can then be entangled with any of the remaining $k-1$ qubits and so on). Finally, one needs to consider what happens with the remaining $m-k$ and $L-m-k$ qubits in the respective subsystems. Those can be an arbitrary valid Bell pair configuration state as they are not entangled to the respective other subsystem, hence giving another $T(m-k)$ and, respectively, $T(L-m-k)$ possibilities. Thus, in total,
\[ N_m(k) = \binom{m}{k} \binom{L-m}{k} \;k!\; T(m-k) \; T(L-m-k), \]
and the average Rényi-$0$ entropy can be expressed for all $m$ as
\begin{equation}\label{eq:finite_telephone_entropy_expression}
    \aventzero_m = \frac{1}{T(L)} \sum_{k=0}^{\min(m,L-m)} k \binom{m}{k} \binom{L-m}{k} \;k!\; T(m-k) \; T(L-m-k).
\end{equation}
Furthermore, since summing $N_m(k)$ over all admissible value of $k$ gives the total number of configurations $T(L)$, we get the normalization identity
\begin{equation}\label{eq:finite_telephone_normalization}
    1 =\frac{1}{T(L)}\sum_{k=0}^{\min(m,L-m)} N_m(k) = \frac{1}{T(L)} \sum_{k=0}^{\min(m,L-m)} \binom{m}{k} \binom{L-m}{k} \;k!\; T(m-k) \; T(L-m-k),
\end{equation} for each value of $m$. \par\medskip

In the following, we show the relation
\begin{equation} \label{eq:important_recursion_in_telephone_proof}
\sum_{k=0}^{\min(m,L-m)} (L-m-k) N_m(k) = (L-m)\big( T(L-1) + (L-1-m) T(L-2)\big).
\end{equation}
One then gets
\begin{align*}
\aventzero_m &= \frac{1}{T(L)} \sum_{k=0}^{\min(m,L-m)} k N_m(k) \\ 
&= \frac{L-m}{T(L)} \sum_{k=0}^{\min(m,L-m)} N_m(k) \quad-\quad \frac{1}{T(L)} \sum_{k=0}^{\min(m,L-m)} (L-m-k) N_m(k) \\
&= \frac{L-m}{T(L)}\big(T(L) - T(L-1) - (L-1-m) T(L-2) \big),
\end{align*}
when using Eqs.~\eqref{eq:finite_telephone_normalization} and~\eqref{eq:important_recursion_in_telephone_proof}. Finally, inserting the recurrence relation for $T(L)$, Eq.~\eqref{eq:telephone_recursion}, one obtains the desired expression,
\[
\aventzero_m = m(L-m) \frac{T(L-2)}{T(L)},
\]
which would complete the proof. \par \medskip

The missing part is hence a proof of Eq.~\eqref{eq:important_recursion_in_telephone_proof}, which we provide here. In the following, we restrict to $m\leq\frac{L}{2}$. Due to symmetry, we can repeat all arguments for $m> \frac{L}{2}$ by replacing $m \mapsto L - m$. We need to consider three cases, labeled, 1 to 3, one of which will always hold:
\begin{enumerate}
    \item $m \leq L/2 -1$,
    \item $L$ is even, and $m = L/2$,
    \item $L$ is odd, and $m = (L-1)/2$.
\end{enumerate}
Firstly, in all of the three cases, we show
\[
\sum_{k=0}^{\min(m,L-m)} (L-m-k) N_m(k) = \sum_{k=0}^{\min(m,L-m-1)} (L-m-k) N_m(k).
\]
Seeing why this is true requires to consider the three cases separately:
\begin{enumerate}
    \item Either, one has $\min(m,L-m) = m = \min(m,L-m-1)$, or
    \item $\min(m,L-m-1) = m - 1 = \min(m,L-m)-1$, but the contribution to the sum with $k=m$ drops out, or
    \item one has $\min(m,L-m) = (L-1)/2 = \min(m,L-m-1)$.
\end{enumerate}
Having established this, we have that $L-m-k\geq 1$ for each value of the summation index $k$. When inserting the corresponding expression for $N_m(k)$ and using $\binom{L-m}{k} (L-m-k)= \binom{L-1-m}{k}(L-m)$, one obtains 
\[
    \sum_{k=0}^{\min(m,L-m-1)} (L-m-k) N_m(k) = (L-m) \sum_{k=0}^{\min(m,L-m-1)} k \binom{m}{k} \binom{L-1-m}{k} \;k!\; T(m-k) \; T(L-m-k).
\]
The next step is to insert the recurrence relation for $T(L-m-k)$, which is possible whenever $L-m-k\geq 2$. Hence, we need to investigate for which values of the summation index $k$ it is possible. Considering again the three cases, one finds:
\begin{enumerate}
    \item Since $m\leq L/2-1$ and $k\leq m$, one always has $L-m-k\geq 2$ and the recurrence can be applied for all values of $k$. Furthermore, $\min(m,L-m-1) = m = \min(m,L-m-2)$.
    \item The maximal value of $k$ appearing in the sum is $k = L/2 -1$, in which case the recurrence relation cannot be applied. However, for this $k$, one has $T(L-m-k)=T(1) = T(0) = T(L-m-k -1)$. For all other values of $k$, $0 \leq k \leq \min(m,L-m-2)$, the recurrence holds.
    \item Similarly as above, one cannot apply the recurrence relation for the maximal value of $k$, given by $k = (L-1)/2$. Again, it holds that $T(L-m-k)=T(1) = T(0) = T(L-m-k -1)$, and that for all other values of $k$, $0 \leq k \leq \min(m,L-m-2)$, the recurrence can be applied.
\end{enumerate}
In all the three cases, one can thus write
\begin{align*} 
\sum_{k=0}^{\min(m,L-m-1)} &k \binom{m}{k} \binom{L-1-m}{k} \;k!\; T(m-k) \; T(L-m-k) \\
=&\quad \sum_{k=0}^{\min(m,L-m-1)} k \binom{m}{k} \binom{L-1-m}{k} \;k!\; T(m-k) \; T(L-1-m-k) \\ 
&+\quad\sum_{k=0}^{\min(m,L-m-2)} k \binom{m}{k} \binom{L-1-m}{k} \;k!\; T(m-k) \; T(L-2-m-k) (L-1-m-k) \\
=&\quad \sum_{k=0}^{\min(m,L-m-1)} k \binom{m}{k} \binom{L-1-m}{k} \;k!\; T(m-k) \; T(L-1-m-k) \\ 
&+\quad (L-1-m)\sum_{k=0}^{\min(m,L-m-2)} k \binom{m}{k} \binom{L-2-m}{k} \;k!\; T(m-k) \; T(L-2-m-k) \\
=&\quad T(L-1) + (L-1-m) T(L-2),
\end{align*}
where to get the last line, we have used Eq.~\eqref{eq:finite_telephone_normalization}. This finally proves Eq.~\eqref{eq:important_recursion_in_telephone_proof}.

\end{proof}

\twocolumngrid

\end{appendix}
\bibliographystyle{quantum}
\bibliography{references}

\end{document}

%% file: figures_in_paper_defs.tex
\input{figures}

\newcommand{\drawIntroductionFigure}{

	\begin{scope}[scale=0.76]
		\coordinate (BBotLeft) at (0,1);
		\coordinate (BBotRight) at ($(BBotLeft) + (10,0)$);
		\coordinate (BTopRight) at ($(BBotRight) + (0,7.5)$);
		\coordinate (BTopLeft) at (BTopRight -| BBotLeft);

		\begin{scope}[yshift = 8.1cm, xshift=0cm]
		\node[anchor=north west] at (0,0) {(a)};
			\begin{scope}[xshift = 1.2cm, yshift=-0.2cm, rotate=90]
				\circuitInitXAt{-2.3}
				\foreach\p in {0,1,2,3,4,5,6}{
					\circuitQubitLine{-2.6}{0}{-\p}
					\draw[fill] ($0.5*(-5.2, -\p)$) circle[radius=1.5pt];
					}
				\circuitMultiGate[\flatgatesize][entcolbg][entcolfg]{}{-1}{-2}{}\circuitAdvanceXBy{\flatgateadvance}
				\circuitMultiGate[\flatgatesize][discolbg][discolfg]{}{-2}{-3}{}\circuitAdvanceXBy{\flatgateadvance}
				\circuitMultiGate[\flatgatesize][entcolbg][entcolfg]{}{-3}{-4}{}\circuitAdvanceXBy{\flatgateadvance}
				\circuitMultiGate[\flatgatesize][entcolbg][entcolfg]{}{-0}{-1}{}\circuitAdvanceXBy{\flatgateadvance}
				\circuitMultiGate[\flatgatesize][discolbg][discolfg]{}{-3}{-4}{}\circuitAdvanceXBy{\flatgateadvance}
				\circuitMultiGate[\flatgatesize][discolbg][discolfg]{}{-0}{-1}{}\circuitAdvanceXBy{\flatgateadvance}
				\circuitMultiGate[\flatgatesize][entcolbg][entcolfg]{}{-5}{-6}{}\circuitAdvanceXBy{\flatgateadvance}
				\circuitMultiGate[\flatgatesize][discolbg][discolfg]{}{-4}{-5}{}\circuitAdvanceXBy{\flatgateadvance}
			\end{scope}
			\begin{scope}[xshift = 5.5cm, yshift=-0.8cm, rotate=90]
				\circuitInitXAt{0.3}
				\node[anchor=south west] at (0,0.-0.7) {entangler};
				\circuitMultiGate[\flatgatesize][entcolbg][entcolfg]{}{0}{-1}{}\circuitAdvanceXBy{-0.7}
				\node[anchor=south west] at (-0.7,-0.7) {disentangler};
				\circuitMultiGate[\flatgatesize][discolbg][discolfg]{}{0}{-1}{}
			\end{scope}
		\end{scope}

		\begin{scope}[yshift = 4.5cm, xshift=0cm]
		\node[anchor=north west] at (0,0) {(b) Right standard form};
			\begin{scope}[xshift = 0.95cm, yshift=-0.6cm, rotate=90]
				\circuitInitXAt{-2.3}
				\foreach\p in {0,1,2,3,4,5,6,7}{
					\circuitQubitLine{-2.6}{-0.2}{-\p}
					\draw[fill] ($0.5*(-5.2, -\p)$) circle[radius=1.5pt];
					}
				\circuitMultiGate[\flatgatesize][gencolbg][gencolfg]{}{-0}{-1}{}
				\circuitMultiGate[\flatgatesize][gencolbg][gencolfg]{}{-2}{-3}{}
				\circuitMultiGate[\flatgatesize][gencolbg][gencolfg]{}{-4}{-5}{}
				\circuitMultiGate[\flatgatesize][gencolbg][gencolfg]{}{-6}{-7}{}\circuitAdvanceXBy{\flatgateadvance}
				\circuitMultiGate[\flatgatesize][gencolbg][gencolfg]{}{-1}{-2}{}
				\circuitMultiGate[\flatgatesize][gencolbg][gencolfg]{}{-3}{-4}{}
				\circuitMultiGate[\flatgatesize][gencolbg][gencolfg]{}{-5}{-6}{}\circuitAdvanceXBy{\flatgateadvance}
				\circuitMultiGate[\flatgatesize][gencolbg][gencolfg]{}{-2}{-3}{}
				\circuitMultiGate[\flatgatesize][gencolbg][gencolfg]{}{-4}{-5}{}
				\circuitMultiGate[\flatgatesize][gencolbg][gencolfg]{}{-6}{-7}{}\circuitAdvanceXBy{\flatgateadvance}
				\circuitMultiGate[\flatgatesize][gencolbg][gencolfg]{}{-3}{-4}{}
				\circuitMultiGate[\flatgatesize][gencolbg][gencolfg]{}{-5}{-6}{}\circuitAdvanceXBy{\flatgateadvance}
				\circuitMultiGate[\flatgatesize][gencolbg][gencolfg]{}{-4}{-5}{}
				\circuitMultiGate[\flatgatesize][gencolbg][gencolfg]{}{-6}{-7}{}\circuitAdvanceXBy{\flatgateadvance}
				\circuitMultiGate[\flatgatesize][gencolbg][gencolfg]{}{-5}{-6}{}\circuitAdvanceXBy{\flatgateadvance}
				\circuitMultiGate[\flatgatesize][gencolbg][gencolfg]{}{-6}{-7}{}
			\end{scope}
		\end{scope}

		\begin{scope}[yshift = 5.8cm, xshift=5.4cm]
			\node[anchor=north west] at (-0.3,0) {(c)};
			\begin{scope}
				\node[anchor=north] at (2.5,-0.1) {entangler};
				\draw[-{>[length=6pt, width=6pt]}, line width=1.5pt] (2.2,-1.45) -- (2.8,-1.45);
					\begin{scope}[xshift = 0.4cm, yshift=-2.2cm, rotate=90]
						\circuitInitXAt{0.3}
						\foreach\p in {0,1,2,3}{
							\circuitQubitLine{0}{1.5}{-\p}
							\draw[fill] ($0.5*(0, -\p)$) circle[radius=1.5pt];
							}
						\circuitMultiGate[\flatgatesize][gencolbg][gencolfg]{}{-0}{-1}{}\circuitAdvanceXBy{\flatgateadvance}
						\circuitMultiGate[\flatgatesize][gencolbg][gencolfg]{}{-1}{-2}{}\circuitAdvanceXBy{\flatgateadvance}
						\circuitMultiGate[\flatgatesize][gencolbg][gencolfg]{}{-2}{-3}{}\circuitAdvanceXBy{\flatgateadvance}
						\circuitMultiGate[\flatgatesize][entcolbg][entcolfg]{}{-1}{-2}{}
				\end{scope}
				\begin{scope}[xshift = 3cm, yshift=-2.2cm, rotate=90]
						\circuitInitXAt{0.3}
						\foreach\p in {0,1,2,3}{
							\circuitQubitLine{0}{1.5}{-\p}
							\draw[fill] ($0.5*(0, -\p)$) circle[radius=1.5pt];
							}
						\circuitMultiGate[\flatgatesize][gencolbg][gencolfg]{}{-2}{-3}{}
						\circuitMultiGate[\flatgatesize][gencolbg][gencolfg]{}{-0}{-1}{}\circuitAdvanceXBy{\flatgateadvance}
						\circuitMultiGate[\flatgatesize][gencolbg][gencolfg]{}{-1}{-2}{}\circuitAdvanceXBy{\flatgateadvance}
						\circuitMultiGate[\flatgatesize][gencolbg][gencolfg]{}{-2}{-3}{}\circuitAdvanceXBy{\flatgateadvance}
				\end{scope}
			\end{scope}
			
			\begin{scope}[yshift=-2.3cm]
				\node[anchor=north] at (2.5,-0.1) {disentangler};
				\draw[-{>[length=6pt, width=6pt]}, line width=1.5pt] (2.2,-1.45) -- (2.8,-1.45);
				\begin{scope}[xshift = 0.4cm, yshift=-2.2cm, rotate=90]
				\circuitInitXAt{0.3}
				\foreach\p in {0,1,2,3}{
					\circuitQubitLine{0}{1.5}{-\p}
					\draw[fill] ($0.5*(0, -\p)$) circle[radius=1.5pt];
					}
				\circuitMultiGate[\flatgatesize][gencolbg][gencolfg]{}{-0}{-1}{}\circuitAdvanceXBy{\flatgateadvance}
				\circuitMultiGate[\flatgatesize][gencolbg][gencolfg]{}{-1}{-2}{}\circuitAdvanceXBy{\flatgateadvance}
				\circuitMultiGate[\flatgatesize][gencolbg][gencolfg]{}{-2}{-3}{}\circuitAdvanceXBy{\flatgateadvance}
				\circuitMultiGate[\flatgatesize][discolbg][discolfg]{}{-0}{-1}{}
			\end{scope}
			\begin{scope}[xshift = 3cm, yshift=-2.2cm, rotate=90]
				\circuitInitXAt{0.3}
				\foreach\p in {0,1,2,3}{
					\circuitQubitLine{0}{1.5}{-\p}
					\draw[fill] ($0.5*(0, -\p)$) circle[radius=1.5pt];
					}
				\circuitMultiGate[\flatgatesize][gencolbg][gencolfg]{}{-1}{-2}{}\circuitAdvanceXBy{\flatgateadvance}
				\circuitMultiGate[\flatgatesize][gencolbg][gencolfg]{}{-2}{-3}{}\circuitAdvanceXBy{\flatgateadvance}
				\end{scope}
			\end{scope}
		
		\end{scope}
	
	\end{scope}
}

\newcommand{\drawMarkovChainFigure}{

	\begin{scope}[scale=0.76]
		\coordinate (BBotLeft) at (0,-0.5);
		\coordinate (BBotRight) at ($(BBotLeft) + (10,0)$);
		\coordinate (BTopRight) at ($(BBotRight) + (0,6.5)$);
		\coordinate (BTopLeft) at (BTopRight -| BBotLeft);

		\draw[line width=1.5pt, {<[length=6pt, width=6pt]}-{>[length=6pt, width=6pt]}] (4.3,3.6) -- node[above]{$\frac14$} ++(-1.6,-0.8);
		\draw[line width=1.5pt, {<[length=6pt, width=6pt]}-{>[length=6pt, width=6pt]}] (5.7,3.6) -- node[above]{$\frac14$} ++(1.6,-0.8);
		\draw[line width=1.5pt, {<[length=6pt, width=6pt]}-{>[length=6pt, width=6pt]}] (4.3,1.4) -- node[below]{$\frac14$} ++(-1.6,0.8);
		\draw[line width=1.5pt, {<[length=6pt, width=6pt]}-{>[length=6pt, width=6pt]}] (5.7,1.4) -- node[below]{$\frac14$} ++(1.6,0.8);
		
		\begin{scope}[xshift=5.25cm, yshift=4.2cm]
			\draw[line width=1.5pt, -{>[length=6pt, width=6pt]}] (0,0) -- +(0.,0.2)node[right]{$\frac12$} arc(-0:180:0.3) -- +(0.,-0.2);
		\end{scope}
		\begin{scope}[xshift=8.75cm, yshift=2.2cm, rotate=-90]
			\draw[line width=1.5pt, -{>[length=6pt, width=6pt]}] (0,0) -- +(0.,0.2) node[below]{$\frac12$}arc(-0:180:0.3) -- +(0.,-0.2);
		\end{scope}
		\begin{scope}[xshift=1.25cm, yshift=2.2cm, rotate=-90]
			\draw[line width=1.5pt, -{>[length=6pt, width=6pt]}] (0,0) -- +(0.,-0.2)node[below]{$\frac12$} arc(-0:-180:0.3) -- +(0.,0.2);
		\end{scope}		
		\begin{scope}[xshift=4.75cm, yshift=0.7cm, rotate=180]
			\draw[line width=1.5pt, -{>[length=6pt, width=6pt]}] (0,0) -- +(0.,0.2) node[left]{$\frac12$}arc(-0:180:0.3) -- +(0.,-0.2);
		\end{scope}
			
		\begin{scope}[xshift = 4.5cm, yshift=3.1cm, rotate=90]
			\circuitInitXAt{0.3}
			\foreach\p in {0,1,2}{
				\circuitQubitLine{0}{1}{-\p}
				\draw[fill] ($0.5*(0, -\p)$) circle[radius=1.5pt];
				}
		\end{scope}
		
		\begin{scope}[xshift = 1.5cm, yshift=2cm, rotate=90]
			\circuitInitXAt{0.3}
			\foreach\p in {0,1,2}{
				\circuitQubitLine{0}{1}{-\p}
				\draw[fill] ($0.5*(0, -\p)$) circle[radius=1.5pt];
				}
			\circuitMultiGate[\flatgatesize][gencolbg][gencolfg]{}{-0}{-1}{}
		\end{scope}
	
		\begin{scope}[xshift = 7.5cm, yshift=2cm, rotate=90]
			\circuitInitXAt{0.3}
			\foreach\p in {0,1,2}{
				\circuitQubitLine{0}{1}{-\p}
				\draw[fill] ($0.5*(0, -\p)$) circle[radius=1.5pt];
				}\circuitMultiGate[\flatgatesize][gencolbg][gencolfg]{}{-1}{-2}{}
		\end{scope}
		\begin{scope}[xshift = 4.5cm, yshift=0.9cm, rotate=90]
			\circuitInitXAt{0.3}
			\foreach\p in {0,1,2}{
				\circuitQubitLine{0}{1}{-\p}
				\draw[fill] ($0.5*(0, -\p)$) circle[radius=1.5pt];
				}
			\circuitMultiGate[\flatgatesize][gencolbg][gencolfg]{}{-0}{-1}{}\circuitAdvanceXBy{\flatgateadvance}
			\circuitMultiGate[\flatgatesize][gencolbg][gencolfg]{}{-1}{-2}{}
		\end{scope}
	\end{scope}

}

\newcommand{\drawYBMoveOnly}{
\begin{tikzpicture}[scale=0.76]
			\node at (2.2,-0.5) {$=$};
				\begin{scope}[xshift = 0.7cm, yshift=-1cm, rotate=90]
					\circuitInitXAt{0.2}
					\foreach\p in {0,1,2}{
						\circuitQubitLine{0}{1}{-\p}
						}
					\circuitMultiGate[\flatgatesize][gencolbg][gencolfg]{}{-0}{-1}{}\circuitAdvanceXBy{\flatgateadvance}
					\circuitMultiGate[\flatgatesize][gencolbg][gencolfg]{}{-1}{-2}{}\circuitAdvanceXBy{\flatgateadvance}
					\circuitMultiGate[\flatgatesize][gencolbg][gencolfg]{}{-0}{-1}{}
				\end{scope}
			\begin{scope}[xshift = 2.7cm, yshift=-1cm, rotate=90]
				\circuitInitXAt{0.2}
				\foreach\p in {0,1,2}{
					\circuitQubitLine{0}{1}{-\p}
					}
				\circuitMultiGate[\flatgatesize][gencolbg][gencolfg]{}{-1}{-2}{}\circuitAdvanceXBy{\flatgateadvance}
				\circuitMultiGate[\flatgatesize][gencolbg][gencolfg]{}{-0}{-1}{}\circuitAdvanceXBy{\flatgateadvance}
				\circuitMultiGate[\flatgatesize][gencolbg][gencolfg]{}{-1}{-2}{}
			\end{scope}
\end{tikzpicture}}

\newcommand{\drawMPSMoveOnly}{
\begin{tikzpicture}[scale=0.76]
		\node at (2.2,-0.5) {$=$};
				\begin{scope}[xshift = 0.7cm, yshift=-1cm, rotate=90]
					\circuitInitXAt{0.3}
					\foreach\p in {0,1,2}{
						\circuitQubitLine{0}{1}{-\p}
						\draw[fill] ($0.5*(0, -\p)$) circle[radius=1.5pt];
						}
					\circuitMultiGate[\flatgatesize][gencolbg][gencolfg]{}{-0}{-1}{}\circuitAdvanceXBy{\flatgateadvance}
					\circuitMultiGate[\flatgatesize][gencolbg][gencolfg]{}{-1}{-2}{}
			\end{scope}
			\begin{scope}[xshift = 2.7cm, yshift=-1cm, rotate=90]
				\circuitInitXAt{0.3}
				\foreach\p in {0,1,2}{
					\circuitQubitLine{0}{1}{-\p}
					\draw[fill] ($0.5*(0, -\p)$) circle[radius=1.5pt];
					}
				\circuitMultiGate[\flatgatesize][gencolbg][gencolfg]{}{-1}{-2}{}\circuitAdvanceXBy{\flatgateadvance}
				\circuitMultiGate[\flatgatesize][gencolbg][gencolfg]{}{-0}{-1}{}
			\end{scope}
\end{tikzpicture}}

\newcommand{\drawAbsorbingone}{

	\begin{scope}[scale=0.76]
		\coordinate (BBotLeft) at (0,-3);
		\coordinate (BBotRight) at ($(BBotLeft) + (10,0)$);
		\coordinate (BTopRight) at ($(BBotRight) + (0,8.5)$);
		\coordinate (BTopLeft) at (BTopRight -| BBotLeft);

		\begin{scope}[yshift = 4cm, xshift = 0.5cm]
			\node at (2.2,-0.5) {$\rightarrow$};
				\begin{scope}[xshift = 0.7cm, yshift=-1cm, rotate=90]
					\circuitInitXAt{0.2}
					\foreach\p in {0,1,2}{
						\circuitQubitLine{0}{1}{-\p}
						}
					\circuitMultiGate[\flatgatesize][gencolbg][gencolfg]{}{-0}{-1}{}\circuitAdvanceXBy{\flatgateadvance}
					\circuitMultiGate[\flatgatesize][gencolbg][gencolfg]{}{-1}{-2}{}\circuitAdvanceXBy{\flatgateadvance}
					\circuitMultiGate[\flatgatesize][entcolbg][entcolfg]{}{-0}{-1}{}
				\end{scope}
			\begin{scope}[xshift = 2.7cm, yshift=-1cm, rotate=90]
				\circuitInitXAt{0.2}
				\foreach\p in {0,1,2}{
					\circuitQubitLine{0}{1}{-\p}
					}
				\circuitMultiGate[\flatgatesize][entcolbg][entcolfg]{}{-1}{-2}{}\circuitAdvanceXBy{\flatgateadvance}
				\circuitMultiGate[\flatgatesize][gencolbg][gencolfg]{}{-0}{-1}{}\circuitAdvanceXBy{\flatgateadvance}
				\circuitMultiGate[\flatgatesize][gencolbg][gencolfg]{}{-1}{-2}{}
			\end{scope}
		\end{scope}
		
	\end{scope}

}

\newcommand{\drawAbsorbingtwo}{

	\begin{scope}[scale=0.76]
		\coordinate (BBotLeft) at (0,-3);
		\coordinate (BBotRight) at ($(BBotLeft) + (10,0)$);
		\coordinate (BTopRight) at ($(BBotRight) + (0,8.5)$);
		\coordinate (BTopLeft) at (BTopRight -| BBotLeft);

		\begin{scope}[yshift = 4cm, xshift = 5.5cm]
			\node at (2.2,-0.5) {$\rightarrow$};
				\begin{scope}[xshift = 0.7cm, yshift=-1cm, rotate=90]
					\circuitInitXAt{0.3}
					\foreach\p in {0,1,2}{
						\circuitQubitLine{0}{1}{-\p}
						\draw[fill] ($0.5*(0, -\p)$) circle[radius=1.5pt];
						}
					\circuitMultiGate[\flatgatesize][gencolbg][gencolfg]{}{-1}{-2}{}\circuitAdvanceXBy{\flatgateadvance}
				\circuitMultiGate[\flatgatesize][entcolbg][entcolfg]{}{-0}{-1}{}
			\end{scope}
			\begin{scope}[xshift = 2.7cm, yshift=-1cm, rotate=90]
				\circuitInitXAt{0.3}
				\foreach\p in {0,1,2}{
					\circuitQubitLine{0}{1}{-\p}
					\draw[fill] ($0.5*(0, -\p)$) circle[radius=1.5pt];
					}
                    \circuitMultiGate[\flatgatesize][entcolbg][entcolfg]{}{-0}{-1}{}\circuitAdvanceXBy{\flatgateadvance}
					\circuitMultiGate[\flatgatesize][gencolbg][gencolfg]{}{-1}{-2}{}
				
			\end{scope}
		\end{scope}
		
	\end{scope}

}

\newcommand{\drawAbsorbingthree}{

	\begin{scope}[scale=0.76]
		\coordinate (BBotLeft) at (0,-3);
		\coordinate (BBotRight) at ($(BBotLeft) + (10,0)$);
		\coordinate (BTopRight) at ($(BBotRight) + (0,8.5)$);
		\coordinate (BTopLeft) at (BTopRight -| BBotLeft);

		\begin{scope}[yshift = 4cm, xshift = 5.5cm]
			\node at (2.2,-0.5) {$\rightarrow$};
				\begin{scope}[xshift = 0.7cm, yshift=-1cm, rotate=90]
					\circuitInitXAt{0.3}
					\foreach\p in {0,1,2}{
						\circuitQubitLine{0}{1}{-\p}
						\draw[fill] ($0.5*(0, -\p)$) circle[radius=1.5pt];
						}
					\circuitMultiGate[\flatgatesize][entcolbg][entcolfg]{}{-1}{-2}{}\circuitAdvanceXBy{\flatgateadvance}
				\circuitMultiGate[\flatgatesize][gencolbg][gencolfg]{}{-0}{-1}{}
			\end{scope}
			\begin{scope}[xshift = 2.7cm, yshift=-1cm, rotate=90]
				\circuitInitXAt{0.3}
				\foreach\p in {0,1,2}{
					\circuitQubitLine{0}{1}{-\p}
					\draw[fill] ($0.5*(0, -\p)$) circle[radius=1.5pt];
					}
                    \circuitMultiGate[\flatgatesize][gencolbg][gencolfg]{}{-0}{-1}{}\circuitAdvanceXBy{\flatgateadvance}
					\circuitMultiGate[\flatgatesize][entcolbg][entcolfg]{}{-1}{-2}{}
				
			\end{scope}
		\end{scope}
		
	\end{scope}

}

\newcommand{\drawAbsorbingfour}{

	\begin{scope}[scale=0.76]
		\coordinate (BBotLeft) at (0,-3);
		\coordinate (BBotRight) at ($(BBotLeft) + (10,0)$);
		\coordinate (BTopRight) at ($(BBotRight) + (0,8.5)$);
		\coordinate (BTopLeft) at (BTopRight -| BBotLeft);

		\begin{scope}[yshift = 4cm, xshift = 0.5cm]
			\node at (2.2,-0.5) {$\rightarrow$};
				\begin{scope}[xshift = 0.7cm, yshift=-1cm, rotate=90]
					\circuitInitXAt{0.2}
					\foreach\p in {0,1,2}{
						\circuitQubitLine{0}{1}{-\p}
						}
					\circuitMultiGate[\flatgatesize][entcolbg][entcolfg]{}{-0}{-1}{}\circuitAdvanceXBy{\flatgateadvance}
					\circuitMultiGate[\flatgatesize][gencolbg][gencolfg]{}{-1}{-2}{}\circuitAdvanceXBy{\flatgateadvance}
					\circuitMultiGate[\flatgatesize][gencolbg][gencolfg]{}{-0}{-1}{}
				\end{scope}
			\begin{scope}[xshift = 2.7cm, yshift=-1cm, rotate=90]
				\circuitInitXAt{0.2}
				\foreach\p in {0,1,2}{
					\circuitQubitLine{0}{1}{-\p}
					}
				\circuitMultiGate[\flatgatesize][gencolbg][gencolfg]{}{-1}{-2}{}\circuitAdvanceXBy{\flatgateadvance}
				\circuitMultiGate[\flatgatesize][gencolbg][gencolfg]{}{-0}{-1}{}\circuitAdvanceXBy{\flatgateadvance}
				\circuitMultiGate[\flatgatesize][entcolbg][entcolfg]{}{-1}{-2}{}
			\end{scope}
		\end{scope}
		
	\end{scope}

}

\newcommand{\drawSpecialCaseOne}{

	\begin{scope}[scale=0.76]
		\begin{scope}[yshift = 4cm, xshift = 5.5cm]
				\begin{scope}[xshift = 0.7cm, yshift=-1cm, rotate=90]
					\circuitInitXAt{0.3}
					\foreach\p in {0,1,2}{
						\circuitQubitLine{0}{1.2}{-\p}
						\draw[fill] ($0.5*(0, -\p)$) circle[radius=1.5pt];
						}
					\circuitMultiGate[\flatgatesize][gray!10][gray!80]{}{-0}{-1}{}\circuitAdvanceXBy{\flatgateadvance}
				\circuitMultiGate[\flatgatesize][gencolbg][gencolfg]{}{-1}{-2}{}\circuitAdvanceXBy{\flatgateadvance}
				\circuitMultiGate[\flatgatesize][discolbg][discolfg]{}{-0}{-1}{}
			\end{scope}
		\end{scope}
		
	\end{scope}
}

\newcommand{\drawExampleActionFigure}{
	\begin{scope}[scale=0.72]
		\coordinate (BBotLeft) at (0,0);
		\coordinate (BBotRight) at ($(BBotLeft) + (20,0)$);
		\coordinate (BTopRight) at ($(BBotRight) + (0,7.5)$);
		\coordinate (BTopLeft) at (BTopRight -| BBotLeft);

		\begin{scope}[yshift = 3cm]
            \node[anchor = north] at (0.3,2.5) {(a)};
			\draw[line width=1pt, -{>[length=6pt, width=6pt]}] (3.5,1.4) -- node[above](NNODE){1} +(1,0); \draw[line width=0.7pt] (NNODE) circle(0.22cm);
			\draw[line width=1pt, -{>[length=6pt, width=6pt]}] (7.5,1.4) -- node[above](NNODE){2} +(1,0); \draw[line width=0.7pt] (NNODE) circle(0.22cm);
			\draw[line width=1pt, -{>[length=6pt, width=6pt]}] (11.5,1.4) -- node[above](NNODE){3} +(1,0); \draw[line width=0.7pt] (NNODE) circle(0.22cm);
			\draw[line width=1pt, -{>[length=6pt, width=6pt]}] (15.5,1.4) -- node[above](NNODE){4} +(1,0); \draw[line width=0.7pt] (NNODE) circle(0.22cm);
            \draw[line width=1pt, -{>[length=6pt, width=6pt]}] (19.5,1.4) -- node[above](NNODE){} +(1,0); %
			\begin{scope}[xshift = 1cm, yshift=3cm, rotate=90]
				\circuitInitXAt{-2.35}
				\foreach\p in {0,1,2,3,4}{
					\circuitQubitLine{-2.6}{-0.6}{-\p}
					\draw[fill] ($0.5*(-5.2, -\p)$) circle[radius=1.5pt];
					}
				\circuitMultiGate[\flatgatesize][gencolbg][gencolfg]{}{-0}{-1}{}
				\circuitMultiGate[\flatgatesize][gencolbg][gencolfg]{}{-2}{-3}{}\circuitAdvanceXBy{\flatgateadvance}
				\circuitMultiGate[\flatgatesize][gencolbg][gencolfg]{}{-1}{-2}{}
				\circuitMultiGate[\flatgatesize][gencolbg][gencolfg]{}{-3}{-4}{}\circuitAdvanceXBy{\flatgateadvance}
				\circuitMultiGate[\flatgatesize][gencolbg][gencolfg]{}{-2}{-3}{}\circuitAdvanceXBy{\flatgateadvance}
				\circuitMultiGate[\flatgatesize][gencolbg][gencolfg]{}{-3}{-4}{}\circuitAdvanceXBy{\flatgateadvance}
				\circuitMultiGate[\flatgatesize][entcolbg][entcolfg]{}{-0}{-1}{}
			\end{scope}
		
			\begin{scope}[xshift = 5cm, yshift=3cm, rotate=90]
				\circuitInitXAt{-2.35}
				\foreach\p in {0,1,2,3,4}{
					\circuitQubitLine{-2.6}{-0.6}{-\p}
					\draw[fill] ($0.5*(-5.2, -\p)$) circle[radius=1.5pt];
					}
				\circuitMultiGate[\flatgatesize][gencolbg][gencolfg]{}{-2}{-3}{}\circuitAdvanceXBy{\flatgateadvance}
				\circuitMultiGate[\flatgatesize][entcolbg][entcolfg]{}{-1}{-2}{}
				\circuitMultiGate[\flatgatesize][gencolbg][gencolfg]{}{-3}{-4}{}\circuitAdvanceXBy{\flatgateadvance}
				\circuitMultiGate[\flatgatesize][gencolbg][gencolfg]{}{-0}{-1}{}\circuitAdvanceXBy{\flatgateadvance}
				\circuitMultiGate[\flatgatesize][gencolbg][gencolfg]{}{-1}{-2}{}\circuitAdvanceXBy{\flatgateadvance}
				\circuitMultiGate[\flatgatesize][gencolbg][gencolfg]{}{-2}{-3}{}\circuitAdvanceXBy{\flatgateadvance}
				\circuitMultiGate[\flatgatesize][gencolbg][gencolfg]{}{-3}{-4}{}\circuitAdvanceXBy{\flatgateadvance}
			\end{scope}
		
			\begin{scope}[xshift = 9cm, yshift=3cm, rotate=90]
				\circuitInitXAt{-2.35}
				\foreach\p in {0,1,2,3,4}{
					\circuitQubitLine{-2.6}{-0.6}{-\p}
					\draw[fill] ($0.5*(-5.2, -\p)$) circle[radius=1.5pt];
					}
				\circuitMultiGate[\flatgatesize][entcolbg][entcolfg]{}{-1}{-2}{}\circuitAdvanceXBy{\flatgateadvance}
				\circuitMultiGate[\flatgatesize][gencolbg][gencolfg]{}{-0}{-1}{}
				\circuitMultiGate[\flatgatesize][gencolbg][gencolfg]{}{-2}{-3}{}\circuitAdvanceXBy{\flatgateadvance}
				\circuitMultiGate[\flatgatesize][gencolbg][gencolfg]{}{-3}{-4}{}
				\circuitMultiGate[\flatgatesize][gencolbg][gencolfg]{}{-1}{-2}{}\circuitAdvanceXBy{\flatgateadvance}
				\circuitMultiGate[\flatgatesize][gencolbg][gencolfg]{}{-2}{-3}{}\circuitAdvanceXBy{\flatgateadvance}
				\circuitMultiGate[\flatgatesize][gencolbg][gencolfg]{}{-3}{-4}{}\circuitAdvanceXBy{\flatgateadvance}
			\end{scope}
		
			\begin{scope}[xshift = 13cm, yshift=3cm, rotate=90]
				\circuitInitXAt{-2.35}
				\foreach\p in {0,1,2,3,4}{
					\circuitQubitLine{-2.6}{-0.6}{-\p}
					\draw[fill] ($0.5*(-5.2, -\p)$) circle[radius=1.5pt];
					}
				\circuitMultiGate[\flatgatesize][gencolbg][gencolfg]{}{-0}{-1}{}\circuitAdvanceXBy{\flatgateadvance}
				\circuitMultiGate[\flatgatesize][entcolbg][entcolfg]{}{-1}{-2}{}\circuitAdvanceXBy{\flatgateadvance}
				\circuitMultiGate[\flatgatesize][gencolbg][gencolfg]{}{-2}{-3}{}\circuitAdvanceXBy{\flatgateadvance}
				\circuitMultiGate[\flatgatesize][gencolbg][gencolfg]{}{-3}{-4}{}
				\circuitMultiGate[\flatgatesize][gencolbg][gencolfg]{}{-1}{-2}{}\circuitAdvanceXBy{\flatgateadvance}
				\circuitMultiGate[\flatgatesize][gencolbg][gencolfg]{}{-2}{-3}{}\circuitAdvanceXBy{\flatgateadvance}
				\circuitMultiGate[\flatgatesize][gencolbg][gencolfg]{}{-3}{-4}{}\circuitAdvanceXBy{\flatgateadvance}
			\end{scope}
		
			\begin{scope}[xshift = 17cm, yshift=3cm, rotate=90]
				\circuitInitXAt{-2.35}
				\foreach\p in {0,1,2,3,4}{
					\circuitQubitLine{-2.6}{-0.6}{-\p}
					\draw[fill] ($0.5*(-5.2, -\p)$) circle[radius=1.5pt];
					}
				\circuitMultiGate[\flatgatesize][gencolbg][gencolfg]{}{-0}{-1}{}
				\circuitMultiGate[\flatgatesize][gencolbg][gencolfg]{}{-2}{-3}{}\circuitAdvanceXBy{\flatgateadvance}
				\circuitMultiGate[\flatgatesize][gencolbg][gencolfg]{}{-1}{-2}{}
				\circuitMultiGate[\flatgatesize][gencolbg][gencolfg]{}{-3}{-4}{}\circuitAdvanceXBy{\flatgateadvance}
				\circuitMultiGate[\flatgatesize][gencolbg][gencolfg]{}{-2}{-3}{}\circuitAdvanceXBy{\flatgateadvance}
				\circuitMultiGate[\flatgatesize][entcolbg][entcolfg]{}{-3}{-4}{}\circuitAdvanceXBy{\flatgateadvance}
				\circuitMultiGate[\flatgatesize][gencolbg][gencolfg]{}{-3}{-4}{}
			\end{scope}
			\begin{scope}[xshift = 21cm, yshift=3cm, rotate=90]
				\circuitInitXAt{-2.35}
				\foreach\p in {0,1,2,3,4}{
					\circuitQubitLine{-2.6}{-0.6}{-\p}
					\draw[fill] ($0.5*(-5.2, -\p)$) circle[radius=1.5pt];
					}
				\circuitMultiGate[\flatgatesize][gencolbg][gencolfg]{}{-0}{-1}{}
				\circuitMultiGate[\flatgatesize][gencolbg][gencolfg]{}{-2}{-3}{}\circuitAdvanceXBy{\flatgateadvance}
				\circuitMultiGate[\flatgatesize][gencolbg][gencolfg]{}{-1}{-2}{}
				\circuitMultiGate[\flatgatesize][gencolbg][gencolfg]{}{-3}{-4}{}\circuitAdvanceXBy{\flatgateadvance}
				\circuitMultiGate[\flatgatesize][gencolbg][gencolfg]{}{-2}{-3}{}\circuitAdvanceXBy{\flatgateadvance}
				\circuitMultiGate[\flatgatesize][gencolbg][gencolfg]{}{-3}{-4}{}
			\end{scope}
		\end{scope}

		\begin{scope}[yshift = 0cm]
            \node[anchor = north] at (0.3,2.5) {(b)};
			\draw[line width=1pt, -{>[length=6pt, width=6pt]}] (3.5,1.4) -- node[above](NNODE){tracing} +(1,0); %
			\draw[line width=1pt, -{>[length=6pt, width=6pt]}] (7.5,1.4) -- node[above](NNODE){2} +(1,0); \draw[line width=0.7pt] (NNODE) circle(0.22cm);
			\draw[line width=1pt, -{>[length=6pt, width=6pt]}] (11.5,1.4) -- node[above](NNODE){1} +(1,0); \draw[line width=0.7pt] (NNODE) circle(0.22cm);
			\draw[line width=1pt, -{>[length=6pt, width=6pt]}] (15.5,1.4) -- node[above](NNODE){} +(1,0); %
			\begin{scope}[xshift = 1cm, yshift=3cm, rotate=90]
				\circuitInitXAt{-2.35}
				\foreach\p in {0,1,2,3,4}{
					\circuitQubitLine{-2.6}{-0.6}{-\p}
					\draw[fill] ($0.5*(-5.2, -\p)$) circle[radius=1.5pt];
					}
				\circuitMultiGate[\flatgatesize][gencolbg][gencolfg]{}{-0}{-1}{}
				\circuitMultiGate[\flatgatesize][gencolbg][gencolfg]{}{-2}{-3}{}\circuitAdvanceXBy{\flatgateadvance}
				\circuitMultiGate[\flatgatesize][gencolbg][gencolfg]{}{-1}{-2}{}
				\circuitMultiGate[\flatgatesize][gencolbg][gencolfg]{}{-3}{-4}{}\circuitAdvanceXBy{\flatgateadvance}
				\circuitMultiGate[\flatgatesize][gencolbg][gencolfg]{}{-2}{-3}{}\circuitAdvanceXBy{\flatgateadvance}
				\circuitMultiGate[\flatgatesize][gencolbg][gencolfg]{}{-3}{-4}{}\circuitAdvanceXBy{\flatgateadvance}
				\circuitMultiGate[\flatgatesize][discolbg][discolfg]{}{-1}{-2}{}
			\end{scope}
		
			\begin{scope}[xshift = 5cm, yshift=3cm, rotate=90]
				\circuitInitXAt{-2.35}
				\foreach\p in {0,1,2,3,4}{
					\circuitQubitLine{-2.6}{-0.6}{-\p}
					\draw[fill] ($0.5*(-5.2, -\p)$) circle[radius=1.5pt];
					}
				\circuitMultiGate[\flatgatesize][gencolbg][gencolfg]{}{-0}{-1}{}
				\circuitMultiGate[\flatgatesize][gray!10][gray!80]{}{-2}{-3}{}\circuitAdvanceXBy{\flatgateadvance}
				\circuitMultiGate[\flatgatesize][gencolbg][gencolfg]{}{-1}{-2}{}
				\circuitMultiGate[\flatgatesize][gencolbg][gencolfg]{}{-3}{-4}{}\circuitAdvanceXBy{\flatgateadvance}
				\circuitMultiGate[\flatgatesize][gencolbg][gencolfg]{}{-2}{-3}{}\circuitAdvanceXBy{\flatgateadvance}
				\circuitMultiGate[\flatgatesize][gencolbg][gencolfg]{}{-3}{-4}{}\circuitAdvanceXBy{\flatgateadvance}
				\circuitMultiGate[\flatgatesize][discolbg][discolfg]{}{-1}{-2}{}
			\end{scope}
		
			\begin{scope}[xshift = 9cm, yshift=3cm, rotate=90]
				\circuitInitXAt{-2.35}
				\foreach\p in {0,1,2,3,4}{
					\circuitQubitLine{-2.6}{-0.6}{-\p}
					\draw[fill] ($0.5*(-5.2, -\p)$) circle[radius=1.5pt];
					}
				\circuitMultiGate[\flatgatesize][gencolbg][gencolfg]{}{-3}{-4}{}\circuitAdvanceXBy{\flatgateadvance}
				\circuitMultiGate[\flatgatesize][gencolbg][gencolfg]{}{-0}{-1}{}
				\circuitMultiGate[\flatgatesize][gray!10][gray!80]{}{-2}{-3}{}\circuitAdvanceXBy{\flatgateadvance}
				\circuitMultiGate[\flatgatesize][gencolbg][gencolfg]{}{-1}{-2}{}\circuitAdvanceXBy{\flatgateadvance}
				\circuitMultiGate[\flatgatesize][gencolbg][gencolfg]{}{-2}{-3}{}\circuitAdvanceXBy{\flatgateadvance}
				\circuitMultiGate[\flatgatesize][gencolbg][gencolfg]{}{-3}{-4}{}\circuitAdvanceXBy{\flatgateadvance}
				\circuitMultiGate[\flatgatesize][discolbg][discolfg]{}{-1}{-2}{}
			\end{scope}
		
			\begin{scope}[xshift = 13cm, yshift=3cm, rotate=90]
				\circuitInitXAt{-2.35}
				\foreach\p in {0,1,2,3,4}{
					\circuitQubitLine{-2.6}{-0.6}{-\p}
					\draw[fill] ($0.5*(-5.2, -\p)$) circle[radius=1.5pt];
					}
				\circuitMultiGate[\flatgatesize][gencolbg][gencolfg]{}{-3}{-4}{}
				\circuitMultiGate[\flatgatesize][gencolbg][gencolfg]{}{-0}{-1}{}\circuitAdvanceXBy{\flatgateadvance}
				\circuitMultiGate[\flatgatesize][gencolbg][gencolfg]{}{-1}{-2}{}\circuitAdvanceXBy{\flatgateadvance}
				\circuitMultiGate[\flatgatesize][gencolbg][gencolfg]{}{-2}{-3}{}\circuitAdvanceXBy{\flatgateadvance}
				\circuitMultiGate[\flatgatesize][gencolbg][gencolfg]{}{-3}{-4}{}\circuitAdvanceXBy{\flatgateadvance}
				\circuitMultiGate[\flatgatesize][gray!10][gray!80]{}{-1}{-2}{}\circuitAdvanceXBy{\flatgateadvance}
				\circuitMultiGate[\flatgatesize][discolbg][discolfg]{}{-1}{-2}{}
			\end{scope}
			\begin{scope}[xshift = 17cm, yshift=3cm, rotate=90]
				\circuitInitXAt{-2.35}
				\foreach\p in {0,1,2,3,4}{
					\circuitQubitLine{-2.6}{-0.6}{-\p}
					\draw[fill] ($0.5*(-5.2, -\p)$) circle[radius=1.5pt];
					}
				\circuitMultiGate[\flatgatesize][gencolbg][gencolfg]{}{-3}{-4}{}
				\circuitMultiGate[\flatgatesize][gencolbg][gencolfg]{}{-0}{-1}{}\circuitAdvanceXBy{\flatgateadvance}
				\circuitMultiGate[\flatgatesize][gencolbg][gencolfg]{}{-1}{-2}{}\circuitAdvanceXBy{\flatgateadvance}
				\circuitMultiGate[\flatgatesize][gencolbg][gencolfg]{}{-2}{-3}{}\circuitAdvanceXBy{\flatgateadvance}
				\circuitMultiGate[\flatgatesize][gencolbg][gencolfg]{}{-3}{-4}{}
			\end{scope}
		\end{scope}

	\end{scope}
}

\newcommand{\drawExampleStringState}{
    \begin{tikzpicture}
\draw[black, thick] (0,0) .. controls (1.333,1) and (2.666,1) .. (4,0);

\draw[black, thick] (1,0) .. controls (1.333,0.5) and (1.666,0.5) .. (2,0);
\draw[black, thick] (3,0) .. controls (4,1) and (5,1) .. (6,0);

\filldraw[black] (0,0) circle (2pt);
\filldraw[black] (1,0) circle (2pt);
\filldraw[black] (2,0) circle (2pt);
\filldraw[black] (3,0) circle (2pt);
\filldraw[black] (4,0) circle (2pt);
\filldraw[black] (5,0) circle (2pt);
\filldraw[black] (6,0) circle (2pt);
\filldraw[black] (7,0) circle (2pt);
\end{tikzpicture}
}

\newcommand{
    \figureExplanationOfStrings}{

\begin{scope}[scale=0.666]
\node at (-0.7,0) {1.};
\begin{scope}[xshift=3cm]
\draw[black, thick] (0,0) .. controls (0.3,0.5) and (0.7,0.5) .. (1,0);
\filldraw[black] (0,0) circle (3pt);
\filldraw[black] (1,0) circle (3pt);
\end{scope}
\draw[arrows=-{>[harpoon,length=5pt, width=6pt]}, entcolfg, line width=1.3pt] (1.6,0.075) -- (2.4,0.075);
\draw[arrows=-{>[harpoon,length=5pt, width=6pt]}, discolfg, line width=1.3pt] (2.4,-0.075) -- (1.6,-0.075);
\begin{scope}[xshift = -3cm]
\filldraw[black] (3,0) circle (3pt);
\filldraw[black] (4,0) circle (3pt);
\end{scope}
\end{scope}

\begin{scope}[yshift = -1cm, scale=0.666]
\node at (-0.7,0) {2.};
\begin{scope}[xshift=3cm]
\draw[black, thick] (0,0) .. controls (0.3,0.45) and (0.7,0.45) .. (1,0.5);
\filldraw[black] (0,0) circle (3pt);
\filldraw[black] (1,0) circle (3pt);
\end{scope}
\draw[arrows=-{>[harpoon,length=5pt, width=6pt]}, entcolfg, line width=1.3pt] (1.6,0.075) -- (2.4,0.075);
\draw[arrows=-{>[harpoon,length=5pt, width=6pt]}, discolfg, line width=1.3pt] (2.4,-0.075) -- (1.6,-0.075);
\begin{scope}[xshift=-3cm]
\draw[black, thick] (4,0) .. controls (4.1,0.3) and (4.3,0.35) .. (4.5,0.4);
\filldraw[black] (3,0) circle (3pt);
\filldraw[black] (4,0) circle (3pt);
\end{scope}
\end{scope}

\begin{scope}[yshift = -2cm, scale=0.666]
\node at (-0.7,0) {3.};
\begin{scope}[xshift=3cm]
\draw[black, thick] (0,0.5) .. controls (0.3,0.45) and (0.7,0.45) .. (1,0);
\filldraw[black] (0,0) circle (3pt);
\filldraw[black] (1,0) circle (3pt);
\end{scope}
\draw[arrows=-{>[harpoon,length=5pt, width=6pt]}, entcolfg, line width=1.3pt] (1.6,0.075) -- (2.4,0.075);
\draw[arrows=-{>[harpoon,length=5pt, width=6pt]}, discolfg, line width=1.3pt] (2.4,-0.075) -- (1.6,-0.075);
\begin{scope}[xshift=-3cm]
\draw[black, thick] (3,0) .. controls (2.9,0.3) and (2.7,0.35) .. (2.5,0.4);
\filldraw[black] (3,0) circle (3pt);
\filldraw[black] (4,0) circle (3pt);
\end{scope}
\end{scope}

\begin{scope}[xshift=4.5cm, scale=0.666]
\node at (-0.7,0) {4.};
\begin{scope}[xshift=3cm]
\draw[black, thick] (0,0.5) .. controls (0.3,0.45) and (0.7,0.45) .. (1,0);
\draw[black, thick] (0,0) .. controls (0.3,0.45) and (0.7,0.45) .. (1,0.5);
\filldraw[black] (0,0) circle (3pt);
\filldraw[black] (1,0) circle (3pt);
\end{scope}
\draw[arrows=-{>[harpoon,length=5pt, width=6pt]}, entcolfg, line width=1.3pt] (1.6,0.075) -- (2.4,0.075);
\draw[arrows=-{>[harpoon,length=5pt, width=6pt]}, discolfg, line width=1.3pt] (2.4,-0.075) -- (1.6,-0.075);
\begin{scope}[xshift=-3cm]
\draw[black, thick] (3,0) .. controls (2.9,0.3) and (2.7,0.35) .. (2.5,0.4);
\draw[black, thick] (4,0) .. controls (4.1,0.3) and (4.3,0.35) .. (4.5,0.4);
\filldraw[black] (3,0) circle (3pt);
\filldraw[black] (4,0) circle (3pt);
\end{scope}
\end{scope}

\begin{scope}[xshift=4.5cm,yshift = -1cm, scale=0.666]
\node at (-0.7,0) {5.};
\begin{scope}[xshift=4cm,xscale=-1]
\begin{scope}[xshift=0cm]
\draw[black, thick] (-0.5,0.6) .. controls (0.1,0.5) and (0.7,0.3) .. (1,0);
\draw[black, thick] (0,0) .. controls (-0.1,0.3) and (-0.3,0.35) .. (-0.5,0.4);
\end{scope}
\begin{scope}[xshift=0cm]
\draw[black, thick] (2.5,0.4) .. controls (3.1,0.35) and (3.7,0.25) .. (4,0);
\draw[black, thick] (3,0) .. controls (2.9,0.3) and (2.7,0.55) .. (2.5,0.6);
\end{scope}
\end{scope}
\filldraw[black] (0,0) circle (3pt);
\filldraw[black] (1,0) circle (3pt);
\draw[arrows=-{>[harpoon,length=5pt, width=6pt]}, entcolfg, line width=1.3pt] (1.6,0.075) -- (2.4,0.075);
\draw[arrows=-{>[harpoon,length=5pt, width=6pt]}, discolfg, line width=1.3pt] (2.4,-0.075) -- (1.6,-0.075);
\filldraw[black] (3,0) circle (3pt);
\filldraw[black] (4,0) circle (3pt);
\end{scope}

\begin{scope}[xshift=4.5cm,yshift = -2cm, scale=0.666]
\node at (-0.7,0) {6.};
\begin{scope}[xshift=3cm]
\draw[black, thick] (-0.5,0.6) .. controls (0.1,0.5) and (0.7,0.3) .. (1,0);
\draw[black, thick] (0,0) .. controls (-0.1,0.3) and (-0.3,0.35) .. (-0.5,0.4);
\end{scope}
\begin{scope}[xshift=-3cm]
\draw[black, thick] (2.5,0.4) .. controls (3.1,0.35) and (3.7,0.25) .. (4,0);
\draw[black, thick] (3,0) .. controls (2.9,0.3) and (2.7,0.55) .. (2.5,0.6);
\end{scope}
\filldraw[black] (0,0) circle (3pt);
\filldraw[black] (1,0) circle (3pt);
\draw[arrows=-{>[harpoon,length=5pt, width=6pt]}, entcolfg, line width=1.3pt] (1.6,0.075) -- (2.4,0.075);
\draw[arrows=-{>[harpoon,length=5pt, width=6pt]}, discolfg, line width=1.3pt] (2.4,-0.075) -- (1.6,-0.075);
\filldraw[black] (3,0) circle (3pt);
\filldraw[black] (4,0) circle (3pt);
\end{scope}
    
    }

\newcommand{\drawAbsorbProofCaseA}{
        \begin{tikzpicture}[scale=0.76]
        	\begin{scope}[rotate=90]
				\circuitInitXAt{0.3}
				\foreach\p in {0,1,2,3,4,5}{
					\circuitQubitLine{0}{1.4}{-\p} }
                    \draw[fill] (0,0) circle[radius=1.5pt]; 
                    \draw[fill] (0,-0.5) circle[radius=1.5pt]; 
				\circuitMultiGate[\flatgatesize][gencolbg][gencolfg]{}{-0}{-1}{}
                    \circuitAdvanceXBy{\flatgateadvance}
				\circuitMultiGate[\flatgatesize][gencolbg][gencolfg]{}{-1}{-2}{}
                    \circuitAdvanceXBy{\flatgateadvance}
				\circuitMultiGate[\flatgatesize][gencolbg][gencolfg]{}{-2}{-3}{}
                    \circuitAdvanceXBy{\flatgateadvance}
				\circuitMultiGate[\flatgatesize][entcolbg][entcolfg]{}{-4}{-5}{}
			\end{scope}
            \draw[line width=1pt, -{>[length=6pt, width=6pt]}] (3,0.7) -- (4,0.7);
            \begin{scope}[xshift=4.5cm, rotate=90]
				\circuitInitXAt{0.3}
				\foreach\p in {0,1,2,3,4,5}{
					\circuitQubitLine{0}{1.4}{-\p} }
                    \draw[fill] (0,0) circle[radius=1.5pt]; 
                    \draw[fill] (0,-0.5) circle[radius=1.5pt]; 
				\circuitMultiGate[\flatgatesize][gencolbg][gencolfg]{}{-0}{-1}{}
				\circuitMultiGate[\flatgatesize][entcolbg][entcolfg]{}{-4}{-5}{}
                    \circuitAdvanceXBy{\flatgateadvance}
				\circuitMultiGate[\flatgatesize][gencolbg][gencolfg]{}{-1}{-2}{}
                    \circuitAdvanceXBy{\flatgateadvance}
				\circuitMultiGate[\flatgatesize][gencolbg][gencolfg]{}{-2}{-3}{}
                    \circuitAdvanceXBy{\flatgateadvance}
			\end{scope}
        \end{tikzpicture}}

\newcommand{\drawAbsorbProofCaseB}{

        \begin{tikzpicture}[scale=0.76]
        	\begin{scope}[rotate=90]
				\circuitInitXAt{0.3}
				\foreach\p in {0,1,2,3,4}{
					\circuitQubitLine{0}{1.4}{-\p} }
                    \draw[fill] (0,0) circle[radius=1.5pt]; 
                    \draw[fill] (0,-0.5) circle[radius=1.5pt]; 
				\circuitMultiGate[\flatgatesize][gencolbg][gencolfg]{}{-0}{-1}{}
                    \circuitAdvanceXBy{\flatgateadvance}
				\circuitMultiGate[\flatgatesize][gencolbg][gencolfg]{}{-1}{-2}{}
                    \circuitAdvanceXBy{\flatgateadvance}
				\circuitMultiGate[\flatgatesize][gencolbg][gencolfg]{}{-2}{-3}{}
                    \circuitAdvanceXBy{\flatgateadvance}
				\circuitMultiGate[\flatgatesize][entcolbg][entcolfg]{}{-3}{-4}{}
			\end{scope}
        \end{tikzpicture}

}

\newcommand{\drawAbsorbProofCaseC}{

        \begin{tikzpicture}[scale=0.76]
        	\begin{scope}[rotate=90]
				\circuitInitXAt{0.3}
				\foreach\p in {0,1,2,3}{
					\circuitQubitLine{0}{1.4}{-\p} }
                    \draw[fill] (0,0) circle[radius=1.5pt]; 
                    \draw[fill] (0,-0.5) circle[radius=1.5pt]; 
				\circuitMultiGate[\flatgatesize][gencolbg][gencolfg]{}{-0}{-1}{}
                    \circuitAdvanceXBy{\flatgateadvance}
				\circuitMultiGate[\flatgatesize][gencolbg][gencolfg]{}{-1}{-2}{}
                    \circuitAdvanceXBy{\flatgateadvance}
				\circuitMultiGate[\flatgatesize][gencolbg][gencolfg]{}{-2}{-3}{}
                    \circuitAdvanceXBy{\flatgateadvance}
				\circuitMultiGate[\flatgatesize][entcolbg][entcolfg]{}{-2}{-3}{}
			\end{scope}
            \draw[line width=1pt, -{>[length=6pt, width=6pt]}] (2,0.7) -- (3,0.7);
            \begin{scope}[xshift=3.5cm, rotate=90]
				\circuitInitXAt{0.3}
				\foreach\p in {0,1,2,3}{
					\circuitQubitLine{0}{1.4}{-\p} }
                    \draw[fill] (0,0) circle[radius=1.5pt]; 
                    \draw[fill] (0,-0.5) circle[radius=1.5pt]; 
				\circuitMultiGate[\flatgatesize][gencolbg][gencolfg]{}{-0}{-1}{}
                    \circuitAdvanceXBy{\flatgateadvance}
				\circuitMultiGate[\flatgatesize][gencolbg][gencolfg]{}{-1}{-2}{}
                    \circuitAdvanceXBy{\flatgateadvance}
				\circuitMultiGate[\flatgatesize][gencolbg][gencolfg]{}{-2}{-3}{}
			\end{scope}
        \end{tikzpicture}

}

\newcommand{\drawAbsorbProofCaseD}{

        \begin{tikzpicture}[scale=0.76]
        	\begin{scope}[rotate=90]
				\circuitInitXAt{0.3}
				\foreach\p in {0,1,2,3}{
					\circuitQubitLine{0}{1.4}{-\p} }
                    \draw[fill] (0,0) circle[radius=1.5pt]; 
                    \draw[fill] (0,-0.5) circle[radius=1.5pt]; 
				\circuitMultiGate[\flatgatesize][gencolbg][gencolfg]{}{-0}{-1}{}
                    \circuitAdvanceXBy{\flatgateadvance}
				\circuitMultiGate[\flatgatesize][gencolbg][gencolfg]{}{-1}{-2}{}
                    \circuitAdvanceXBy{\flatgateadvance}
				\circuitMultiGate[\flatgatesize][gencolbg][gencolfg]{}{-2}{-3}{}
                    \circuitAdvanceXBy{\flatgateadvance}
				\circuitMultiGate[\flatgatesize][entcolbg][entcolfg]{}{-1}{-2}{}
			\end{scope}
            \draw[line width=1pt, -{>[length=6pt, width=6pt]}] (2,0.7) -- (3,0.7);
            \begin{scope}[xshift=3.5cm, rotate=90]
				\circuitInitXAt{0.3}
				\foreach\p in {0,1,2,3}{
					\circuitQubitLine{0}{1.4}{-\p} }
                    \draw[fill] (0,0) circle[radius=1.5pt]; 
                    \draw[fill] (0,-0.5) circle[radius=1.5pt]; 
				\circuitMultiGate[\flatgatesize][gencolbg][gencolfg]{}{-0}{-1}{}
				\circuitMultiGate[\flatgatesize][entcolbg][entcolfg]{}{-2}{-3}{}
                    \circuitAdvanceXBy{\flatgateadvance}
				\circuitMultiGate[\flatgatesize][gencolbg][gencolfg]{}{-1}{-2}{}
                    \circuitAdvanceXBy{\flatgateadvance}
				\circuitMultiGate[\flatgatesize][gencolbg][gencolfg]{}{-2}{-3}{}
			\end{scope}
        \end{tikzpicture}

}

\newcommand{\drawAbsorbProofCaseE}{

        \begin{tikzpicture}[scale=0.76]
        	\begin{scope}[rotate=90]
				\circuitInitXAt{0.3}
                \clip (-0.2,0.2) rectangle (1.5,-1.7);
				\foreach\p in {0,1,2,3}{
					\circuitQubitLine{0}{1.4}{-\p} }
                    \draw[fill] (0,0) circle[radius=1.5pt]; 
                    \draw[fill] (0,-0.5) circle[radius=1.5pt]; 
                    \draw[fill] (0,-1) circle[radius=1.5pt]; 
				\circuitMultiGate[\flatgatesize][gencolbg][gencolfg]{}{-0}{-1}{}
				\circuitMultiGate[\flatgatesize][gencolbg][gencolfg]{}{-3}{-4}{}
                    \circuitAdvanceXBy{\flatgateadvance}
				\circuitMultiGate[\flatgatesize][gencolbg][gencolfg]{}{-1}{-2}{}
                    \circuitAdvanceXBy{\flatgateadvance}
				\circuitMultiGate[\flatgatesize][gencolbg][gencolfg]{}{-2}{-3}{}
                    \circuitAdvanceXBy{\flatgateadvance}
				\circuitMultiGate[\flatgatesize][entcolbg][entcolfg]{}{-0}{-1}{}
			\end{scope}
            \draw[line width=1pt, -{>[length=6pt, width=6pt]}] (1.75,0.7) -- (2.25,0.7);
            \begin{scope}[xshift=2.5cm, rotate=90]
				\circuitInitXAt{0.3}
                \clip (-0.2,0.2) rectangle (1.5,-1.7);
				\foreach\p in {0,1,2,3}{
					\circuitQubitLine{0}{1.4}{-\p} }
                    \draw[fill] (0,0) circle[radius=1.5pt]; 
                    \draw[fill] (0,-0.5) circle[radius=1.5pt]; 
                    \draw[fill] (0,-1) circle[radius=1.5pt]; 
				\circuitMultiGate[\flatgatesize][entcolbg][entcolfg]{}{-1}{-2}{}
				\circuitMultiGate[\flatgatesize][gencolbg][gencolfg]{}{-3}{-4}{}
                    \circuitAdvanceXBy{\flatgateadvance}
				\circuitMultiGate[\flatgatesize][gencolbg][gencolfg]{}{-0}{-1}{}
                    \circuitAdvanceXBy{\flatgateadvance}
				\circuitMultiGate[\flatgatesize][gencolbg][gencolfg]{}{-1}{-2}{}
                    \circuitAdvanceXBy{\flatgateadvance}
				\circuitMultiGate[\flatgatesize][gencolbg][gencolfg]{}{-2}{-3}{}
			\end{scope}
            \draw[line width=1pt, -{>[length=6pt, width=6pt]}] (4.25,0.7) -- (4.75,0.7);
            \begin{scope}[xshift=5cm, rotate=90]
				\circuitInitXAt{0.3}
                \clip (-0.2,0.2) rectangle (1.5,-1.7);
				\foreach\p in {0,1,2,3}{
					\circuitQubitLine{0}{1.4}{-\p} }
                    \draw[fill] (0,0) circle[radius=1.5pt]; 
                    \draw[fill] (0,-0.5) circle[radius=1.5pt]; 
                    \draw[fill] (0,-1) circle[radius=1.5pt]; 
				\circuitMultiGate[\flatgatesize][gencolbg][gencolfg]{}{-0}{-1}{}
				\circuitMultiGate[\flatgatesize][gencolbg][gencolfg]{}{-3}{-4}{}
                    \circuitAdvanceXBy{\flatgateadvance}
				\circuitMultiGate[\flatgatesize][entcolbg][entcolfg]{}{-1}{-2}{}
                    \circuitAdvanceXBy{\flatgateadvance}
				\circuitMultiGate[\flatgatesize][gencolbg][gencolfg]{}{-1}{-2}{}
                    \circuitAdvanceXBy{\flatgateadvance}
				\circuitMultiGate[\flatgatesize][gencolbg][gencolfg]{}{-2}{-3}{}
			\end{scope}
            \draw[line width=1pt, -{>[length=6pt, width=6pt]}] (6.75,0.7) -- (7.25,0.7);
            \begin{scope}[xshift=7.5cm, rotate=90]
				\circuitInitXAt{0.3}
                \clip (-0.2,0.2) rectangle (1.5,-1.7);
				\foreach\p in {0,1,2,3}{
					\circuitQubitLine{0}{1.4}{-\p} }
                    \draw[fill] (0,0) circle[radius=1.5pt]; 
                    \draw[fill] (0,-0.5) circle[radius=1.5pt]; 
                    \draw[fill] (0,-1) circle[radius=1.5pt]; 
				\circuitMultiGate[\flatgatesize][gencolbg][gencolfg]{}{-0}{-1}{}
				\circuitMultiGate[\flatgatesize][gencolbg][gencolfg]{}{-3}{-4}{}
                    \circuitAdvanceXBy{\flatgateadvance}
				\circuitMultiGate[\flatgatesize][gencolbg][gencolfg]{}{-1}{-2}{}
                    \circuitAdvanceXBy{\flatgateadvance}
				\circuitMultiGate[\flatgatesize][gencolbg][gencolfg]{}{-2}{-3}{}
			\end{scope}
        \end{tikzpicture}

}

\newcommand{\drawAbsorbProofCaseF}{

        \begin{tikzpicture}[scale=0.76]
        	\begin{scope}[rotate=90]
				\circuitInitXAt{0.3}
                \clip (-0.2,0.2) rectangle (2,-1.7);
				\foreach\p in {0,1,2,3}{
					\circuitQubitLine{0}{1.7}{-\p} 
					\draw[fill] ($0.5*(0, -\p)$) circle[radius=1.5pt]; }
				\circuitMultiGate[\flatgatesize][gencolbg][gencolfg]{}{-0}{-1}{}
				\circuitMultiGate[\flatgatesize][gencolbg][gencolfg]{}{-2}{-3}{}
                    \circuitAdvanceXBy{\flatgateadvance}
				\circuitMultiGate[\flatgatesize][gencolbg][gencolfg]{}{-1}{-2}{}
				\circuitMultiGate[\flatgatesize][gencolbg][gencolfg]{}{-3}{-4}{}
                    \circuitAdvanceXBy{\flatgateadvance}
				\circuitMultiGate[\flatgatesize][gencolbg][gencolfg]{}{-2}{-3}{}
                    \circuitAdvanceXBy{\flatgateadvance}
				\circuitMultiGate[\flatgatesize][entcolbg][entcolfg]{}{-0}{-1}{}
			\end{scope}
            \draw[line width=1pt, -{>[length=6pt, width=6pt]}] (1.75,0.85) -- (2.25,0.85);
            \begin{scope}[xshift=2.5cm, rotate=90]
				\circuitInitXAt{0.3}
                \clip (-0.2,0.2) rectangle (2,-1.7);
				\foreach\p in {0,1,2,3}{
					\circuitQubitLine{0}{1.7}{-\p} 
					\draw[fill] ($0.5*(0, -\p)$) circle[radius=1.5pt]; }
				\circuitMultiGate[\flatgatesize][gencolbg][gencolfg]{}{-2}{-3}{}
                    \circuitAdvanceXBy{\flatgateadvance}
				\circuitMultiGate[\flatgatesize][entcolbg][entcolfg]{}{-1}{-2}{}
				\circuitMultiGate[\flatgatesize][gencolbg][gencolfg]{}{-3}{-4}{}
                    \circuitAdvanceXBy{\flatgateadvance}
				\circuitMultiGate[\flatgatesize][gencolbg][gencolfg]{}{-0}{-1}{}
                    \circuitAdvanceXBy{\flatgateadvance}
				\circuitMultiGate[\flatgatesize][gencolbg][gencolfg]{}{-1}{-2}{}
                    \circuitAdvanceXBy{\flatgateadvance}
				\circuitMultiGate[\flatgatesize][gencolbg][gencolfg]{}{-2}{-3}{}
			\end{scope}
            \draw[line width=1pt, -{>[length=6pt, width=6pt]}] (4.25,0.85) -- (4.75,0.85);
            \begin{scope}[xshift=5cm, rotate=90]
				\circuitInitXAt{0.3}
                \clip (-0.2,0.2) rectangle (2,-1.7);
				\foreach\p in {0,1,2,3}{
					\circuitQubitLine{0}{1.7}{-\p} 
					\draw[fill] ($0.5*(0, -\p)$) circle[radius=1.5pt]; }
				\circuitMultiGate[\flatgatesize][entcolbg][entcolfg]{}{-1}{-2}{}
                    \circuitAdvanceXBy{\flatgateadvance}
				\circuitMultiGate[\flatgatesize][gencolbg][gencolfg]{}{-0}{-1}{}
				\circuitMultiGate[\flatgatesize][gencolbg][gencolfg]{}{-2}{-3}{}
                    \circuitAdvanceXBy{\flatgateadvance}
				\circuitMultiGate[\flatgatesize][gencolbg][gencolfg]{}{-1}{-2}{}
				\circuitMultiGate[\flatgatesize][gencolbg][gencolfg]{}{-3}{-4}{}
                    \circuitAdvanceXBy{\flatgateadvance}
				\circuitMultiGate[\flatgatesize][gencolbg][gencolfg]{}{-2}{-3}{}
			\end{scope}
            \draw[line width=1pt, -{>[length=6pt, width=6pt]}] (6.75,0.85) -- (7.25,0.85);
            \begin{scope}[xshift=7.5cm, rotate=90]
				\circuitInitXAt{0.3}
                \clip (-0.2,0.2) rectangle (2,-1.7);
				\foreach\p in {0,1,2,3}{
					\circuitQubitLine{0}{1.7}{-\p} 
					\draw[fill] ($0.5*(0, -\p)$) circle[radius=1.5pt]; }
				\circuitMultiGate[\flatgatesize][gencolbg][gencolfg]{}{-0}{-1}{}
                    \circuitAdvanceXBy{\flatgateadvance}
				\circuitMultiGate[\flatgatesize][entcolbg][entcolfg]{}{-1}{-2}{}
                    \circuitAdvanceXBy{\flatgateadvance}
				\circuitMultiGate[\flatgatesize][gencolbg][gencolfg]{}{-2}{-3}{}
                    \circuitAdvanceXBy{\flatgateadvance}
				\circuitMultiGate[\flatgatesize][gencolbg][gencolfg]{}{-1}{-2}{}
				\circuitMultiGate[\flatgatesize][gencolbg][gencolfg]{}{-3}{-4}{}
                    \circuitAdvanceXBy{\flatgateadvance}
				\circuitMultiGate[\flatgatesize][gencolbg][gencolfg]{}{-2}{-3}{}
			\end{scope}
        \end{tikzpicture}
}

\newcommand{\drawAbsorbProofCaseG}{

        \begin{tikzpicture}[scale=0.76]
            \begin{scope}[xshift=0cm, rotate=90]
				\circuitInitXAt{0.3}
				\foreach\p in {0,1,2,3}{
					\circuitQubitLine{0}{2.}{-\p} 
					\draw[fill] ($0.5*(0, -\p)$) circle[radius=1.5pt]; } \circuitQubitLine{0}{2}{-4}
				\circuitMultiGate[\flatgatesize][gencolbg][gencolfg]{}{-0}{-1}{}
                    \circuitAdvanceXBy{\flatgateadvance}
				\circuitMultiGate[\flatgatesize][entcolbg][entcolfg]{}{-1}{-2}{}
                    \circuitAdvanceXBy{\flatgateadvance}
				\circuitMultiGate[\flatgatesize][gencolbg][gencolfg]{}{-2}{-3}{}
                    \circuitAdvanceXBy{\flatgateadvance}
				\circuitMultiGate[\flatgatesize][gencolbg][gencolfg]{}{-1}{-2}{}
				\circuitMultiGate[\flatgatesize][gencolbg][gencolfg]{}{-3}{-4}{}
                    \circuitAdvanceXBy{\flatgateadvance}
				\circuitMultiGate[\flatgatesize][gencolbg][gencolfg]{}{-2}{-3}{}
                    \circuitAdvanceXBy{\flatgateadvance}
				\circuitMultiGate[\flatgatesize][gencolbg][gencolfg]{}{-3}{-4}{}
			\end{scope}
            \draw[line width=1pt, -{>[length=6pt, width=6pt]}] (2.4,1) -- (3.1,1);
            \begin{scope}[xshift=3.5cm, rotate=90]
				\circuitInitXAt{0.3}
				\foreach\p in {0,1,2,3}{
					\circuitQubitLine{0}{2.}{-\p} 
					\draw[fill] ($0.5*(0, -\p)$) circle[radius=1.5pt]; } \circuitQubitLine{0}{2}{-4}
				\circuitMultiGate[\flatgatesize][gencolbg][gencolfg]{}{-0}{-1}{}
                    \circuitAdvanceXBy{\flatgateadvance}
				\circuitMultiGate[\flatgatesize][gencolbg][gencolfg]{}{-2}{-3}{}
                    \circuitAdvanceXBy{\flatgateadvance}
				\circuitMultiGate[\flatgatesize][gencolbg][gencolfg]{}{-1}{-2}{}
				\circuitMultiGate[\flatgatesize][gencolbg][gencolfg]{}{-3}{-4}{}
                    \circuitAdvanceXBy{\flatgateadvance}
				\circuitMultiGate[\flatgatesize][gencolbg][gencolfg]{}{-2}{-3}{}
                    \circuitAdvanceXBy{\flatgateadvance}
				\circuitMultiGate[\flatgatesize][entcolbg][entcolfg]{}{-3}{-4}{}
                    \circuitAdvanceXBy{\flatgateadvance}
				\circuitMultiGate[\flatgatesize][gencolbg][gencolfg]{}{-3}{-4}{}
			\end{scope}
            \draw[line width=1pt, -{>[length=6pt, width=6pt]}] (5.9,1) -- (6.6,1);
            \begin{scope}[xshift=7cm, rotate=90]
				\circuitInitXAt{0.3}
				\foreach\p in {0,1,2,3}{
					\circuitQubitLine{0}{2.}{-\p} 
					\draw[fill] ($0.5*(0, -\p)$) circle[radius=1.5pt]; } \circuitQubitLine{0}{2}{-4}
				\circuitMultiGate[\flatgatesize][gencolbg][gencolfg]{}{-0}{-1}{}
                    \circuitAdvanceXBy{\flatgateadvance}
				\circuitMultiGate[\flatgatesize][gencolbg][gencolfg]{}{-2}{-3}{}
                    \circuitAdvanceXBy{\flatgateadvance}
				\circuitMultiGate[\flatgatesize][gencolbg][gencolfg]{}{-1}{-2}{}
				\circuitMultiGate[\flatgatesize][gencolbg][gencolfg]{}{-3}{-4}{}
                    \circuitAdvanceXBy{\flatgateadvance}
				\circuitMultiGate[\flatgatesize][gencolbg][gencolfg]{}{-2}{-3}{}
                    \circuitAdvanceXBy{\flatgateadvance}
				\circuitMultiGate[\flatgatesize][gencolbg][gencolfg]{}{-3}{-4}{}
			\end{scope}
        \end{tikzpicture}

}

\newcommand{\drawAbsorbProofCaseH}{

        \begin{tikzpicture}[scale=0.76]
            \begin{scope}[xshift=0cm, rotate=90]
				\circuitInitXAt{0.3}
				\foreach\p in {0,1,2,3}{
					\circuitQubitLine{0}{1.7}{-\p} 
					\draw[fill] ($0.5*(0, -\p)$) circle[radius=1.5pt]; } \circuitQubitLine{0}{1.7}{-4}\circuitQubitLine{0}{1.7}{-5}
				\circuitMultiGate[\flatgatesize][gencolbg][gencolfg]{}{-0}{-1}{}
                    \circuitAdvanceXBy{\flatgateadvance}
				\circuitMultiGate[\flatgatesize][entcolbg][entcolfg]{}{-1}{-2}{}
                    \circuitAdvanceXBy{\flatgateadvance}
				\circuitMultiGate[\flatgatesize][gencolbg][gencolfg]{}{-2}{-3}{}
                    \circuitAdvanceXBy{\flatgateadvance}
				\circuitMultiGate[\flatgatesize][gencolbg][gencolfg]{}{-1}{-2}{}
				\circuitMultiGate[\flatgatesize][gencolbg][gencolfg]{}{-3}{-4}{}
                    \circuitAdvanceXBy{\flatgateadvance}
				\circuitMultiGate[\flatgatesize][gencolbg][gencolfg]{}{-2}{-3}{}
				\circuitMultiGate[\flatgatesize][gencolbg][gencolfg]{}{-4}{-5}{}
			\end{scope}
            \draw[line width=1pt, -{>[length=6pt, width=6pt]}] (3,0.85) -- (4,0.85);
            \begin{scope}[xshift=4.5cm, rotate=90]
				\circuitInitXAt{0.3}
				\foreach\p in {0,1,2,3}{
					\circuitQubitLine{0}{1.7}{-\p} 
					\draw[fill] ($0.5*(0, -\p)$) circle[radius=1.5pt]; } \circuitQubitLine{0}{1.7}{-4}\circuitQubitLine{0}{1.7}{-5}
				\circuitMultiGate[\flatgatesize][gencolbg][gencolfg]{}{-0}{-1}{}
				\circuitMultiGate[\flatgatesize][gencolbg][gencolfg]{}{-2}{-3}{}
                    \circuitAdvanceXBy{\flatgateadvance}
				\circuitMultiGate[\flatgatesize][gencolbg][gencolfg]{}{-1}{-2}{}
				\circuitMultiGate[\flatgatesize][gencolbg][gencolfg]{}{-3}{-4}{}
                    \circuitAdvanceXBy{\flatgateadvance}
				\circuitMultiGate[\flatgatesize][gencolbg][gencolfg]{}{-2}{-3}{}
                    \circuitAdvanceXBy{\flatgateadvance}
				\circuitMultiGate[\flatgatesize][entcolbg][entcolfg]{}{-3}{-4}{}
                    \circuitAdvanceXBy{\flatgateadvance}
				\circuitMultiGate[\flatgatesize][gencolbg][gencolfg]{}{-4}{-5}{}
			\end{scope}
        \end{tikzpicture}

}

\newcommand{\drawAbsorbProofCaseI}{

        \begin{tikzpicture}[scale=0.76]
        	\begin{scope}[rotate=90]
				\circuitInitXAt{0.3}
				\foreach\p in {-1,0,1,2,3}{
					\circuitQubitLine{0}{1.4}{-\p} }
                    \draw[fill] (0,0.5) circle[radius=1.5pt]; 
                    \draw[fill] (0,0) circle[radius=1.5pt]; 
                    \draw[fill] (0,-0.5) circle[radius=1.5pt]; 
				\circuitMultiGate[\flatgatesize][gencolbg][gencolfg]{}{-0}{-1}{}
                    \circuitAdvanceXBy{\flatgateadvance}
				\circuitMultiGate[\flatgatesize][gencolbg][gencolfg]{}{-1}{-2}{}
                    \circuitAdvanceXBy{\flatgateadvance}
				\circuitMultiGate[\flatgatesize][gencolbg][gencolfg]{}{-2}{-3}{}
                    \circuitAdvanceXBy{\flatgateadvance}
				\circuitMultiGate[\flatgatesize][entcolbg][entcolfg]{}{1}{0}{}
			\end{scope}
            \draw[line width=1pt, -{>[length=6pt, width=6pt]}] (2,0.7) -- (3,0.7);
            \begin{scope}[xshift=4cm, rotate=90]
				\circuitInitXAt{0.3}
				\foreach\p in {-1,0,1,2,3}{
					\circuitQubitLine{0}{1.4}{-\p} }
                    \draw[fill] (0,0.5) circle[radius=1.5pt]; 
                    \draw[fill] (0,0) circle[radius=1.5pt]; 
                    \draw[fill] (0,-0.5) circle[radius=1.5pt]; 
				\circuitMultiGate[\flatgatesize][entcolbg][entcolfg]{}{1}{0}{}
                    \circuitAdvanceXBy{\flatgateadvance}
				\circuitMultiGate[\flatgatesize][gencolbg][gencolfg]{}{0}{-1}{}
                    \circuitAdvanceXBy{\flatgateadvance}
				\circuitMultiGate[\flatgatesize][gencolbg][gencolfg]{}{-1}{-2}{}
                    \circuitAdvanceXBy{\flatgateadvance}
				\circuitMultiGate[\flatgatesize][gencolbg][gencolfg]{}{-2}{-3}{}
			\end{scope}
        \end{tikzpicture}
}

\newcommand{\drawAbsorbProofCaseJ}{

        \begin{tikzpicture}[scale=0.76]
        	\begin{scope}[rotate=90]
				\circuitInitXAt{0.3}
				\foreach\p in {-2,-1,0,1,2,3}{
					\circuitQubitLine{0}{1.1}{-\p}
					 }
                     \draw[fill] ($0.5*(0, 2)$) circle[radius=1.5pt];
                     \draw[fill] ($0.5*(0, 1)$) circle[radius=1.5pt];
                     \draw[fill] ($0.5*(0, 0)$) circle[radius=1.5pt];
                     \draw[fill] ($0.5*(0, -1)$) circle[radius=1.5pt];
				\circuitMultiGate[\flatgatesize][entcolbg][entcolfg]{}{1}{2}{}
				\circuitMultiGate[\flatgatesize][gencolbg][gencolfg]{}{-0}{-1}{}
                    \circuitAdvanceXBy{\flatgateadvance}
				\circuitMultiGate[\flatgatesize][gencolbg][gencolfg]{}{-1}{-2}{}
                    \circuitAdvanceXBy{\flatgateadvance}
				\circuitMultiGate[\flatgatesize][gencolbg][gencolfg]{}{-2}{-3}{}
			\end{scope}
        \end{tikzpicture}

}

\newcommand{\drawDisentanglingExampleI}{

        \begin{tikzpicture}[scale=0.76]
        	\begin{scope}[rotate=90]
				\circuitInitXAt{0.3}
				\foreach\p in {0,1,2,3}{
					\circuitQubitLine{0}{1.4}{-\p}
					 }
                     \draw[fill] ($0.5*(0, -3)$) circle[radius=1.5pt];
                     \draw[fill] ($0.5*(0, -2)$) circle[radius=1.5pt];
                     \draw[fill] ($0.5*(0, 0)$) circle[radius=1.5pt];
                     \draw[fill] ($0.5*(0, -1)$) circle[radius=1.5pt];
				\circuitMultiGate[\flatgatesize][gencolbg][gencolfg]{}{-2}{-3}{}
				\circuitMultiGate[\flatgatesize][gencolbg][gencolfg]{}{-0}{-1}{}
                    \circuitAdvanceXBy{\flatgateadvance}
				\circuitMultiGate[\flatgatesize][gencolbg][gencolfg]{}{-1}{-2}{}
                    \circuitAdvanceXBy{\flatgateadvance}
				\circuitMultiGate[\flatgatesize][gencolbg][gencolfg]{}{-2}{-3}{}

                    \circuitAdvanceXBy{\flatgateadvance}

                \circuitMultiGate[\flatgatesize][discolbg][discolfg]{}{-0}{-1}{}
                
			\end{scope}
        \end{tikzpicture}

}

\newcommand{\drawDisentanglingExampleII}{

        \begin{tikzpicture}[scale=0.76]
        	\begin{scope}[rotate=90]
				\circuitInitXAt{0.3}
				\foreach\p in {0,1,2,3}{
					\circuitQubitLine{0}{0.9}{-\p}
					 }
                     \draw[fill] ($0.5*(0, -3)$) circle[radius=1.5pt];
                     \draw[fill] ($0.5*(0, -2)$) circle[radius=1.5pt];
                     \draw[fill] ($0.5*(0, 0)$) circle[radius=1.5pt];
                     \draw[fill] ($0.5*(0, -1)$) circle[radius=1.5pt];
				\circuitMultiGate[\flatgatesize][gencolbg][gencolfg]{}{-2}{-3}{}
				\circuitMultiGate[\flatgatesize][gencolbg][gencolfg]{}{-0}{-1}{}
                    \circuitAdvanceXBy{\flatgateadvance}
				\circuitMultiGate[\flatgatesize][gencolbg][gencolfg]{}{-1}{-2}{}

			\end{scope}
        \end{tikzpicture}

}

\newcommand{\drawLoopFigure}{

        \begin{tikzpicture}[scale=0.76]
        	\begin{scope}[rotate=-90]
                \def\linew{1.3pt}
                \def\dotr{2.5pt}
                \def\coordxa{1}
                \def\coordxb{1.7}
                \def\coordxc{2.7}
                \def\coordxd{3.4}
                \def\coordxe{4.4}
                \def\coordxf{5.1}
                
                \def\coordgyas{0.2}
                \def\coordgyae{2}
                \def\coordgybs{2.2}
                \def\coordgybe{4}
                \def\coordgycs{4.2}
                \def\coordgyce{6}
                \def\coordgyds{6.2}
                \def\coordgyde{8}
                \def\coordgyes{8.2}

                \colorlet{paira}{black!20}
                \colorlet{pairb}{black!50}
                \colorlet{pairc}{black}

                \def\gatedw{1.3pt}
                \def\gatedc{5pt}
                \def\gatexe{0.3}
                \draw[blue!60, fill=blue!10, line width=\gatedw, rounded corners=\gatedc] %
                        ($(\coordxa, \coordgyas) - (\gatexe,0)$) rectangle 
                        ($(\coordxd, \coordgyae) + (\gatexe,0)$);
                \draw[blue!60, fill=blue!10, line width=\gatedw, rounded corners=\gatedc] %
                        ($(\coordxc, \coordgybs) - (\gatexe,0)$) rectangle 
                        ($(\coordxf, \coordgybe) + (\gatexe,0)$);
                \draw[discolfg, fill=discolbg, line width=\gatedw, rounded corners=\gatedc] %
                        ($(\coordxa, \coordgycs) - (\gatexe,0)$) rectangle 
                        ($(\coordxd, \coordgyce) + (\gatexe,0)$);
                \draw[discolfg, fill=discolbg, line width=\gatedw, rounded corners=\gatedc] %
                        ($(\coordxc, \coordgyds) - (\gatexe,0)$) rectangle 
                        ($(\coordxf, \coordgyde) + (\gatexe,0)$);

                     \draw[paira, fill] (\coordxa,0) circle[radius=\dotr];
                     \draw[paira, fill] (\coordxb,0) circle[radius=\dotr];
                     \draw[pairb, fill] (\coordxc,0) circle[radius=\dotr];
                     \draw[pairb, fill] (\coordxd,0) circle[radius=\dotr];
                     \draw[pairc, fill] (\coordxe,0) circle[radius=\dotr];
                     \draw[pairc, fill] (\coordxf,0) circle[radius=\dotr];

                \draw[paira, line width=\linew] (\coordxa, 0) -- 
                        (\coordxa, \coordgyas) to[out=90,in=-90] (\coordxb, \coordgyae) -- %
                        (\coordxb, \coordgybs) to[out=90,in=-90] (\coordxb, \coordgybe) -- %
                        (\coordxb, \coordgycs) to[out=90,in=-90] (\coordxd, \coordgyce) -- %
                        (\coordxd, \coordgyds) to[out=90,in=-90] (\coordxd, \coordgyde) -- %
                        (\coordxd, \coordgyes);
                \draw[paira, line width=\linew] (\coordxb, 0) -- 
                        (\coordxb, \coordgyas) to[out=90,in=-90] (\coordxc, \coordgyae) -- %
                        (\coordxc, \coordgybs) to[out=90,in=-90] (\coordxe, \coordgybe) -- %
                        (\coordxe, \coordgycs) to[out=90,in=-90] (\coordxe, \coordgyce) -- %
                        (\coordxe, \coordgyds) to[out=90,in=-90] (\coordxc, \coordgyde) -- %
                        (\coordxc, \coordgyes);
                \draw[pairb, line width=\linew] (\coordxc, 0) -- 
                        (\coordxc, \coordgyas) to[out=90,in=-90] (\coordxa, \coordgyae) -- %
                        (\coordxa, \coordgybs) to[out=90,in=-90] (\coordxa, \coordgybe) -- %
                        (\coordxa, \coordgycs) to[out=90,in=-90] (\coordxc, \coordgyce) -- %
                        (\coordxc, \coordgyds) to[out=90,in=-90] (\coordxe, \coordgyde) -- %
                        (\coordxe, \coordgyes);
                \draw[pairb, line width=\linew] (\coordxd, 0) -- 
                        (\coordxd, \coordgyas) to[out=90,in=-90] (\coordxd, \coordgyae) -- %
                        (\coordxd, \coordgybs) to[out=90,in=-90] (\coordxf, \coordgybe) -- %
                        (\coordxf, \coordgycs) to[out=90,in=-90] (\coordxf, \coordgyce) -- %
                        (\coordxf, \coordgyds) to[out=90,in=-90] (\coordxf, \coordgyde) -- %
                        (\coordxf, \coordgyes);
                \draw[pairc, line width=\linew] (\coordxe, 0) -- 
                        (\coordxe, \coordgyas) to[out=90,in=-90] (\coordxe, \coordgyae) -- %
                        (\coordxe, \coordgybs) to[out=90,in=-90] (\coordxc, \coordgybe) -- %
                        (\coordxc, \coordgycs) to[out=90,in=-90] (\coordxa, \coordgyce) -- %
                        (\coordxa, \coordgyds) to[out=90,in=-90] (\coordxa, \coordgyde) -- %
                        (\coordxa, \coordgyes);
                \draw[pairc, line width=\linew] (\coordxf, 0) -- 
                        (\coordxf, \coordgyas) to[out=90,in=-90] (\coordxf, \coordgyae) -- %
                        (\coordxf, \coordgybs) to[out=90,in=-90] (\coordxd, \coordgybe) -- %
                        (\coordxd, \coordgycs) to[out=90,in=-90] (\coordxb, \coordgyce) -- %
                        (\coordxb, \coordgyds) to[out=90,in=-90] (\coordxb, \coordgyde) -- %
                        (\coordxb, \coordgyes);

                \draw[paira, line width=\linew] (\coordxa, 0) to[out=-90, in=-90] (\coordxb,0);
                \draw[pairb, line width=\linew] (\coordxc, 0) to[out=-90, in=-90] (\coordxd,0);
                \draw[pairc, line width=\linew] (\coordxe, 0) to[out=-90, in=-90] (\coordxf,0);

			\end{scope}
        \end{tikzpicture}

}

\newcommand{\drawFourQubitRSF}{

        \begin{tikzpicture}[scale=0.76]
        	\begin{scope}[rotate=90]
				\circuitInitXAt{0.3}
				\foreach\p in {0,1,2,3}{
					\circuitQubitLine{0}{1.4}{-\p}
					 }
                     \draw[fill] ($0.5*(0, -3)$) circle[radius=1.5pt];
                     \draw[fill] ($0.5*(0, -2)$) circle[radius=1.5pt];
                     \draw[fill] ($0.5*(0, 0)$) circle[radius=1.5pt];
                     \draw[fill] ($0.5*(0, -1)$) circle[radius=1.5pt];
				\circuitMultiGate[\flatgatesize][gencolbg][gencolfg]{}{-2}{-3}{}
				\circuitMultiGate[\flatgatesize][gencolbg][gencolfg]{}{-0}{-1}{}
                    \circuitAdvanceXBy{\flatgateadvance}
				\circuitMultiGate[\flatgatesize][gencolbg][gencolfg]{}{-1}{-2}{}
                    \circuitAdvanceXBy{\flatgateadvance}
				\circuitMultiGate[\flatgatesize][gencolbg][gencolfg]{}{-2}{-3}{}

			\end{scope}
        \end{tikzpicture}

}

\newcommand{\Bellpairmodelexample}{
    \begin{tikzpicture}[scale=0.76]
\begin{scope}[yshift = 4.5cm, xshift=0cm]
			\begin{scope}[xshift = 0.95cm, yshift=-0.6cm, rotate=90]
				\circuitInitXAt{-2.3}
				\foreach\p in {0,1,2,3,4,5,6,7}{
					\circuitQubitLine{-2.6}{-1.1}{-\p}
					\draw[fill] ($0.5*(-5.2, -\p)$) circle[radius=1.5pt];
					}
				\circuitMultiGate[\flatgatesize][gencolbg][gencolfg]{}{-0}{-1}{}
				\circuitMultiGate[\flatgatesize][gencolbg][gencolfg]{}{-2}{-3}{}
				\circuitMultiGate[\flatgatesize][gencolbg][gencolfg]{}{-4}{-5}{}
				{}\circuitAdvanceXBy{\flatgateadvance}
				\circuitMultiGate[\flatgatesize][gencolbg][gencolfg]{}{-1}{-2}{}
	
				\circuitMultiGate[\flatgatesize][gencolbg][gencolfg]{}{-5}{-6}{}\circuitAdvanceXBy{\flatgateadvance}
				\circuitMultiGate[\flatgatesize][gencolbg][gencolfg]{}{-2}{-3}{}
				{}\circuitAdvanceXBy{\flatgateadvance}
				\circuitMultiGate[\flatgatesize][gencolbg][gencolfg]{}{-3}{-4}{}

			\end{scope}
		\end{scope}
    \end{tikzpicture}
}

\newcommand{\drawIncompleteRSFExampleACircuitOnly}{

			\begin{scope}[rotate=90]
				\circuitInitXAt{-2.3}
				\foreach\p in {0,1,2,3,4,5,6,7}{
					\circuitQubitLine{-2.6}{-0.8}{-\p}
					}
				\circuitMultiGate[\flatgatesize][gencolbg][gencolfg]{}{-0}{-1}{}
				\circuitMultiGate[\flatgatesize][gencolbg][gencolfg]{}{-3}{-4}{}
				\circuitMultiGate[\flatgatesize][gencolbg][gencolfg]{}{-5}{-6}{}
				{}\circuitAdvanceXBy{\flatgateadvance}
				\circuitMultiGate[\flatgatesize][gencolbg][gencolfg]{}{-1}{-2}{}
	
				\circuitMultiGate[\flatgatesize][gencolbg][gencolfg]{}{-4}{-5}{}
                \circuitAdvanceXBy{\flatgateadvance}
				\circuitMultiGate[\flatgatesize][gencolbg][gencolfg]{}{-2}{-3}{}
				\circuitMultiGate[\flatgatesize][gencolbg][gencolfg]{}{-5}{-6}{}
				{}\circuitAdvanceXBy{\flatgateadvance}
				\circuitMultiGate[\flatgatesize][gencolbg][gencolfg]{}{-3}{-4}{}
				\circuitMultiGate[\flatgatesize][gencolbg][gencolfg]{}{-6}{-7}{}
				{}\circuitAdvanceXBy{\flatgateadvance}
				\circuitMultiGate[\flatgatesize][gencolbg][gencolfg]{}{-4}{-5}{}

			\end{scope}
}

\newcommand{\drawIncompleteRSFExampleALabeled}{
        \begin{tikzpicture}[scale=2]
            \def\flatgatesize{0.225cm}
            \def\circuitRoundedCorners{5pt}
            \draw[black!15!blue!15,fill=black!15!blue!15, line width=2pt, rounded corners=1pt] (-0.1, -2.3) -- (0.6,-2.3) -- (2.6, -1.1) -- (1.9,-1.1) -- cycle;

            \draw[black!15!blue!15,fill=black!15!blue!15, line width=2pt, rounded corners=1pt] (1.4, -2.3) -- (2.1,-2.3) -- (3.6, -1.4) -- (2.9,-1.4) -- cycle;

            \draw[black!15!blue!15,fill=black!15!blue!15, line width=2pt, rounded corners=1pt] (2.4, -2.4) -- (3.1,-2.4) -- (3.1,-2.2) -- (2.4, -2.2) -- cycle;
            \drawIncompleteRSFExampleACircuitOnly
            \node at (0.28,-2.7) {$D^{(1)}$};
            \node at (1.78,-2.7) {$D^{(2)}$};
            \node at (2.78,-2.7) {$D^{(3)}$};

            \node at (0.28, -2.3) {\small $U^{(11)}$};
            \node at ($(0.78, -2.3) + 1*(0, 0.3)$) {\small $U^{(12)}$};
            \node at ($(1.28, -2.3) + 2*(0, 0.3)$) {\small $U^{(13)}$};
            \node at ($(1.78, -2.3) + 3*(0, 0.3)$) {\small $U^{(14)}$};
            \node at ($(2.28, -2.3) + 4*(0, 0.3)$) {\small $U^{(15)}$};

            \node at (1.78, -2.3) {\small $U^{(21)}$};
            \node at ($(2.28, -2.3) + 1*(0, 0.3)$) {\small $U^{(22)}$};
            \node at ($(2.78, -2.3) + 2*(0, 0.3)$) {\small $U^{(23)}$};
            \node at ($(3.28, -2.3) + 3*(0, 0.3)$) {\small $U^{(24)}$};
            
            \node at (2.78, -2.3) {\small $U^{(31)}$};
        \end{tikzpicture}
}

\newcommand{\drawIncompleteRSFExampleAMainText}{
        \begin{tikzpicture}[scale=0.76]
        \begin{scope}[yshift = 4.5cm, xshift=0cm]
                \draw[black!15!blue!15,fill=black!15!blue!15, line width=2pt, rounded corners=1pt] (-0.1, -2.3) -- (0.6,-2.3) -- (2.6, -1.1) -- (1.9,-1.1) -- cycle;

                \draw[black!15!blue!15,fill=black!15!blue!15, line width=2pt, rounded corners=1pt] (1.4, -2.3) -- (2.1,-2.3) -- (3.6, -1.4) -- (2.9,-1.4) -- cycle;

                \draw[black!15!blue!15,fill=black!15!blue!15, line width=2pt, rounded corners=1pt] (2.4, -2.4) -- (3.1,-2.4) -- (3.1,-2.2) -- (2.4, -2.2) -- cycle;
            	\begin{scope}[rotate=90]
				\circuitInitXAt{-2.3}
				\foreach\p in {0,1,2,3,4,5,6,7}{
					\circuitQubitLine{-2.6}{-0.8}{-\p}
					}
				\circuitMultiGate[\flatgatesize][gencolbg][gencolfg]{}{-0}{-1}{}
				\circuitMultiGate[\flatgatesize][gencolbg][gencolfg]{}{-3}{-4}{}
				\circuitMultiGate[\flatgatesize][gencolbg][gencolfg]{}{-5}{-6}{}
				{}\circuitAdvanceXBy{\flatgateadvance}
				\circuitMultiGate[\flatgatesize][gencolbg][gencolfg]{}{-1}{-2}{}
	
				\circuitMultiGate[\flatgatesize][gencolbg][gencolfg]{}{-4}{-5}{}
                \circuitAdvanceXBy{\flatgateadvance}
				\circuitMultiGate[\flatgatesize][gencolbg][gencolfg]{}{-2}{-3}{}
				\circuitMultiGate[\flatgatesize][gencolbg][gencolfg]{}{-5}{-6}{}
				{}\circuitAdvanceXBy{\flatgateadvance}
				\circuitMultiGate[\flatgatesize][gencolbg][gencolfg]{}{-3}{-4}{}
				\circuitMultiGate[\flatgatesize][gencolbg][gencolfg]{}{-6}{-7}{}
				{}\circuitAdvanceXBy{\flatgateadvance}
				\circuitMultiGate[\flatgatesize][gencolbg][gencolfg]{}{-4}{-5}{}

			\end{scope}
		\end{scope}
        \end{tikzpicture}
}

%% file: figures.tex
\newcommand\circuitResetGlobalParameters{

        \definecolor{circuitcolorbg}{RGB}{255,255,255}
        \definecolor{circuitcolorfg}{RGB}{0,0,0}

        \def\circuitlinewidth{1pt}
        \def\circuitCNOTControl{2.2pt}
        \def\circuitCNOTTarget{3.7pt}
        \def\circuitRoundedCorners{2pt}
        \def\circuitMultiLabelMargin{0.1}
        \def\circuitMultiLabelLineWidth{0.5pt}
        \def\circuitMeasureInnerLineWidth{0.7pt}

        \def\circuitlinespacing{0.5}

        \def\circuitGateStandardWidth{\circuitlinespacing*1.2} %
        \def\circuitGateHeight{\circuitlinespacing*0.8} %
        \def\circuitMultiGateStandardWidth{\circuitlinespacing*2.2}
        \def\circuitMultiGateHeight{\circuitlinespacing*0.4} %
        \def\circuitStandardGateDistance{\circuitlinespacing*0.4}
        \def\circuitMeasureSize{0.4}
        \def\circuitMeasureArcOffset{0.85} %
        \def\circuitMeasureArrowLength{1.05} %
        \def\circuitMeasureArrowAngle{37}
        \def\circuitMeasureArrowHeadLength{4pt}
        \def\circuitMeasureArrowHeadWidth{3pt}
}
\circuitResetGlobalParameters %

\newcommand\circuitInitXAt[1]{\coordinate (circuitTheX) at (#1,0);}

\NewDocumentCommand\circuitAdvanceXBy{O{0} m}{ %
	\coordinate (circuitTheX) at ($(circuitTheX) + (#2, 0)$);
	\ifthenelse{\isempty{#1} \OR #1=0}{}{\coordinate (circuitTheX) at ($(circuitTheX) + (\circuitStandardGateDistance, 0)$);}
}

\NewDocumentCommand\circuitCNOT{O{circuitcolorfg} O{myCenterNode} m m m}{ 
	\ifthenelse{\isempty{#1}}{\def\mycolorfg{circuitcolorfg}}{\def\mycolorfg{#1}}
	\ifthenelse{\isempty{#3}}{\coordinate (myXCoord) at (circuitTheX);}{\coordinate (myXCoord) at (#3, 0);}

	\ifthenelse{\isempty{#2}}{\def\mynodename{myCenterCoord}}{\def\mynodename{#2}}
	
	\coordinate (myControlCoord) at ($(myXCoord) + (0, #4*\circuitlinespacing)$);
	\coordinate (myTargetCoord) at ($(myXCoord) + (0, #5*\circuitlinespacing)$);

	\coordinate (\mynodename) at ($0.5*(myControlCoord) + 0.5*(myTargetCoord)$);
	
	\draw[draw=\mycolorfg, fill=\mycolorfg] (myControlCoord) circle (\circuitCNOTControl); %
	\draw[line width=\circuitlinewidth, draw=\mycolorfg] (myTargetCoord) circle (\circuitCNOTTarget); %
	\draw[line width=\circuitlinewidth, draw=\mycolorfg] (myControlCoord) -- (myTargetCoord); %
	\draw[line width=\circuitlinewidth, draw=\mycolorfg] ($(myTargetCoord) + (0,-\circuitCNOTTarget)$) -- ($(myTargetCoord) + (0,\circuitCNOTTarget)$); 
	\draw[line width=\circuitlinewidth, draw=\mycolorfg] ($(myTargetCoord) + (-\circuitCNOTTarget,0)$) -- ($(myTargetCoord) + (\circuitCNOTTarget,0)$);
}

\NewDocumentCommand\circuitCZ{O{circuitcolorfg} O{myCenterNode} m m m}{ 
	\ifthenelse{\isempty{#1}}{\def\mycolorfg{circuitcolorfg}}{\def\mycolorfg{#1}}
	\ifthenelse{\isempty{#3}}{\coordinate (myXCoord) at (circuitTheX);}{\coordinate (myXCoord) at (#3, 0);}

	\ifthenelse{\isempty{#2}}{\def\mynodename{myCenterCoord}}{\def\mynodename{#2}}
	
	\coordinate (myControlCoord) at ($(myXCoord) + (0, #4*\circuitlinespacing)$);
	\coordinate (myTargetCoord) at ($(myXCoord) + (0, #5*\circuitlinespacing)$);

	\coordinate (\mynodename) at ($0.5*(myControlCoord) + 0.5*(myTargetCoord)$);
	
	\draw[draw=\mycolorfg, fill=\mycolorfg] (myControlCoord) circle (\circuitCNOTControl); %
	\draw[draw=\mycolorfg, fill=\mycolorfg] (myTargetCoord) circle (\circuitCNOTControl); %
	\draw[line width=\circuitlinewidth, draw=\mycolorfg] (myControlCoord) -- (myTargetCoord); %
}

\NewDocumentCommand\circuitSingleGate{O{\circuitGateStandardWidth} O{circuitcolorbg} O{circuitcolorfg} O{myCenterCoord} m m m}{
	\ifthenelse{\isempty{#1}}{\def\mywidth{\circuitGateStandardWidth}}{\def\mywidth{#1}}
	\ifthenelse{\isempty{#2}}{\def\mycolorbg{circuitcolorbg}}{\def\mycolorbg{#2}}
	\ifthenelse{\isempty{#3}}{\def\mycolorfg{circuitcolorfg}}{\def\mycolorfg{#3}}
	\ifthenelse{\isempty{#5}}{\coordinate (myXCoord) at (circuitTheX);}{\coordinate (myXCoord) at (#5, 0);}
	\ifthenelse{\isempty{#4}}{\def\mynodename{myCenterCoord}}{\def\mynodename{#4}}
	
	\coordinate (\mynodename) at ($(myXCoord) + (0,#6*\circuitlinespacing)$);

	\draw[line width=\circuitlinewidth, fill=\mycolorbg, draw=\mycolorfg, rounded corners=\circuitRoundedCorners]
		($(\mynodename) - 0.5*(\mywidth, \circuitGateHeight)$) 
		rectangle ($(\mynodename) + 0.5*(\mywidth, \circuitGateHeight)$) node[midway, \mycolorfg] {#7};
}

\NewDocumentCommand\circuitMultiGate{O{\circuitMultiGateStandardWidth} O{circuitcolorbg} O{circuitcolorfg} O{myCenterCoord} m m m m}{
	\ifthenelse{\isempty{#1}}{\def\mywidth{\circuitMultiGateStandardWidth}}{\def\mywidth{#1}}
	\ifthenelse{\isempty{#2}}{\def\mycolorbg{circuitcolorbg}}{\def\mycolorbg{#2}}
	\ifthenelse{\isempty{#3}}{\def\mycolorfg{circuitcolorfg}}{\def\mycolorfg{#3}}
	\ifthenelse{\isempty{#4}}{\def\mynodename{myCenterCoord}}{\def\mynodename{#4}}
	\ifthenelse{\isempty{#5}}{\coordinate (myXCoord) at (circuitTheX);}{\coordinate (myXCoord) at (#5, 0);}
	
	\ifthenelse{#6 > #7}{\def \yUpper{#6*\circuitlinespacing} \def\yLower{#7*\circuitlinespacing}}{\def \yUpper{#7*\circuitlinespacing} \def\yLower{#6*\circuitlinespacing}}
	\coordinate (myLowerCoord) at ($(myXCoord) + (0,\yLower)$);
	\coordinate (myUpperCoord) at ($(myXCoord) + (0,\yUpper)$);
	
	\coordinate (\mynodename) at ($0.5*(myLowerCoord) + 0.5*(myUpperCoord)$);
	
	\draw[line width=\circuitlinewidth, fill=\mycolorbg, draw=\mycolorfg, rounded corners=\circuitRoundedCorners]
		($(myLowerCoord) - 0.5*(\mywidth, \circuitMultiGateHeight)$) 
		rectangle ($(myUpperCoord) + 0.5*(\mywidth, \circuitMultiGateHeight)$) node[midway, \mycolorfg] {#8};
	
}

\NewDocumentCommand\circuitMeasure{O{circuitcolorbg} O{circuitcolorfg} O{1,0} O{myCenterCoord} m m m}{
	\ifthenelse{\isempty{#1}}{\def\mycolorbg{circuitcolorbg}}{\def\mycolorbg{#1}}
	\ifthenelse{\isempty{#2}}{\def\mycolorfg{circuitcolorfg}}{\def\mycolorfg{#2}}
	\ifthenelse{\isempty{#3}}{\coordinate (myLabelOffset) at (\circuitMeasureSize,0);}{\coordinate (myLabelOffset) at ($\circuitMeasureSize*(#3)$);}
	\ifthenelse{\isempty{#5}}{\coordinate (myXCoord) at (circuitTheX);}{\coordinate (myXCoord) at (#5, 0);}
	\ifthenelse{\isempty{#4}}{\def\mynodename{myCenterCoord}}{\def\mynodename{#4}}
	
	\coordinate (\mynodename) at ($(myXCoord) + (0,#6*\circuitlinespacing)$);
	
	\coordinate (myLowerCoord) at ($(\mynodename) - 0.5*(\circuitMeasureSize,\circuitMeasureSize)$);
	
	\draw[line width=\circuitlinewidth, fill=\mycolorbg, draw=\mycolorfg] (myLowerCoord)
		rectangle ($(\mynodename) + 0.5*(\circuitMeasureSize,\circuitMeasureSize)$);
	
	\draw[line width=\circuitMeasureInnerLineWidth, draw=\mycolorfg] 
		($(myLowerCoord) + (0, \circuitMeasureSize*\circuitMeasureArcOffset)$) to[out=0, in=90]
		($(myLowerCoord) + (\circuitMeasureSize*\circuitMeasureArcOffset,0)$);
	
	\draw[-{Latex[length=\circuitMeasureArrowHeadLength, width=\circuitMeasureArrowHeadWidth]}, line width=\circuitMeasureInnerLineWidth, draw=\mycolorfg] 
		(myLowerCoord) -- +(\circuitMeasureArrowAngle:\circuitMeasureArrowLength*\circuitMeasureSize);
		
	\ifthenelse{\isempty{#7}}{}{\node[\mycolorfg] at ($(\mynodename) + (myLabelOffset)$) {#7};}

}

\NewDocumentCommand\circuitQubitLine{O{} O{} O{circuitcolorfg} m m m}{
	\ifthenelse{\isempty{#1}}{\def\myincommand{}}{\def\myincommand{node[left] {#1}}}
	\ifthenelse{\isempty{#2}}{\def\myoutcommand{}}{\def\myoutcommand{node[right] {#2}}}
	\ifthenelse{\isempty{#3}}{\def\mycolorfg{circuitcolorfg}}{\def\mycolorfg{#3}}
	
	\draw[line width=\circuitlinewidth, \mycolorfg] (#4, #6*\circuitlinespacing) \myincommand -- (#5,#6*\circuitlinespacing) \myoutcommand;
}

\NewDocumentCommand\circuitMultiInput{O{circuitcolorfg} m m m m}{
	\ifthenelse{\isempty{#1}}{\def\mycolorfg{circuitcolorfg}}{\def\mycolorfg{#1}}
	
	\ifthenelse{#3 > #4}{\def \yUpper{#3*\circuitlinespacing} \def\yLower{#4*\circuitlinespacing}}{\def \yUpper{#4*\circuitlinespacing} \def\yLower{#3*\circuitlinespacing}}
	
	\draw[line width=\circuitMultiLabelLineWidth, \mycolorfg] ($(#2, \yUpper) + (\circuitMultiLabelMargin, \circuitMultiLabelMargin)$)
		 -- ($(#2, \yUpper) + (- \circuitMultiLabelMargin, \circuitMultiLabelMargin)$)
		  -- node[left] {#5} ($(#2, \yLower) + (-\circuitMultiLabelMargin, -\circuitMultiLabelMargin)$)
		 -- ($(#2, \yLower) + (\circuitMultiLabelMargin, -\circuitMultiLabelMargin)$);
}

\NewDocumentCommand\circuitMultiOutput{O{circuitcolorfg} m m m m}{
	\ifthenelse{\isempty{#1}}{\def\mycolorfg{circuitcolorfg}}{\def\mycolorfg{#1}}
	
	\ifthenelse{#3 > #4}{\def \yUpper{#3*\circuitlinespacing} \def\yLower{#4*\circuitlinespacing}}{\def \yUpper{#4*\circuitlinespacing} \def\yLower{#3*\circuitlinespacing}}
	
	\draw[line width=\circuitMultiLabelLineWidth, \mycolorfg] ($(#2, \yUpper) + (-\circuitMultiLabelMargin, \circuitMultiLabelMargin)$)
		 -- ($(#2, \yUpper) + ( \circuitMultiLabelMargin, \circuitMultiLabelMargin)$)
		  -- node[right] {#5} ($(#2, \yLower) + (\circuitMultiLabelMargin, -\circuitMultiLabelMargin)$)
		 -- ($(#2, \yLower) + (-\circuitMultiLabelMargin, -\circuitMultiLabelMargin)$);
}